\documentclass[11pt]{article}
\usepackage[utf8]{inputenc}
\usepackage{textcomp}
\usepackage{amsfonts}
\usepackage{amsmath}
\usepackage{amssymb}
\usepackage{mathrsfs}
\usepackage{bbold}
\usepackage[margin=2.2cm]{geometry}
\usepackage{longtable}
\usepackage{lscape,graphicx}
\usepackage{graphics}
\usepackage{color}
\usepackage{array}
\usepackage{makecell}
\usepackage{bm}
\usepackage{comment}
\usepackage{multicol}
\usepackage{blindtext}
\usepackage{float}

\usepackage{hyperref}
\usepackage{cleveref}

\usepackage[numbers, sort&compress]{natbib}

\usepackage{slashed}

\usepackage{sidecap}
\usepackage{caption}
\usepackage{subcaption}

\newcommand{\ba}{\begin{array}{c}}
\newcommand{\ea}{\end{array}}

\newcommand{\be}{\beta}

\usepackage[dvipsnames]{xcolor}

\def\be{\begin{equation}}
\def\ee{\end{equation}}
\def\beq{\begin{equation}}
\def\eeq{\end{equation}}
\def\bc{\begin{center}}
\def\ec{\end{center}}
\def\bea{\begin{eqnarray}}
\def\eea{\end{eqnarray}}

\begin{document}
\begin{titlepage}
\vspace*{-1cm}
\phantom{hep-ph/***}
\flushright

\vskip 1.5cm
\begin{center}
\mathversion{bold}
{\LARGE\bf
Charged lepton flavour violation from inverse seesaw\\[0.05in] with flavour and CP symmetries
\mathversion{normal}
}
\vskip .3cm
\end{center}
\vskip 0.5  cm
\begin{center}
{\large F.~P.~Di Meglio} and
{\large C.~Hagedorn}
\\
\vskip .7cm
{\footnotesize
Instituto de F\'isica Corpuscular, Universidad de Valencia and CSIC,
Edificio Institutos Investigaci\'on, Catedr\'atico Jos\'e Beltr\'an 2, 46980 Paterna, Spain
\vskip .5cm
\begin{minipage}[l]{.9\textwidth}
\begin{center}
\textit{E-mail:}
\tt{francescopaolo.dimeglio@ific.uv.es}, \tt{claudia.hagedorn@ific.uv.es}
\end{center}
\end{minipage}
}
\end{center}
\vskip 1cm
\begin{abstract}
We study charged lepton flavour violation in a scenario in which light neutrino masses are generated via the inverse seesaw mechanism
with 3+3 gauge singlet fermions, $N_i$ and $S_j$, $i,j=1,2,3$.
Lepton mixing is predicted with the help of the flavour symmetries $\Delta (3 \, n^2)$ and $\Delta (6 \, n^2)$ combined with CP. In the neutral lepton sector, the non-trivial
flavour structure is only encoded in the Dirac neutrino Yukawa matrix (the coupling relating left-handed lepton doublets and gauge singlets $N_i$). 
Current experimental bounds on the processes
 $\mu \to e \, \gamma$, $\mu \to 3 \, e$, $\mu-e$ conversion in nuclei and the tau lepton decays $\tau \to \ell \, \gamma$ and $\tau \to 3 \, \ell$, $\ell=e, \mu$,  
 do not constrain the considered parameter space of this scenario.
Prospective limits on the decay $\mu\to 3 \, e$ and $\mu-e$ conversion in aluminium instead can markedly reduce the available parameter space. 
We also comment on the effects of the heavy sterile states on light neutrino masses and lepton mixing. 
\end{abstract}
\end{titlepage}
\setcounter{footnote}{0}

%%%%%%%%%%%%%%%%%%%%%%%%%%%%%%%%%%%%%%%%%%%%%%%%%%%%%%%
\section{Introduction}
\label{intro}
%%%%%%%%%%%%%%%%%%%%%%%%%%%%%%%%%%%%%%%%%%%%%%%%%%%%%%%

The Standard Model (SM) successfully describes many experimental results.
However, important observations such as the smallness of neutrino masses and the peculiar flavour structure in the lepton and quark sectors
remain unexplained.

Many extensions of the SM offer a generation mechanism for neutrino masses. Among these are the
well-known type-I~\cite{typeIseesaw1,typeIseesaw2,typeIseesaw3,typeIseesaw4,typeIseesaw5}, 
type-II~\cite{typeIIseesaw1,typeIIseesaw2,typeIIseesaw3,typeIIseesaw4,typeIIseesaw5,typeIIseesaw6} 
and type-III seesaw mechanisms~\cite{typeIIIseesaw} which typically require new particles with masses
much larger than the electroweak scale, if couplings are of order one (and no symmetry is implemented, see e.g.~\cite{Shaposhnikov:2006nn,Kersten:2007vk,Moffat:2017feq}). Alternatives such as the inverse seesaw (ISS)
mechanism~\cite{Mohapatra:1986bd,GonzalezGarcia:1988rw,Mohapatra:1986aw,Bernabeu:1987gr} can reconcile new particles with masses in the TeV-range with not too small couplings, while
the smallness of neutrino masses is due to a small breaking of lepton number (in the form of a Majorana mass term for the gauge singlets). 

The organising principle of the flavour sector responsible for the observed pattern of fermion masses and mixing is yet to be found.
Nevertheless, discrete flavour symmetries, especially when combined with CP, that are broken in a certain way among e.g.~charged leptons and 
neutral states are promising candidates for correctly describing the observed lepton mixing angles and making testable predictions
for the leptonic CP phases~\cite{GfCP,GfCPothers1,GfCPothers2} (for earlier works see~\cite{GfCPearly1,GfCPearly2,GfCPearly3,GfCPearly4,GfCPearly5,GfCPearly6}); 
reviews can be found, for example, in~\cite{Gfreview1,Gfreview2,Gfreview3,Gfreview4,Gfreview_math}. 
In this work, we employ the series of discrete groups $\Delta (3 \, n^2)$
and $\Delta (6 \, n^2)$, $n$ integer,~\cite{Luhn:2007uq,Escobar:2008vc} and 
CP symmetries, which correspond to automorphisms of these~\cite{GfCP,GfCPothers1,GfCPothers2}.
As is known~\cite{Hagedorn:2014wha}, four different types of lepton mixing patterns are found, called Case 1), Case 2), Case 3 a) and Case 3 b.1),
if the flavour symmetry $G_f$ and CP are broken to a residual $Z_3$ symmetry among charged leptons and to $Z_2 \times CP$
among the neutral states. For further studies of lepton mixing from the groups $\Delta (3 \, n^2)$
and $\Delta (6 \, n^2)$ and CP, 
see e.g.~\cite{DeltaCPothers1,DeltaCPothers2,DeltaCPothers3,DeltaCPothers4,DeltaCPothers5,DeltaCPothers6,DeltaCPothers7,DeltaCPothers8,DeltaCPothers9,DeltaCPothers10,DeltaCPothers11}.

In the current study, we analyse a scenario in which light neutrino masses are generated by the ISS mechanism with 3+3 gauge singlet
fermions and the flavour symmetries $\Delta (3 \, n^2)$ and $\Delta (6 \, n^2)$ and CP are broken non-trivially.~\footnote{We do not specify the breaking mechanism here.}
Concretely, the non-trivial flavour structure among the neutral states is encoded in the Dirac neutrino Yukawa matrix, while
the mass matrices (Dirac and Majorana) of the gauge singlet fermions do neither break $G_f$ nor CP.
This possibility has been called option 2 in~\cite{Hagedorn:2021ldq}.
We scrutinise analytically and numerically the Pontecorvo-Maki-Nakagawa-Sakata (PMNS) mixing matrix 
and different charged lepton flavour violating (cLFV) processes for Case 1) through Case 3 b.1). The main findings are
that the effects of non-unitarity of the PMNS mixing matrix induced by the existence of the heavy sterile states have a more general form than 
those found in the scenario analysed in~\cite{Hagedorn:2021ldq} (called option 1) and that signals of $\mu-e$ transitions can be sizeable in
the studied parameter space, while those involving the tau lepton are more suppressed. 

The remainder of the paper is organised as follows: in section~\ref{sec2} we outline the scenario, show the form of the different mass and coupling matrices,
comment on the masses of the different neutral states, discuss the mixing matrices and the effects of the heavy sterile states on the light neutrinos.
Section~\ref{sec:leptonmixing} contains information on lepton mixing including the characteristics of the mixing patterns of Case 1) through Case 3 b.1) and 
 the effects of non-unitarity due to the mixing of the light neutrinos and the heavy sterile states.
In section~\ref{sec:muetrans} we present analytic estimates as well as numerical results for the branching ratios (BRs) $\mathrm{BR} (\mu\to e \gamma)$ and $\mathrm{BR} (\mu \to 3 \, e)$
and the $\mu-e$ conversion rate (CR) in aluminium, $\mathrm{CR} (\mu-e, \mathrm{Al})$, for examples of all cases, Case 1) through Case 3 b.1).
We comment on the expected size of the signals of the cLFV tau lepton decays $\tau \to \mu \gamma$, $\tau \to e \gamma$, $\tau \to 3 \, \mu$ and $\tau \to 3 \, e$ in section~\ref{sec:taudecays}
as well as on the possible impact of experimental constraints from the study of Tritium beta ($\beta$) decay and searches for neutrinoless double beta ($0\nu\beta\beta$) decay in different nuclei in section~\ref{beta0nubbdecay}.
 Eventually, we summarise in section~\ref{summ}.
In appendix~\ref{app:grouptheory} basics of the group theory of $\Delta (3 \, n^2)$ and $\Delta (6 \, n^2)$ can be found, while we collect the form of different relevant matrices in appendix~\ref{app:matrices}.
One can find information on the residual symmetries, the CP transformations and the mixing patterns for each case, Case 1) through Case 3 b.1), in appendix~\ref{app:residualsCPUPMNS}.
Appendix~\ref{app:numerics} comprises a concise description of the numerical scan and supplementary plots displaying results of the studied $\mu-e$ transitions for Case 2) and Case 3 b.1) are given in appendix~\ref{app:Case2Case3b1}. 

%%%%%%%%%%%%%%%%%%%%%%%%%%%%%%%%%%%%%%%%%%%%%%%%%%%%%%%
\section{Scenario}
\label{sec2}
%%%%%%%%%%%%%%%%%%%%%%%%%%%%%%%%%%%%%%%%%%%%%%%%%%%%%%%

In the following, we focus on the lepton sector. We add 3+3 gauge singlet fermions $N_i$ and $S_j$, $i,j=1,2,3$, 
to the SM in order to generate light neutrino masses via the ISS mechanism. 

The flavour structure of the scenario is controlled by a flavour and a CP symmetry that are assumed to be broken to different non-trivial subgroups in the charged
lepton, $G_e$, and in the neutral lepton sectors, $G_\nu$, respectively. As flavour symmetry we choose one of the groups $\Delta (3 \, n^2)$ and $\Delta (6 \, n^2)$ with $n \geq 2$, see appendix~\ref{app:grouptheory} for 
relevant details of these groups. The CP symmetry
 corresponds to an automorphism of the flavour group and is selected according to the findings of~\cite{Hagedorn:2014wha}.

Left-handed (LH) lepton doublets $L_\alpha$, $\alpha=e, \mu, \tau$, are assigned to a triplet ${\bf 3}$. This representation is taken to be irreducible, faithful and complex.
The gauge singlets $N_i$ and $S_j$ instead each transform as an irreducible, unfaithful, real triplet ${\bf 3^\prime}$. The existence of such a representation requires that 
the flavour group has an even index $n$. On the contrary, right-handed (RH) charged leptons $\ell_{\alpha R}$ are singlets, all in ${\bf 1}$, under the flavour group.
In order to distinguish these, we employ an additional $Z_3$ symmetry $Z_3^{\mathrm{(aux)}}$,
i.e.~$\ell_{e R} \sim 1$, $\ell_{\mu R} \sim \omega$ and  $\ell_{\tau R} \sim \omega^2$ with $\omega= e^{\frac{2 \, \pi i}{3}}$. LH lepton doublets and the gauge singlets
are neutral under this additional group. 

The group $G_e$ determines the structure of the charged lepton mass matrix $m_\ell$. We use a residual $Z_3$ symmetry 
that is a diagonal subgroup of the $Z_3$ symmetry, generated by the generator $a$ of the flavour group (which is represented by a diagonal matrix in the chosen basis, compare appendix~\ref{app:grouptheory}), and the additional $Z_3$ symmetry $Z_3^{\mathrm{(aux)}}$.
 Consequently, $m_\ell$ is diagonal and contains three free parameters corresponding to the masses of the electron, muon and tau lepton.\footnote{At the level of unbroken flavour and CP symmetries, the Yukawa couplings responsible for the 
 charged lepton masses are expected to be of the form e.g.~$\frac{y_e}{\Lambda} \, \overline{L}_\alpha \, H \, \phi_e \, \ell_{e R} $ with the phase of $y_e$ constrained by the imposed CP symmetry, $\Lambda$ being the cutoff scale of the theory 
 and $\phi_e$ an SM gauge singlet that transforms as $({\bf 3}, 1)$ under the flavour symmetry and $Z_3^{\mathrm{(aux)}}$. If this scalar acquires a VEV that preserves the residual symmetry $G_e$, i.e.~it has the form $\langle \phi_e \rangle = v_e \, \left(
\begin{array}{c}
1 \\ 0 \\ 0
\end{array}
\right)$ with $v_e$ in general complex, we obtain $\frac{y_e}{\Lambda} \, \overline{L}_e \, H \, v_e \, \ell_{e R}$ and after electroweak symmetry breaking the electron mass reads $m_e = \Big| \frac{y_e}{\Lambda} \, v_e \Big| \, \langle H \rangle$. 
The muon and tau lepton masses can be generated in an analogous way. For further information regarding this point, see, for example,~\cite{King:2011ab} and appendix D of~\cite{Hagedorn:2016lva}.} 
 We take these to be ordered canonically
  such that the contribution $U_\ell$ to lepton mixing is $U_\ell= \mathbb{1}$.

The masses of the neutral states arise from the terms\footnote{For simplicity, we do not take into account a possible coupling between LH lepton doublets and the gauge singlets $S_j$ nor do we consider a Majorana mass
term for the singlets $N_i$. Additional symmetries and/or an extension of the gauge group can motivate this choice.}
\begin{equation}
\label{eq:Lneutral}
- (y_D)_{\alpha i} \, \overline{L}^c_\alpha \, H \, N^c_i - (M_{NS})_{ij} \, \overline{N}_i \, S_j - \frac 12\, (\mu_S)_{kl} \, \overline{S}^c_k \, S_l + \mathrm{h.c.}
\end{equation}
with $H$ being the Higgs field, $y_D$ the Dirac neutrino Yukawa matrix, $M_{NS}$ the matrix connecting the gauge singlets $N_i$ and $S_j$ and $\mu_S$ the Majorana mass matrix of the singlets $S_i$.
 From eq.~(\ref{eq:Lneutral}), we can define the Dirac neutrino mass matrix $m_D$ as $m_D=y_D \, \langle H \rangle$
with $\langle H \rangle$ being the vacuum expectation value (VEV) of the Higgs, $\langle H \rangle \approx 174 \, \mbox{GeV}$. The masses of all neutral states, light and heavy ones, come from the nine-by-nine mass matrix 
\begin{equation}
\label{eq:MMaj}
{\cal M}_{\mathrm{Maj}} = \left(
\begin{array}{ccc}
\mathbb{0} & m_D & \mathbb{0} \\
m_D^T & \mathbb{0} & M_{NS} \\
\mathbb{0} & M_{NS}^T & \mu_S
\end{array}
\right) \; ,
\end{equation}
which is given in the basis $(\nu_{\alpha L}, N_i^c, S_j)$. 
This matrix can be diagonalised as 
\begin{equation}
\label{eq:UMdiag}
{\cal U}^T \, {\cal M}_{\mathrm{Maj}} \, {\cal U} = {\cal M}_{\mathrm{Maj}}^{\mathrm{diag}}  
\end{equation}
with
\begin{equation}
\label{eq:formU}
{\cal U} = \left(
\begin{array}{cc}
\tilde{U}_\nu & S \\
T & V
\end{array}
\right) \;\; \mbox{and} \;\;  {\cal M}_{\mathrm{Maj}}^{\mathrm{diag}}  = \mathrm{diag} \left( m_1, m_2, m_3, m_4, \dots, m_9 \right) \; ,
\end{equation}
where $\tilde{U}_\nu$ is a three-by-three, $S$ a three-by-six, $T$ a six-by-three and $V$ a six-by-six matrix. The masses $m_{1,2,3}$ are the light neutrino masses and $m_4$ to $m_9$ those of the heavy states.
The matrix ${\cal U}$ is unitary, while none of the matrices $\tilde{U}_\nu$, $S$, $T$ and $V$ has this property. We assume in the following $|\mu_S| \ll |m_D| \ll |M_{NS}|$.
As shown in~\cite{Hettmansperger:2011bt}, the light neutrino mass matrix at leading, $m_\nu$, and at sub-leading order, $m_\nu^1$, in $(|m_D|/|M_{NS}|)^2$ read
\begin{equation}
m_\nu = m_D \, \Big( M_{NS}^{-1} \Big)^T \, \mu_S \, M_{NS}^{-1} \, m_D^T
\end{equation}
and
\begin{eqnarray}
\nonumber
\!\!\!\!\!\!\!\!\!m_\nu^1 &=& -\frac 12 \, m_D \, \Big( M_{NS}^{-1} \Big)^T \, \left( \mu_S \, M_{NS}^{-1} \, m_D^T \, m_D^\star \, \Big( M_{NS}^{-1} \Big)^\dagger + \Big( M_{NS}^{-1} \Big)^\star \, m_D^\dagger \, m_D \, \Big( M_{NS}^{-1} \Big)^T \, \mu_S \right) \, M_{NS}^{-1} \, m_D^T \\
\label{eq:mnu1}
&=& - \left( m_\nu \, \eta + \eta^\star \, m_\nu \right)  \; ,
\end{eqnarray}
using the matrix $\eta$ defined in eq.~(\ref{eq:eta}).
The matrix $m_\nu$ is (approximately) diagonalised by $\tilde{U}_\nu$,
\begin{equation}
\tilde{U}_\nu^T \, m_\nu \tilde{U}_\nu \approx \mathrm{diag} (m_1, m_2, m_3) \; .
\end{equation}
We can define the (non-unitary) lepton mixing matrix
\begin{equation}
\tilde{U}_{\mathrm{PMNS}} = U_\ell^\dagger \, \tilde{U}_\nu = \tilde{U}_\nu \; ,
\end{equation}
where the second equality takes into account that $U_\ell= \mathbb{1}$ in the chosen basis.
This non-unitarity is due to the mixing of the light neutrinos with the heavy states and it can be encoded in the Hermitian matrix $\eta$,
\begin{equation}
\label{eq:UPMNStilde}
\tilde{U}_{\mathrm{PMNS}} = \Big( \mathbb{1} - \eta \Big) \, U_0
\end{equation} 
with $U_0$ being unitary. Note that in the chosen basis eq.~(\ref{eq:UPMNStilde}) also holds for $\tilde{U}_\nu$. The form of the matrix $\eta$ is at leading order given by
\begin{equation}
\label{eq:eta}
\eta = \frac 12 \, m_D^\star \, \Big( M_{NS}^{-1} \Big)^\dagger \, M_{NS}^{-1} \, m_D^T \; .
\end{equation}
As $N_i$ and $S_j$ transform both as the real representation ${\bf 3^\prime}$, the matrices $M_{NS}$ and $\mu_S$ are non-vanishing in the limit of unbroken flavour and CP symmetries. In particular, using the 
basis given in appendix~\ref{app:grouptheory}, the form of the matrices is
\begin{equation}
\label{eq:MNSmuS}
M_{NS} = M_0 \, 
\left(
\begin{array}{ccc}
1 & 0 & 0\\
0 & 1 & 0\\
0 & 0 & 1
\end{array}
\right)
\;\;\; \mbox{and} \;\;\;
\mu_S = \mu_0 \, 
\left(
\begin{array}{ccc}
1 & 0 & 0\\
0 & 0 & 1\\
0 & 1 & 0
\end{array}
\right) 
\end{equation}
with $M_0$ and $\mu_0$ being positive and having dimension of mass. The non-trivial flavour information is encoded in the Dirac neutrino Yukawa matrix $y_D$ whose 
 form is determined by the following equations
\begin{equation}
Z ({\bf 3})^T \, y_D \, Z ({\bf 3^\prime})^\star = y_D \;\;\; \mbox{and} \;\;\; X ({\bf 3}) \, y_D \, X ({\bf 3^\prime})^\star = y_D^\star \; ,
\end{equation}
where the matrices $Z ({\bf 3})$ and $Z ({\bf 3^\prime})$ are the representation matrices of the generator of the preserved $Z_2$ group, belonging to the residual 
symmetry $G_\nu$, in the representations ${\bf 3}$ (for the LH lepton doublets) and ${\bf 3^\prime}$ (for the gauge singlets $N_i$), respectively. The matrices
$X ({\bf 3})$ and $X ({\bf 3^\prime})$ represent the CP transformation in these representations that also belongs to the group $G_\nu$. The explicit form of these matrices can be 
found in~\cite{Drewes:2022kap}. As a result, the matrix $y_D$
 depends on five real parameters in total: three Yukawa couplings, $y_1$, $y_2$ and $y_3$, and two angles, called $\theta_L$
and $\theta_R$. Its form is 
\begin{equation}
\label{eq:yD}
y_D = \Omega ({\bf 3})^\star \, R_{ij} (\theta_L) \, \mathrm{diag} (y_1, y_2, y_3) \, P^{ij}_{kl} \, R_{kl} (-\theta_R) \, \Omega ({\bf 3^\prime})^T \; .
\end{equation}
The different matrices are fixed as follows: the unitary matrices $\Omega ({\bf 3})$ and $\Omega ({\bf 3^\prime})$ are determined by the form of the matrices
$X ({\bf 3})$ and $X ({\bf 3^\prime})$, respectively, 
\begin{equation}
\Omega ({\bf 3}) \, \Omega ({\bf 3})^T = X ({\bf 3}) \;\;\; \mbox{and} \;\;\; \Omega ({\bf 3^\prime}) \, \Omega ({\bf 3^\prime})^T = X ({\bf 3^\prime}) \; ,
\end{equation}
the planes of the rotation matrices $R_{ij} (\theta_L)$ and $R_{kl} (\theta_R)$ are given by the plane of the two degenerate eigenvalues of the matrices $Z ({\bf 3})$
and $Z ({\bf 3^\prime})$ in the basis transformed by $\Omega ({\bf 3})$ and $\Omega ({\bf 3^\prime})$, respectively. If these two planes do not coincide, the 
permutation matrix $P^{ij}_{kl}$ is necessary. The form of these matrices depends on the mixing pattern which we consider, i.e.~Case 1) through Case 3 b.1). 
It can be found in~\cite{Drewes:2022kap} and is repeated for convenience in appendix~\ref{app:matrices}.  
This scenario has been first mentioned in the context of the ISS mechanism in~\cite{Hagedorn:2021ldq} and called option 2. However, neither the results for fermion masses and mixing
nor for the signal strength of different cLFV processes have been discussed there. This analysis is performed in the current study.

With the given information, we can compute both the masses of the neutral states as well as their mixing. We begin with the mass matrix of the heavy states and their mass spectrum, since this structure is very simple.
It can be derived (to very good approximation) from the matrix
\begin{equation}
\left( 
\begin{array}{cc}
\mathbb{0} & M_{NS}\\
 M_{NS}^T& \mu_S
 \end{array}
 \right)
\end{equation}
which is (approximately) diagonalised by the six-by-six matrix $V$, i.e.~
\begin{equation}
V^T\, \left( 
\begin{array}{cc}
\mathbb{0} & M_{NS}\\
 M_{NS}^T& \mu_S
 \end{array}
 \right) \, V \approx \mathrm{diag} \left( m_4, \dots, m_9 \right) \; .
\end{equation}
For $M_{NS}$ and $\mu_S$ as in eq.~(\ref{eq:MNSmuS}), one finds for the masses of the heavy states that these are all degenerate to very high degree and form three pairs of pseudo-Dirac neutrinos, splitted by $\mu_0$, 
\begin{equation}
\label{eq:massesheavy}
m_4 \approx m_5 \approx m_6 \approx M_0 - \frac{\mu_0}{2} \;\;\; \mbox{and} \;\;\; m_7 \approx m_8 \approx m_9 \approx M_0 + \frac{\mu_0}{2}\; ,
\end{equation}
as well as the approximate form of $V$
\begin{equation}
\label{eq:formV}
V = \frac{1}{\sqrt{2}} \, \left(
\begin{array}{cc}
i \, U_S^\star & U_S^\star \\
-i \, U_S & U_S 
\end{array}
\right)
\end{equation}
with $U_S$ being chosen as 
\begin{equation}
U_S = \frac{1}{\sqrt{2}} \, \left(
\begin{array}{ccc}
\sqrt{2} & 0 & 0\\
0 & 1 & i \\
0 & 1 & -i
\end{array}
\right) 
\end{equation}
such that 
\begin{equation}
 U_S^T \, \mu_S \, U_S = \mu_0 \, \left( 
 \begin{array}{ccc}
 1 & 0 & 0\\
 0 & 1 & 0\\
 0 & 0 & 1
 \end{array}
 \right) \; ,
\end{equation}
compare~\cite{Hagedorn:2021ldq,Drewes:2022kap}. The approximate form of the matrix $S$ is related to the form of $V$, i.e.~
\begin{equation}
\label{eq:formS}
S =  \left( \begin{array}{ccc} 
 \mathbb{0} &,&m_D^\star \, \Big(M_{NS}^{-1} \Big)^\dagger
\end{array}
\right) \, V =  \frac{\langle H \rangle}{\sqrt{2} \, M_0} \, \left( 
\begin{array}{ccc} 
-  i \, y_D^\star \, U_S &,& y_D^\star \, U_S
\end{array}
\right) \; , 
\end{equation}
see also~\cite{Hettmansperger:2011bt,Hagedorn:2021ldq}; for the implicit definition of $S$ see eq.~(\ref{eq:formU}).

Using the general formulae, found in~\cite{Hettmansperger:2011bt,Hagedorn:2021ldq}, we see that
the light neutrino mass matrix $m_\nu$ at leading order is of the form
\begin{eqnarray}
\nonumber
m_\nu&=& \left( \frac{\mu_0 \, \langle H \rangle^2}{M_0^2} \right) \, U_0 (\theta_L)^\star \, \mathrm{diag} (y_1, y_2, y_3) 
\\
&&\nonumber \;\, \times \left[ P^{ij}_{kl} \, R_{kl} (-\theta_R) \, \Omega ({\bf 3^\prime})^T \, \left(
\begin{array}{ccc}
1 & 0 & 0\\
0 & 0 & 1\\
0 & 1 & 0
\end{array}
\right) \, \Omega ({\bf 3^\prime}) \, R_{kl} (\theta_R) \, (P^{ij}_{kl})^T  \right] 
\\
&&\label{eq:mnuform} \;\;\; \, \times \, \mathrm{diag} (y_1, y_2, y_3) \, U_0 (\theta_L)^\dagger
\end{eqnarray}
with the definition of the unitary matrix $U_0 (\theta)$
\begin{equation}
\label{eq:U0}
U_0 (\theta) = \Omega ({\bf 3}) \, R_{ij} (\theta)
\end{equation}
and the form of the Hermitian matrix $\eta$ is given by
\begin{equation}
\label{eq:etaU0}
\eta = \eta_0^\prime\, U_0 (\theta_L) \, \mathrm{diag} (y_1^2, y_2^2, y_3^2) \, U_0 (\theta_L)^\dagger \;\;\; \mbox{with} \;\;\; \eta_0^\prime =  \frac{\langle H \rangle^2}{2 \, M_0^2} \; .
\end{equation}
The quantity $\eta_0^\prime$ is similar to the quantity $\eta_0$ defined in~\cite{Hagedorn:2021ldq}.

In order to continue we have to have knowledge about the form of the matrix combination in square brackets in eq.~(\ref{eq:mnuform}).
Indeed, it is known~\cite{Drewes:2022kap} that this expression\footnote{In~\cite{Drewes:2022kap}, the complex conjugate of this matrix combination is found.} is either diagonal
or has a block-diagonal form that can be diagonalised by a rotation in the $(ij)$-plane, i.e.~the same plane in which the rotation $R_{ij} (\theta_L)$
acts. In the following, we first discuss the situation in which this expression is diagonal and then turn to the instance in which an additional
rotation is necessary. 

If the expression is diagonal, we find that the matrix diagonalising $m_\nu$ at leading order is of the form 
\begin{equation}
\label{eq:UtildenuthL}
 \tilde{U}_\nu \approx U_0 (\theta_L) \, K_\nu = \Omega ({\bf 3}) \, R_{ij} (\theta_L) \; K_\nu \; ,
\end{equation}
where the role of $K_\nu$ is to ensure that the light neutrino masses are positive semi-definite.\footnote{We set $K_\nu$ to the identity matrix in analytic results, but take it properly into 
account in the numerical study.} The explicit forms of mixing patterns can be found, for convenience, in appendix~\ref{app:residualsCPUPMNS}.
Interestingly, then the leading order values of the 
light neutrino masses $m_f$ are simply given by
\begin{equation}
\label{eq:mfyf}
m_f = \left( \frac{\mu_0 \, \langle H \rangle^2}{M_0^2} \right) \, y_f^2 \;\;\; \mbox{with} \;\;\; f=1,2,3
\end{equation}
for Case 1) through Case 3 a), whereas for Case 3 b.1) we have to take into account an additional permutation $P$ among the light neutrino masses in order to produce
the corresponding mixing pattern, i.e.~
 \begin{equation}
 \label{eq:mfyf_Case3b1}
m_1 = \left( \frac{\mu_0 \, \langle H \rangle^2}{M_0^2} \right)  \, y_3^2 \; , \;\;
m_2 = \left( \frac{\mu_0 \, \langle H \rangle^2}{M_0^2} \right)  \, y_1^2 \; , \;\; 
m_3 = \left( \frac{\mu_0 \, \langle H \rangle^2}{M_0^2} \right)  \, y_2^2 
\end{equation}
and thus
\begin{equation}
\label{eq:UtildenuthL_Case3b1}
 \tilde{U}_\nu \approx U_0 (\theta_L) \, P \, K_\nu = \Omega ({\bf 3}) \, R_{ij} (\theta_L) \, P \; K_\nu \;\; \mbox{with} \;\; 
 P= \left(
 \begin{array}{ccc}
 0 & 1 & 0\\
 0 & 0 & 1\\
 1 & 0 & 0
 \end{array}
 \right)
 \; ,
\end{equation}
see e.g.~\cite{Drewes:2022kap}. 
 Furthermore, we can express $\eta$ in terms of $ \tilde{U}_\nu$ and the light neutrino masses $m_f$
\begin{equation}
\label{eq:etaUnu}
\eta \approx \frac{1}{2 \, \mu_0} \,  \tilde{U}_\nu \, \mathrm{diag} (m_1, m_2, m_3) \,  \tilde{U}_\nu^\dagger \; .
\end{equation}
This has two immediate consequences: firstly, the matrix $m_\nu^1$ is diagonalised by the same matrix as the leading order term $m_\nu$, since we can write the expression in the second line of eq.~(\ref{eq:mnu1}) as 
\begin{equation}
m_\nu^1 \approx - \frac{1}{\mu_0} \, \tilde{U}_\nu^\star \, \mathrm{diag} (m_1^2,  m_2^2, m_3^2) \,  \tilde{U}_\nu^\dagger \; .
\end{equation}
Consequently, lepton mixing is not influenced by this sub-leading term, while the light neutrino masses are (slightly) corrected. Secondly, we can estimate the effect of non-unitarity of the lepton mixing matrix using eq.~(\ref{eq:UPMNStilde})
\begin{equation}
\label{eq:Unudiag}
\!\!\!\!\!\!\tilde U_\nu= U_0 (\theta_L) \, \left( \mathbb{1} - \eta_0^\prime \, \mathrm{diag} (y_1^2, y_2^2, y_3^2)  \right) \;\; \mbox{and} \;\; 
\tilde U_\nu= U_0 (\theta_L) \, P \, \left( \mathbb{1} - \eta_0^\prime \, \mathrm{diag} (y_3^2, y_1^2, y_2^2)  \right)   
\end{equation}
for Case 1) through Case 3 a) and Case 3 b.1), respectively. This can also be written in terms of the light neutrino masses $m_f$ as
\begin{equation}
\label{eq:Unudiag_mf}
\!\!\!\!\!\!\tilde U_\nu= \tilde U_\nu^0 \, \left( \mathbb{1} - \frac{1}{2 \, \mu_0} \, \mathrm{diag} (m_1, m_2, m_3)  \right) \; ,  
\end{equation}
where $\tilde{U}_\nu^0$ indicates the unitary matrix in eq.~(\ref{eq:UtildenuthL}) and~(\ref{eq:UtildenuthL_Case3b1}) for Case 1) through Case 3 a) and Case 3 b.1), respectively. 
A suppression of the columns of the PMNS mixing matrix is induced that is in general different for the different columns, since it depends on the Yukawa couplings $y_f$/the light neutrino masses $m_f$. In particular,
in case of strong normal ordering (NO) or strong inverted ordering (IO), i.e.~the lightest neutrino mass $m_0$ vanishes, 
one of the columns remains unaffected. This is slightly more general than the effect observed in~\cite{Hagedorn:2021ldq}, where a common 
suppression factor for all entries of the mixing matrix is encountered. We discuss the effect of the suppression in eq.~(\ref{eq:Unudiag}) in more detail in section~\ref{sec:leptonmixing}. 

In case the expression in square brackets in eq.~(\ref{eq:mnuform}) is not diagonal, we have to take into account an additional rotation through an angle determined by the Yukawa couplings $y_f$ 
and the angle $\theta_R$ occurring in the $(ij)$-plane such that the angle $\theta_L$ becomes replaced by an effective angle $\overline{\theta}_L$,\footnote{The effective angle $\overline{\theta}_L$ is denoted by $\widetilde{\theta}_L$ in~\cite{Drewes:2022kap}.} i.e.~(disregarding the matrix $K_\nu$)
\begin{equation}
\label{eq:UnutildethetaLbar}
 \tilde{U}_\nu \approx U_0 (\overline{\theta}_L) = \Omega ({\bf 3}) \, R_{ij} (\overline{\theta}_L) \; .
\end{equation}
Only one of the three light neutrino masses, $m_k$ with $k \neq i$ and $k \neq j$, depends on a single Yukawa coupling, as shown in eq.~(\ref{eq:mfyf}), while the others are determined by the remaining Yukawa couplings 
and the angle $\theta_R$, e.g.~for Case 1) (and independent of the choice of the parameter $s$, see section~\ref{sec:leptonmixing}) we have 
\begin{equation}
\label{eq:mfyfthRCase1}
m_2 = \left( \frac{\mu_0 \, \langle H \rangle^2}{M_0^2} \right) \, y_2^2 \; , \;\; m_{1,3} = \left(  \frac{\mu_0 \, \langle H \rangle^2}{2 \, M_0^2} \right) \, \left| (y_1^2 - y_3^2)  \, \cos 2 \, \theta_R \pm \sqrt{4 \, y_1^2 \, y_3^2 + (y_1^2-y_3^2)^2 \, \cos^2 2 \, \theta_R} \right| \; ,
\end{equation}
which in the case of strong NO ($m_1=0$) is reduced to
\begin{equation}
\label{eq:mfyfthRCase1_strongNO}
m_1=0 \; , \;\; m_2 = \left( \frac{\mu_0 \, \langle H \rangle^2}{M_0^2} \right) \, y_2^2 \; , \;\; m_3 = \left( \frac{\mu_0 \, \langle H \rangle^2}{M_0^2} \right) \, y_3^2 \, |\cos 2 \, \theta_R| \; ,
\end{equation}
and for strong IO ($m_3=0$) to
\begin{equation}
\label{eq:mfyfthRCase1_strongIO}
m_1=\left( \frac{\mu_0 \, \langle H \rangle^2}{M_0^2} \right) \, y_1^2 \, |\cos 2 \, \theta_R| \; , \;\; m_2 = \left( \frac{\mu_0 \, \langle H \rangle^2}{M_0^2} \right) \, y_2^2 \; , \;\; m_3 = 0 \; .
\end{equation}
 The form of the matrix $\eta$ in
eq.~(\ref{eq:etaU0}) does not change, but we can no longer identify $U_0 (\theta_L)$ with $\tilde{U}_\nu$. Instead we have 
\begin{equation} 
\label{eq:etaUnuDeltath}
\eta  \approx \eta_0^\prime \,  \tilde{U}_\nu \, R_{ij}(\Delta \theta) \, \mathrm{diag} (y_1^2, y_2^2, y_3^2) \, R_{ij}(-\Delta \theta)  \,  \tilde{U}_\nu^\dagger \, ,
\end{equation}
using the definition of $U_0 (\theta)$ in eq.~(\ref{eq:U0}) and setting $\Delta \theta = \theta_L -\overline{\theta}_L$. Regarding the form of the sub-leading term $m_\nu^1$ we note that it can be written as
\begin{eqnarray}
\nonumber
m_\nu^1 &\approx& - \eta_0^\prime \, \tilde{U}_\nu^\star \, \left[ \mathrm{diag} (m_1, m_2, m_3) \, R_{ij} (\Delta \theta) \,  \mathrm{diag} (y_1^2, y_2^2, y_3^2) \, R_{ij} (-\Delta \theta) \right.
\\  \label{eq:mnu1rot}
\;\;\;\;\; &+& \left. R_{ij} (\Delta \theta) \,  \mathrm{diag} (y_1^2, y_2^2, y_3^2) \, R_{ij} (-\Delta \theta) \,  \mathrm{diag} (m_1, m_2, m_3) \right] \, \tilde{U}_\nu^\dagger \; .
\end{eqnarray}
Hence, this sub-leading term is not diagonalised by $\tilde{U}_\nu$ (the matrix that diagonalises the leading term $m_\nu$), but induces a (small) additional rotation in the $(ij)$-plane. Also, the light neutrino masses
get slightly shifted. For the effects of non-unitarity of lepton mixing, we have in this case
\begin{equation}
\label{eq:Ununondiag}
\tilde U_\nu= U_0 (\overline{\theta}_L) \, \left( \mathbb{1} - \eta_0^\prime \, R_{ij} (\Delta \theta) \, \mathrm{diag} (y_1^2, y_2^2, y_3^2) \, R_{ij} (-\Delta \theta) \right) \; .
\end{equation}
Thus, only one column receives a common suppression determined by one of the light neutrino masses, while the effects on the other two are more involved, see discussion in section~\ref{sec:leptonmixing}.
We note that the formulae given here are valid for Case 1) through Case 3 a), while for Case 3 b.1) the permutation $P$ has to be taken into account which slightly complicates the expressions.

%%%%%%%%%%%%%%%%%%%%%%%%%%%%%%%%%%%%%%%%%%%%%%%%%%%%%%%
\section{Lepton mixing}
\label{sec:leptonmixing}
%%%%%%%%%%%%%%%%%%%%%%%%%%%%%%%%%%%%%%%%%%%%%%%%%%%%%%%

In this section, we comment on the results for the lepton mixing parameters derived from the different cases, Case 1) through Case 3 b.1), and how these
are affected by the presence of the heavy sterile states. There are two effects, already mentioned in section~\ref{sec2}: the sub-leading contribution $m_\nu^1$ can lead to deviations
from the results, determined by the flavour and CP symmetries, and the non-unitarity of the lepton mixing matrix, induced through the heavy sterile states. 

Whether the sub-leading contribution $m_\nu^1$ gives rise to deviations from the lepton mixing obtained at leading order or not depends on whether the expression in 
square brackets in eq.~(\ref{eq:mnuform}) is diagonal or not. If it is diagonal, then the lepton mixing parameters remain unaltered. Otherwise, the form of $m_\nu^1$
leads to an additional rotation, compare eq.~(\ref{eq:mnu1rot}), whose effect is suppressed by $\eta_0^\prime$. The typical size of $\eta_0^\prime$ is expected to be of
the order of the experimental bounds on the entries of the matrix $\eta$, $|\eta_{\alpha\beta}| < 1.0 \times 10^{-5} \div 1.4 \times 10^{-3}$ depending on the flavours $\alpha$ and $\beta$~\cite{Blennow:2023mqx}.
 This effect is, thus, in general small. 

In order to analyse the non-unitarity of the lepton mixing matrix, we extract 
the lepton mixing angles and CP invariants $J_{\mathrm{CP}}$, $I_1$ and $I_2$ from a three-by-three mixing matrix $U$ with elements $U_{\alpha i}$, $\alpha=e,\mu,\tau$
and $i=1,2,3$, as follows\footnote{The formulae in eq.~(\ref{eq:extracting}) correspond to those derived from a unitary mixing matrix that is parametrised by the three mixing angles $\theta_{12}$, $\theta_{13}$ and $\theta_{23}$ as well as the three CP phases $\delta$, $\alpha$ and $\beta$ in the following way
\begin{equation}
\nonumber
\begin{pmatrix}
c_{12} c_{13} & s_{12} c_{13} & s_{13} e^{- i \delta} \\
-s_{12} c_{23} - c_{12} s_{23} s_{13} e^{i \delta} & c_{12} c_{23} - s_{12} s_{23} s_{13} e^{i \delta} & s_{23} c_{13} \\
s_{12} s_{23} - c_{12} c_{23} s_{13} e^{i \delta} & -c_{12} s_{23} - s_{12} c_{23} s_{13} e^{i \delta} & c_{23} c_{13}
\end{pmatrix} \, 
\left( \begin{array}{ccc}
 1  & 0 & 0\\
 0  &  e^{i \alpha/2}  & 0\\
 0 & 0  & e^{i (\beta/2 + \delta)}
\end{array}
\right) 
\end{equation}
with $s_{ij}=\sin \theta_{ij}$ and $c_{ij}=\cos \theta_{ij}$.
}
\begin{eqnarray}
\nonumber
&&\sin^2 \theta_{13} = |U_{e3}|^2 \; , \;\; \sin^2 \theta_{12} = \frac{|U_{e2}|^2}{1-|U_{e3}|^2} \; , \;\; \sin^2 \theta_{23}  = \frac{|U_{\mu 3}|^2}{1-|U_{e3}|^2} \; , 
\\ \nonumber
&&J_{\mathrm{CP}} = \mathrm{Im} \left( U_{e1} \, U_{e3}^\star \, U_{\tau 1}^\star \, U_{\tau 3}\right) = \frac 18 \, \sin (2 \, \theta_{12}) \, \sin (2 \, \theta_{23}) \, \sin (2 \, \theta_{13}) \, \cos \theta_{13} \, \sin \delta \; ,
\\ \nonumber
&&I_1 = \mathrm{Im} \left( U_{e2}^2 (U_{e1}^\star)^2 \right) = \frac 14 \, \sin^2 (2 \, \theta_{12}) \, \cos^4 \theta_{13} \, \sin \alpha \; , 
\\ \label{eq:extracting}
&&I_2 = \mathrm{Im} \left( U_{e3}^2 (U_{e1}^\star)^2 \right) = \frac 14 \, \sin^2 (2 \, \theta_{13}) \, \cos^2 \theta_{12} \, \sin \beta \; .
\end{eqnarray}
The size of these effects crucially depends on the parameter $\eta_0^\prime$, the Yukawa couplings $y_f$ and potentially also the angle $\theta_R$ (i.e.~the difference between the angles $\theta_L$ and $\overline{\theta}_L$). 

Before commenting on the results for the different cases, we discuss the generic effects using as form of the lepton mixing matrix either the one in eq.~(\ref{eq:Unudiag}) or~(\ref{eq:Ununondiag}), depending on whether 
the matrix in square brackets in eq.~(\ref{eq:mnuform}) is diagonal or not. We define the (relative) deviations of the lepton mixing angles and CP invariants in the ISS framework as 
\begin{equation}
\label{eq:Deltadevs}
\Delta \sin^2 \theta_{ij} = \frac{(\sin^2 \theta_{ij})_{\mathrm{ISS}}-(\sin^2 \theta_{ij})_{\mathrm{FCS}}}{(\sin^2 \theta_{ij})_{\mathrm{FCS}}} \; , \;\;
\Delta J_{\mathrm{CP}} = \frac{(J_{\mathrm{CP}})_{\mathrm{ISS}}-(J_{\mathrm{CP}})_{\mathrm{FCS}}}{(J_{\mathrm{CP}})_{\mathrm{FCS}}} \; , \;\;
\Delta I_i = \frac{(I_i)_{\mathrm{ISS}}-(I_i)_{\mathrm{FCS}}}{(I_i)_{\mathrm{FCS}}} \; .
\end{equation}
Here, the quantities labelled with ``ISS" correspond to those extracted from (in general non-unitary) lepton mixing matrices which have a form as found in e.g.~eq.~(\ref{eq:Unudiag}), while the ones carrying the label ``FCS" 
(flavour and CP symmetries) are obtained from  (unitary) lepton mixing matrices whose forms are determined by the mismatch of the residual symmetries $G_e$ and $G_\nu$ (without assuming a specific model for generating
Majorana masses for the light neutrinos), compare the lepton mixing patterns found in e.g.~\cite{Hagedorn:2014wha} (see also appendix~\ref{app:residualsCPUPMNS}). The definition in eq.~(\ref{eq:Deltadevs}) is only valid, if 
 the relevant quantity with ``FCS" is non-vanishing. We note that this definition is identical to the one used in~\cite{Hagedorn:2021ldq}.

As already mentioned in section~\ref{sec2}, from eq.~(\ref{eq:Unudiag}) follows that the elements of one column are suppressed by the same factor. We always observe a suppression, since $\eta_0^\prime \, y_f^2$ is positive semi-definite.  Using eq.~(\ref{eq:extracting}) and the result for Case 1) through Case 3 a) in eq.~(\ref{eq:Unudiag}), the explicit dependence on $\eta_0^\prime$ and $y_f$
of these deviations is 
\begin{eqnarray}
\nonumber
&&\Delta \sin^2 \theta_{13} \approx -2 \, \eta_0^\prime \, y_3^2 \; , \;\; \Delta \sin^2 \theta_{12} \approx -2 \, \eta_0^\prime \, y_2^2 \; , \;\; \Delta \sin^2 \theta_{23} \approx - 2 \, \eta_0^\prime \, y_3^2  \; ,
\\ \label{eq:devgendiag}
&& \Delta J_{\mathrm{CP}} \approx -2 \, \eta_0^\prime \, (y_1^2 + y_3^2) \; , \;\; \Delta I_1 \approx -2 \, \eta_0^\prime \, (y_1^2 + y_2^2) \; , \;\; \Delta I_2 \approx -2 \, \eta_0^\prime \, (y_1^2 + y_3^2) \; .
\end{eqnarray}
In these approximations, we only take into account the leading order in $\eta_0^\prime$ and neglect terms of order $|U_{e3}|^2$ or smaller.
Since $\eta_0^\prime \, y_f^2 \geq 0$, all deviations are negative or zero. Given that experimental data are compatible with at most one vanishing light neutrino mass, only one of the couplings
$y_f$ can be zero. We comment on the consequences below. Obviously, in the (unrealistic) limit in which $y_f$ are equal, we end up with the same expressions as in~\cite{Hagedorn:2021ldq}.

If the matrix in square brackets in eq.~(\ref{eq:mnuform}) is not diagonal, we encounter as form of the non-unitary lepton mixing matrix the one mentioned in eq.~(\ref{eq:Ununondiag}), at least for Case 1) through Case 3 a).
Then, the deviations of the lepton mixing angles and the CP invariants in the ISS framework also depend on the difference, $\Delta \theta$, between the angles $\theta_L$ and $\overline{\theta}_L$, where the latter depends on the 
 Yukawa couplings $y_f$ and the 
free angle $\theta_R$. For example for $(ij)=(13)$, i.e.~the additional rotation diagonalising the matrix in square brackets in eq.~(\ref{eq:mnuform}) acts in the $(13)$-plane, we find at linear order in $\eta_0^\prime$  
and $|U_{e3}|$
\begin{eqnarray}
&&\nonumber  \Delta \sin^2 \theta_{13} \approx \eta_0^\prime \, \left( \frac{\mathrm{Re} \left( U_{e1}^\star \, U_{e3} \right)}{|U_{e3}|^2} \, (y_1^2-y_3^2) \, \sin 2 \, \Delta\theta -2 \, (y_1^2 \, \sin^2 \Delta\theta + y_3^2 \, \cos^2 \Delta\theta) \right) \; ,
\\
&& \Delta \sin^2 \theta_{12} \approx \eta_0^\prime \, \left( -2 \, y_2^2 + \mathrm{Re} \left( U_{e1}^\star \, U_{e3} \right) \, (y_1^2-y_3^2) \, \sin 2 \, \Delta\theta \right) \; ,
\label{eq:devthrot}
\\ \nonumber
&&\Delta \sin^2 \theta_{23} \approx \eta_0^\prime \, \left( -2 \, (y_1^2 \, \sin^2 \Delta \theta + y_3^2 \, \cos^2 \Delta \theta) + \mathrm{Re} \left( U_{e1}^\star \, U_{e3} + \frac{U_{\mu 1}^\star \, U_{\mu 3}}{|U_{\mu 3}|^2} \right) \, (y_1^2 - y_3^2) \, \sin 2 \, \Delta \theta \right)
\end{eqnarray} 
as well as
\begin{eqnarray}
&&\nonumber \Delta J_{\mathrm{CP}} \approx -2 \, \eta_0^\prime \, (y_1^2+y_3^2) 
\\
&& \nonumber \;\;\;\;\;\;\;\;\;\;\;
+ \eta_0^\prime \, (y_1^2-y_3^2) \, \frac{\left( \left( |U_{e1}|^2 +  |U_{e3}|^2 \right) \, \mathrm{Im} (U_{\tau 1}^\star \, U_{\tau 3}) + \left( |U_{\tau 1}|^2 + |U_{\tau 3}|^2 \right) \, \mathrm{Im} (U_{e1} \, U_{e3}^\star) \right)}{2 \, \mathrm{Im} \left( U_{e1} \, U_{e3}^\star \, U_{\tau 1}^\star  \, U_{\tau 3}\right)} \, \sin 2 \, \Delta \theta \; ,
\\
&&\nonumber \Delta I_1 \approx -2 \, \eta_0^\prime \, (y_1^2 \, \cos^2 \Delta \theta + y_2^2 + y_3^2 \, \sin^2 \Delta \theta) + \eta_0^\prime \, (y_1^2-y_3^2) \, \frac{\mathrm{Im} \left( U_{e2}^2 \, U_{e1}^\star U_{e3}^\star \right)}{\mathrm{Im} \left( U_{e 2}^2 \, (U_{e 1}^\star)^2 \right)} \, \sin 2 \, \Delta\theta \; , 
\\
&& \Delta I_2 \approx -2 \, \eta_0^\prime \, (y_1^2+y_3^2) + \eta_0^\prime \, (y_1^2-y_3^2) \, \frac{\left( |U_{e1}|^2 + |U_{e3}|^2\right)}{2 \, \mathrm{Re} \left(  U_{e1}^\star \, U_{e3} \right)} \, \sin 2 \, \Delta \theta \; ,
\label{eq:devCPinvrot}
\end{eqnarray} 
where we refer here with $U_{\alpha i}$ to the elements of the unitary matrix, i.e.~the one obtained from flavour and CP symmetries. Clearly, in the limit in which $\theta_L$ and $\overline{\theta}_L$ coincide, $\Delta \theta=0$, we recover the formulae
found in eq.~(\ref{eq:devgendiag}). Given the more general form of the deviation it is possible that the non-unitary lepton mixing matrix leads to non-trivial CP violation, although flavour and CP symmetries predict certain CP phases to be trivial.

\paragraph{Case 1)} As has been shown in~\cite{Hagedorn:2014wha} (compare also appendix~\ref{app:residualsCPUPMNS}), the choice of residual symmetries is described by one integer, called $s$, varying between $0$ and $n-1$, which corresponds to the employed CP symmetry.
 Two of the three CP phases, $\beta$ and $\delta$, are trivial in this case, i.e.~$I_2=0$ and $J_{\mathrm{CP}}=0$,
while the third one, the Majorana phase $\alpha$, is determined by the choice of the CP symmetry, $|\sin \alpha|=|\sin 6 \, \phi_s|$ with $\phi_s=\frac{\pi \, s}{n}$.\footnote{The sign in the relation for the Majorana phase $\alpha$ depends on the
 explicit form of the matrix $K_\nu$.} In contrast to this, the three lepton mixing angles are independent of this choice and only depend on the free angle $\theta_L$ (or in case the matrix combination in square brackets in eq.~(\ref{eq:mnuform}) is not diagonal this angle is given by $\overline{\theta}_L$). This angle is adjusted such that the reactor mixing angle is very close 
to its experimental best-fit value~\cite{Esteban:2020cvm} and, consequently, the solar and atmospheric mixing angle are given by $\sin^2 \theta_{12} \approx 0.341$ and $\sin^2 \theta_{23} \approx 0.604 \, (5)$ for light neutrino masses with
NO (IO). 

As has been analysed in~\cite{Drewes:2022kap}, only for $y_1=0$ or $y_3=0$ or $\theta_R$ chosen such that $\sin 2 \, \theta_R=0$, the matrix combination in square brackets 
in eq.~(\ref{eq:mnuform}) is diagonal. Then, we can apply the results found in eq.~(\ref{eq:devgendiag}). In particular, we see that in the case of light neutrino masses with strong NO ($y_1=0$) the deviations of the CP invariants also only 
depend on one coupling $y_f$, while for strong IO ($y_3=0$) the deviations of both the reactor and the atmospheric mixing angle are suppressed. Otherwise, we can use the formulae in eqs.~(\ref{eq:devthrot}) and~(\ref{eq:devCPinvrot}), since 
the rotation always occurs in the $(13)$-plane in this case. Although one may presume that one can achieve non-trivial CP phases $\delta$ and $\beta$ then, this does not happen, as can be checked by explicit computation.

\paragraph{Case 2)} From~\cite{Hagedorn:2014wha} it is known that the CP symmetry is fixed by two integer parameters, $s$ and $t$, ranging from $0$ to $n-1$; see appendix~\ref{app:residualsCPUPMNS} for its explicit form. Alternatively, one can define the combinations $u$ and $v$ as 
\begin{equation}
\label{eq:defuv} 
u= 2 \, s -t \;\;\; \mbox{and} \;\;\; v=3 \, t \; .
\end{equation}
The smallness of the reactor mixing angle can be achieved for $\frac un$ ($\phi_u=\frac{\pi \, u}{n}$) being small, $-0.1 \lesssim \frac{u}{n} \lesssim 0.12$ ($-0.31 \lesssim \phi_u \lesssim 0.37$), 
and the free angle, appearing in the lepton mixing matrix, close to zero or $\pi$, see~\cite{Hagedorn:2014wha,Drewes:2022kap}.
The solar mixing angle is, similar to Case 1), constrained to fulfil $\sin^2 \theta_{12} \gtrsim \frac 13$, while the atmospheric mixing angle can lie in its experimentally preferred $3 \, \sigma$ range~\cite{Esteban:2020cvm}. Valid choices of $u$ and $n$
can be found in section~\ref{sec:rescase2}, where the numerical results are presented, as well as in e.g.~\cite{Hagedorn:2014wha,Hagedorn:2021ldq,Drewes:2022kap}.
While the Dirac phase $\delta$ and 
the Majorana phase $\beta$ depend like the lepton mixing angles on the parameter $u$ and the free angle, the Majorana phase $\alpha$ is (mainly) determined by the parameter $v$ and to very good approximation it holds $|\sin \alpha| \approx
|\sin \phi_v|$ with $\phi_v=\frac{\pi \, v}{n}$.\footnote{As for Case 1), the sign in this relation depends on $K_\nu$.} 

In~\cite{Drewes:2022kap} it has been found that for $t$ even the matrix combination in square brackets in eq.~(\ref{eq:mnuform}) is always diagonal, while for $t$ odd this only occurs for $y_1=0$ or $y_3=0$ or $\theta_R$ 
chosen such that $\cos 2 \, \theta_R=0$. We, thus, can apply in these situations the results given in eq.~(\ref{eq:devgendiag}). Like for Case 1), we can consider light neutrino masses with strong NO ($y_1=0$) or strong IO ($y_3=0$) 
and arrive at the same conclusions. Note that setting $y_2$ to zero does not lead to a viable light neutrino mass spectrum, since $y_2$ is always associated with the light neutrino mass $m_2$, compare eq.~(\ref{eq:mfyf}).
 For $t$ odd in general, a rotation in the $(13)$-plane is necessary in order to diagonalise the matrix combination in square brackets in eq.~(\ref{eq:mnuform}). Consequently, we can apply the formulae in eqs.~(\ref{eq:devthrot}) 
 and~(\ref{eq:devCPinvrot}). As has been shown in~\cite{Hagedorn:2014wha}, for the free angle being zero, one can accommodate the lepton mixing angles well for $\frac{u}{n} \approx 0.12$, while both $J_{\mathrm{CP}}$ and $I_2$ vanish. 
 Taking into account the effects of non-unitarity, we see that for $\sin 2 \Delta \theta \neq 0$ $J_{\mathrm{CP}}$ as well as $I_2$ can be non-zero. On the other hand, for $u=0$ the CP invariant $I_2$ is zero and the inclusion of the effects of non-unitarity  
 does not change this result.

\paragraph{Case 3 a)} In this case the residual $Z_2$ symmetry is characterised by the parameter $m$ which can take values between $0$ and $n-1$, while the CP symmetry is fixed by one integer parameter $s$, also varying between $0$ and $n-1$; for details see appendix~\ref{app:residualsCPUPMNS}.
For Case 3 a) the ratio $\frac mn$ has to be either close to zero or to one (or equivalently $\phi_m$, $\phi_m=\frac{\pi \, m}{n}$, small or close to $\pi$) in order to accommodate well the reactor mixing angle. For this reason, $n$ should be at least $16$ and, consequently, $m=1$ or $m=15$.\footnote{Since the index $n$ has to be even in the current scenario, $n=17$ and $m=1$ (or $m=16$) is not eligible, but very similar results can be obtained for $n=34$ and $m=2$ (or $m=32$), see~\cite{Drewes:2022kap}. This 
choice also permits to study even values of $m$.} At the same time, fixing this ratio determines the value of the atmospheric mixing angle. The free angle, contained in the lepton mixing matrix, can take, depending on the choice of $s$ ($\phi_s = \frac{\pi \, s}{n}$), 
up to two distinct values that allow for a good fit of the solar mixing angle. Further details can be found in~\cite{Hagedorn:2014wha,Hagedorn:2021ldq,Drewes:2022kap} as well as in section~\ref{sec:rescase3a}. 
The three CP phases are in general non-trivial; numerical values and analytic estimates are given in the cited works.

Using the results of~\cite{Drewes:2022kap} we know that for the parameters $m$ and $s$ both even or both odd the matrix combination in square brackets in eq.~(\ref{eq:mnuform}) is diagonal and the formulae in 
eq.~(\ref{eq:devgendiag}) can be used. In contrast, for $m$ even and $s$ odd or vice versa this matrix combination is only diagonal for $y_1=0$ or $\sin 2 \, \theta_R=0$. For $y_1=0$ we encounter the same situation as for Case 1) and Case 2),
since light neutrino masses then follow strong NO and each deviation only depends on one Yukawa coupling $y_f$. If both parameters $m$ and $s$ are even or both are odd, setting $y_3=0$ permits us to suppress the deviations of the reactor and the atmospheric
mixing angle, compare eq.~(\ref{eq:devgendiag}). In case the matrix combination in square brackets in eq.~(\ref{eq:mnuform}) is not diagonal, it is diagonalised by a rotation in the $(12)$-plane. For this situation similar formulae
as those in eqs.~(\ref{eq:devthrot}) and~(\ref{eq:devCPinvrot}) can be derived.

\paragraph{Case 3 b.1)} Like for Case 3 a), the integer parameters $m$ and $s$ with $0 \leq m, s \leq n-1$ describe the residual $Z_2$ symmetry and the CP symmetry, respectively. From~\cite{Hagedorn:2014wha}, we know 
that the requirement to accommodate the solar mixing angle well constrains the parameter $m$ to be close to $\frac n2$; consult appendix~\ref{app:residualsCPUPMNS} for the explicit form of the mixing matrix. The free angle, appearing in the lepton mixing matrix, has to be close to $\frac{\pi}{2}$ such that the reactor mixing
 angle can be fitted well. For $\frac mn=\frac 12$, usually two such values exist. Otherwise, it depends on the choice of $m$ and $s$ how many distinct values of the free angle admit a good fit to the 
 experimental data of the lepton mixing angles. The atmospheric mixing angle also depends on the parameter $s$. Likewise, all CP phases are determined by the actual values of the ratios $\frac mn$ and $\frac sn$ as well as the free angle. 
 More information, in particular about analytical estimates and numerical values of the CP phases, is given in~\cite{Hagedorn:2014wha,Hagedorn:2021ldq,Drewes:2022kap}. Explicit choices of $m$ and $s$ are discussed in section~\ref{sec:rescase3b1}.  
 
 Whether or not the matrix combination in square brackets in eq.~(\ref{eq:mnuform}) is diagonal is subject to very similar constraints as for Case 3 a). The only difference is due to the fact that the light neutrino masses are assigned differently
 in Case 3 b.1) than Case 3 a), compare eqs.~(\ref{eq:mfyf_Case3b1}) and~(\ref{eq:mfyf}), i.e.~in order to have a diagonal matrix combination for $m$ even and $s$ odd or vice versa we have to set $y_2=0$ which corresponds to light 
 neutrino masses with strong IO. For Case 3 b.1) the relevant rotation effectively appears in the $(23)$-plane and expressions equivalent to those in eqs.~(\ref{eq:devthrot}) and~(\ref{eq:devCPinvrot})
 can be obtained.
 
%%%%%%%%%%%%%%%%%%%%%%%%%%%%%%%%%%%%%%%%%%%%%%%%%%%%%%%
\mathversion{bold}
\section{$\mu-e$ transitions}
\mathversion{normal}
\label{sec:muetrans}
%%%%%%%%%%%%%%%%%%%%%%%%%%%%%%%%%%%%%%%%%%%%%%%%%%%%%%%

In the discussion of cLFV processes, we focus on different $\mu-e$ transitions, since they are subject to the strongest experimental bounds.
As we show in the following, current experimental limits do not constrain the considered parameter space of this scenario, while prospective bounds,
in particular on $\mu-e$ conversion in aluminium, turn out to be very constraining in general.

%%%%%%%%%%%%%%%%%%%%%%%%%%%%%%%%%%%%%%%%%%%%%%%%%%%%%%%
\subsection{Prerequisites}
\label{sec:prerequisites}
%%%%%%%%%%%%%%%%%%%%%%%%%%%%%%%%%%%%%%%%%%%%%%%%%%%%%%%

We first list the experimental inputs and constraints we apply. Furthermore, we outline the parameter space of the scenario that we scan for viable choices of the group theory parameters (depending on the case) 
 in terms of the free angle $\theta_R$ and the two mass scales $\mu_0$ and $M_0$.
 The angle $\theta_L$ is determined by the requirement to accommodate the measured lepton mixing angles well and the couplings $y_f$ are fixed by the light neutrino mass spectrum, i.e.~the value of the lightest neutrino mass $m_0$
and the mass ordering, either NO or IO. 

%%%%%%%%%%%%%%%%%%%%%%%%%%%%%%%%%%%%%%%%%%%%%%%%%%%%%%%
\subsubsection{Experimental inputs and constraints}
\label{sec:expdata}
%%%%%%%%%%%%%%%%%%%%%%%%%%%%%%%%%%%%%%%%%%%%%%%%%%%%%%%

The lepton mixing angles are constrained by the global fits found in~\cite{Esteban:2020cvm}.\footnote{In the numerical analysis we use version 5.2, but check that the fitted lepton mixing angles are also
compatible with the results given in version 5.3.} Since the experimental preference 
for a certain value of the Dirac phase $\delta$ is still rather weak, we do not include this information here, when fitting
the angle $\theta_L$ in order to accommodate well the lepton mixing parameters. For all displayed points of the numerical scans, e.g.~for Case 1) in fig.~\ref{fig:Case1_scan},
 the lepton mixing angles agree at the $3 \, \sigma$ level or better with the global fit results~\cite{Esteban:2020cvm}.

Taking into account the experimental information on neutrino masses, in particular the results for the mass squared differences
from neutrino oscillations~\cite{Esteban:2020cvm} and the bound on the sum of the light neutrino masses, derived from cosmology~\cite{Planck:2018vyg}, we
use the following benchmark values for the lightest neutrino mass $m_0$: either $m_0=0$ (strong NO or strong IO) or
\begin{equation}
\label{eq:m0bm}
m_0 = 0.03 \, \mathrm{eV} \;\; \mbox{for NO and} \;\;  m_0 = 0.015 \, \mathrm{eV} \;\; \mbox{for IO.}
\end{equation}
The mass squared differences $\Delta m^2_{\mathrm{sol}}$ and $\Delta m^2_{\mathrm{atm}}$ are defined as 
\begin{equation}
\Delta m^2_{\mathrm{sol}} = m_2^2-m_1^2 \;\; \mbox{and} \;\; \Delta m^2_{\mathrm{atm}} = \left\{ \begin{array}{c}m_3^2 - m_1^2 \;\; \mbox{(NO)} \\ \!\!m_3^2 - m_2^2 \;\; \mbox{(IO)} \end{array} \right. \; .
\end{equation}
These are always fixed to their experimental best-fit values~\cite{Esteban:2020cvm}. The light neutrino masses for NO are given by
\begin{equation}
m_1 = m_0 \; , \;\; m_2 = \sqrt{m_0^2 + \Delta m^2_{\mathrm{sol}}} \; , \;\;  m_3 = \sqrt{m_0^2 + \Delta m^2_{\mathrm{atm}}} \; ,
\end{equation}
and for IO they read
\begin{equation}
m_1 =  \sqrt{m_0^2 + |\Delta m^2_{\mathrm{atm}}| - \Delta m^2_{\mathrm{sol}}} \; , \;\; m_2 = \sqrt{m_0^2 + |\Delta m^2_{\mathrm{atm}}|} \; , \;\;  m_3 = m_0 \; .
\end{equation}

\noindent As experimental bounds on the parameters $\eta_{\alpha\beta}$, $\alpha$, $\beta=e,\mu,\tau$, we use~\cite{Blennow:2023mqx}
\begin{equation}
\label{eq:etalimits}
|\eta_{\alpha\beta}| < \left( \begin{array}{ccc}
1.3 \, (1.4) \times 10^{-3} & 1.2 \times 10^{-5} & 9.0 \, (8.0) \times 10^{-4}\\
1.2 \times 10^{-5} & 1.1 \, (1.0) \times 10^{-5} & 5.7 \, (1.8) \times 10^{-5}\\
9.0 \, (8.0) \times 10^{-4} & 5.7 \, (1.8) \times 10^{-5} & 1.0 \, (0.81) \times 10^{-3} 
\end{array}
\right) \;\; \mbox{at} \;\; 95\% \; \mathrm{C.L.}
\end{equation}
for light neutrino masses with NO (IO). Additionally, the trace of $\eta$ is constrained by 
\begin{equation}
\mathrm{tr} (\eta) < 1.9 \, (1.5) \times 10^{-3} \;\; \mbox{at} \;\; 95\% \; \mathrm{C.L.}
\end{equation}
for NO (IO).

\noindent Furthermore, we employ as current bounds on the BRs of the cLFV decays $\mu \to e \gamma$ and $\mu \to 3 \, e$ 
\begin{eqnarray}
\nonumber
&&\mathrm{BR} (\mu \to e \gamma) < 3.1 \times 10^{-13} \;\;\, \mbox{at 90\% C.L.} \;\;\mbox{(MEG II~\cite{MEGII:2023ltw})}\; ,\\
&&\mathrm{BR} (\mu \to 3 \, e) < 1.0 \times 10^{-12} \;\; \mbox{at 90\% C.L.}\;\;\,\mbox{(SINDRUM~\cite{SINDRUM:1987nra})}\; ,
\end{eqnarray}
as well as on the CR of $\mu-e$ conversion in gold and titanium
\begin{eqnarray}
\nonumber
&&\mathrm{CR} (\mu-e, \mathrm{Au}) < 7.0 \times 10^{-13} \;\; \mbox{at 90\% C.L.} \;\; \;\;\mbox{(SINDRUMII~\cite{SINDRUMII:2006dvw})} \; ,\\
&&\mathrm{CR} (\mu-e, \mathrm{Ti}) <  6.1 \times 10^{-13} \;\;\; \mbox{at 90\% C.L.}\;\;\;\;\,\mbox{(SINDRUMII~\cite{SINDRUMIITi})} \; .
\end{eqnarray}
Future bounds on the first two processes are
\begin{eqnarray}\nonumber
&&\mathrm{BR} (\mu \to e \gamma) < 6 \times 10^{-14} \;\;\;\;\;\;\;\;\;\; \mbox{(MEG II~\cite{MEGII:2021fah})} \; ,\\
\label{eq:BRsmuefut}
&&\mathrm{BR} (\mu \to 3 \, e) < 20 \, (1) \times 10^{-16} \;\;\,\mbox{(Phase 1(2) of Mu3E~\cite{Blondel:2013ia})} \; ,
\end{eqnarray}
and for the CR of $\mu-e$ conversion in aluminium we expect
\begin{eqnarray}
\nonumber
&&\mathrm{CR} (\mu-e, \mathrm{Al}) < 7 \times 10^{-17} \;\;\mbox{(COMET~\cite{COMET:2018auw,Jansen:2023ojv})} \; ,\\
\label{eq:CRmuefut}
&&\mathrm{CR} (\mu-e, \mathrm{Al}) < 8 \times 10^{-17} \;\;\mbox{(Mu2e~\cite{Mu2e:2014fns,Artuso:2022ouk})} \; .
\end{eqnarray}

%%%%%%%%%%%%%%%%%%%%%%%%%%%%%%%%%%%%%%%%%%%%%%%%%%%%%%%
\subsubsection{Parameter space}
\label{sec:scan}
%%%%%%%%%%%%%%%%%%%%%%%%%%%%%%%%%%%%%%%%%%%%%%%%%%%%%%%

The parameter space of this scenario is spanned by the following quantities: the group theory parameters which depend on the considered case ($n$ and $s$ for Case 1), $n$, $s$ and $t$ (or $u$ and $v$) for Case 2),
 $n$, $m$ and $s$ for Case 3 a) and Case 3 b.1)), the free angles $\theta_L$ and $\theta_R$, the three couplings $y_f$ and the two scales $\mu_0$ and $M_0$, appearing in the matrices $y_D$, $\mu_S$ and $M_{NS}$,
 see eqs.~(\ref{eq:yD}) and~(\ref{eq:MNSmuS}), respectively. The group theory parameters are fixed to certain values that have been shown to lead to lepton mixing angles that agree well with the experimental data,
 see~\cite{Hagedorn:2014wha,Hagedorn:2021ldq,Drewes:2022kap}. The angle $\theta_R$ can be chosen freely and different values are explored. In general, it is linearly varied in the interval
 \begin{equation}
  \label{eq:rangethetaR}
 0 \leq \theta_R \leq 2 \, \pi \; .
 \end{equation} 
 In several of the following figures, see e.g.~figs.~\ref{fig:Case1_thetaL},~\ref{fig:Case2_un_thetaL_todd} and~\ref{fig:Case3b1_sn_thetaL_m9seven_thR}, we only display results for values of $\theta_R$ that lie in the range 
 $0 \leq \theta_R \leq \frac{\pi}{4}$. Nevertheless, we have checked that we obtain the same results for values outside this range.
 The angle $\theta_L$, on the other hand, is determined by fitting the measured values of the lepton mixing angles as best as possible. Depending on the case, Case 1) through Case 3 b.1), we find one or two such values for $\theta_L$, see 
 section~\ref{sec:leptonmixing}.
 The couplings $y_f$ are fixed such that the neutrino masses $m_i$, $i=1,2,3$, are correctly reproduced, while still being perturbative, i.e.~$y_f \leq \sqrt{8 \, \pi}$~\cite{Allwicher:2021rtd}. Indeed, in the presented scans these range
 in the interval $4 \times 10^{-5} \lesssim y_f \lesssim 1.2$, independent of the ordering of the light neutrino masses. We choose the signs of $y_f$ to be positive. 
 The scales $\mu_0$ and $M_0$ are varied in certain ranges. In particular, we take the lepton number violating parameter $\mu_0$
 to be much smaller than the electroweak scale, as expected in the ISS framework, and vary it logarithmically as follows
 \begin{equation}
  \label{eq:rangemu0}
 100 \, \mbox{eV} \leq \mu_0 \leq 100 \, \mbox{keV} \; ,
 \end{equation} 
 while the scale $M_0$ is in the range
 \begin{equation}
 \label{eq:rangeM0}
 150 \, \mbox{GeV} \leq M_0 \leq 10 \, \mbox{TeV} \; .
 \end{equation}  
Heavy sterile states in such a mass range can have different observational imprints, see e.g.~\cite{delAguila:2008cj,delAguila:2008hw,Chen:2011hc,Das:2012ze,Abada:2014vea,Arganda:2014dta,Abada:2014nwa,Abada:2014kba,Abada:2014cca,Arganda:2015naa,Abada:2015oba,DeRomeri:2016gum,Antusch:2016ejd,Crivellin:2022cve,Abada:2024hpb}.
 In order to optimise the exploration of this range of $M_0$, we divide it into three intervals, i.e.~a lower mass interval, $150 \, \mbox{GeV} \leq M_0 \leq 1 \, \mbox{TeV}$, an intermediate mass interval, $1 \, \mbox{TeV} \leq M_0 \leq 5 \, \mbox{TeV}$, 
 and a higher mass interval, $5 \, \mbox{TeV} \leq M_0 \leq 10 \, \mbox{TeV}$. In each of these intervals we vary $M_0$ uniformly and logarithmically. 
 A concise description of the numerical scan is found in appendix~\ref{app:numerics}.
 
%%%%%%%%%%%%%%%%%%%%%%%%%%%%%%%%%%%%%%%%%%%%%%%%%%%%%%%
\subsection{Analytical considerations}
\label{sec:anacon}
%%%%%%%%%%%%%%%%%%%%%%%%%%%%%%%%%%%%%%%%%%%%%%%%%%%%%%%

In this subsection, we remark features of the discussed BRs and CR that are common to all cases. In the following, we use the formulae found in~\cite{Alonso:2012ji,Ilakovac:1994kj}. 
These can be simplified by taking into account the mass spectrum of the neutral states and the fact that the heavy sterile states form three pairs of pseudo-Dirac neutrinos, compare eqs.~(\ref{eq:formV}) and~(\ref{eq:formS}), 
see details in~\cite{Hagedorn:2021ldq}.
For the radiative cLFV decays $\ell_\beta \to \ell_\alpha \, \gamma$, we have 
\begin{equation}
\mathrm{BR} (\ell_\beta \to \ell_\alpha \, \gamma) \approx \frac{\alpha_w^3 \, s_w^2}{64 \, \pi^2} \, \frac{m_\beta^4}{M_W^4} \, \frac{m_\beta}{\Gamma_\beta} \, \left| \eta_{\alpha \beta} \right|^2 \, G_\gamma (x_0)^2 \, ,
\end{equation} 
where $\alpha_w=\frac{g_w^2}{4 \, \pi}$ is the weak coupling, $s_w$ the sine of the weak mixing angle, $M_W$ the mass of the $W$ boson, $m_\beta$ the mass of the charged lepton of flavour $\beta$ and $\Gamma_\beta$ 
the corresponding total decay width as well as $G_\gamma (x)$ the relevant loop function with the limit $G_\gamma (x) \approx \frac 12$ for $x \gg 1$ and $x_0 = \left( \frac{M_0}{M_W} \right)^2$.
Similarly, we get for the BR of the tri-lepton decays $\ell_\beta \to 3 \, \ell_\alpha$
\begin{eqnarray}
\nonumber
\mathrm{BR} (\ell_\beta \to 3 \, \ell_\alpha) &=& \frac{\alpha_w^4}{24576 \, \pi^3} \, \frac{m_\beta^4}{M_W^4} \, \frac{m_\beta}{\Gamma_\beta} \, \left( 2 \, \Big| \frac 12 \, F^{\beta 3 \alpha}_{\mathrm{box}} + F^{\beta\alpha}_Z
-2 \, s_w^2 \, (F^{\beta\alpha}_Z-F^{\beta\alpha}_\gamma) \Big|^2  \right.\\
\nonumber
&&+ \, 4 \, s_w^4 \, |F^{\beta\alpha}_Z-F^{\beta\alpha}_\gamma|^2 + 16 \, s_w^2 \, \mathrm{Re} \left( (F^{\beta\alpha}_Z - \frac 12 \, F^{\beta 3 \alpha}_{\mathrm{box}}) \, (G^{\beta\alpha}_\gamma)^\star \right) \\
&& \left.- \, 48 \, s_w^4 \, \mathrm{Re} \left( (F^{\beta\alpha}_Z-F^{\beta\alpha}_\gamma) \, (G^{\beta\alpha}_\gamma)^\star \right) + 32 \, s_w^4 \, |G^{\beta\alpha}_\gamma|^2 \, \left( \mathrm{log} \frac{m_\beta^2}{m_\alpha^2} - \frac{11}{4} \right) 
\right)
 \end{eqnarray}
with the form factors reading approximately
\begin{eqnarray}
\nonumber
G^{\beta\alpha}_\gamma &\approx& 2 \, \eta_{\alpha\beta} \, G_\gamma (x_0) \; , \;\; F^{\beta\alpha}_\gamma \approx 2 \, \eta_{\alpha\beta} \, F_\gamma (x_0) \; ,
\\
\nonumber
F^{\beta\alpha}_Z &\approx& 2 \, \eta_{\alpha\beta} \, F_Z (x_0) + 4 \, (\eta_{\alpha\beta} - 2 \, \eta^2_{\alpha\beta}) \, G_Z (x_0, 0) + 4 \, \eta^2_{\alpha\beta} \, G_Z (x_0, x_0) \; ,
\\
\label{eq:GgFZFb3a}
F^{\beta 3 \alpha}_{\mathrm{box}} &\approx& -4 \, (\eta_{\alpha\beta} - 2 \, \eta_{\alpha\alpha} \, \eta_{\alpha\beta}) -8 \, \eta_{\alpha\alpha} \, \eta_{\alpha\beta} \, F_{\mathrm{Xbox}} (x_0, x_0)
\end{eqnarray}
with the limits for $x \gg 1$ being $F_\gamma (x) \approx -\frac{7}{12} - \frac{1}{6} \, \mbox{log} \, x$, 
$F_Z (x) \approx  \frac{5}{2} - \frac{5}{2} \, \mbox{log} \, x$, 
$G_Z (x, 0) = G_Z (0, x) \approx \frac{1}{2} \, \mbox{log} \, x$, 
$G_Z (x, x) \approx -\frac{x}{2} - \frac{x \, \mathrm{log} \, x}{1-x}$ 
and $F_{\mathrm{Xbox}} (x, x) \approx -\frac x4 + \frac 14 \, (13- 6 \, \mbox{log} \, x)$. 
 Furthermore, the CR for $\mu-e$ conversion in nuclei (N) is
\begin{equation}
\mathrm{CR} (\mu-e, \mathrm{N}) = \frac{2 \, G_F^2 \, \alpha_w^2 \, m_\mu^5}{(4 \, \pi)^2 \, \Gamma_{\mathrm{capt}}} \, \Big| 4 \, V^{(p)} \, \left( 2 \, \widetilde{F}^{\mu e}_u + \widetilde{F}^{\mu e}_d \right)  
+4 \, V^{(n)} \, \left(\widetilde{F}^{\mu e}_u + 2 \, \widetilde{F}^{\mu e}_d \right) + s_w^2 \, \frac{G^{\mu e}_\gamma \, D}{2 \, e} \Big|^2 \; ,
\end{equation}
where $G_F$ is the Fermi constant, $\Gamma_{\mathrm{capt}}$ the muon capture rate and $D$, $V^{(p)}$ and $V^{(n)}$ are nuclear form factors.
The expressions for the form factors are
\begin{eqnarray}
\nonumber
\widetilde{F}^{\mu e}_d &=& -\frac 13 \, s_w^2 \, F^{\mu e}_\gamma - \left( \frac 14 - \frac 13 \, s_w^2 \right) \, F^{\mu e}_Z + \frac 14 \, F^{\mu e d d}_{\mathrm{box}} \; ,
\\
\widetilde{F}^{\mu e}_u &=& \frac 23 \, s_w^2 \, F^{\mu e}_\gamma + \left( \frac 14 - \frac 23 \, s_w^2 \right) \, F^{\mu e}_Z + \frac 14 \, F^{\mu e u u}_{\mathrm{box}}
\end{eqnarray}
with $F^{\mu e}_\gamma$ and $F^{\mu e}_Z$ found in eq.~(\ref{eq:GgFZFb3a}), and 
\begin{equation}
F^{\mu e d d}_{\mathrm{box}} \approx 2 \, \eta_{e \mu} \;\;\; \mbox{and} \;\;\;
F^{\mu e u u}_{\mathrm{box}} \approx - 8 \, \eta_{e \mu} 
\end{equation}
in the limit $x_0 \gg 1$. Moreover, one makes use of the unitarity of the quark mixing matrix and takes into account the smallness of $|V_{td}|$.
These formulae can be used to show the following three statements: 
all these observables are mostly proportional to $|\eta_{\alpha \beta}|^2$, both the tri-lepton decays and $\mu-e$ conversion in nuclei are dominated by the contribution due to the $Z$ penguin
(in particular for larger values of $x_0$), see also~\cite{Hirsch:2012ax,Abada:2012cq}, as well as $\mathrm{CR} (\mu-e, \mathrm{N})$ is very suppressed for a specific value of $M_0$ which depends on the actual nucleus $\mathrm{N}$. In the case of aluminium and using the 
(approximate) formula found in~\cite{Alonso:2012ji} together with the numerical values of the nuclear form factors $D$, $V^{(p)}$ and $V^{(n)}$, $D = 0.0362$, $V^{(p)} = 0.0161$, $V^{(n)} = 0.0173$, 
and the capture rate $\Gamma_{\mathrm{capt}}=0.7054 \times 10^6 \, \mbox{s}^{-1}$, all given in~\cite{Kitano:2002mt}, we arrive at
\begin{equation}
\label{eq:M0cancelCR}
x_0 \approx  6470 \;\;\; \mbox{corresponding to} \;\;\; M_0 \approx 6.5 \, \mbox{TeV}
\end{equation}
for $s_w^2=0.23$. This cancellation is clearly visible in e.g.~fig.~\ref{fig:Case1_M0}.

%%%%%%%%%%%%%%%%%%%%%%%%%%%%%%%%%%%%%%%%%%%%%%%%%%%%%%%
\subsection{Results for Case 1)}
\label{sec:rescase1}
%%%%%%%%%%%%%%%%%%%%%%%%%%%%%%%%%%%%%%%%%%%%%%%%%%%%%%%

We discuss in detail the results for Case 1). For this, we choose as exemplary values
\begin{equation}
n=26 \;\;\; \mbox{and} \;\;\; 0 \leq s \leq 25 \; ,
\end{equation}
see also~\cite{Hagedorn:2021ldq}. We inspect the form of the matrix $U_0 (\theta)$ in eq.~(\ref{eq:U0}) for Case 1), using $\Omega(s) ({\bf 3})$ as given in eq.~(\ref{eq:Case1Omega3s}) in appendix~\ref{app:matrices} 
and that the rotation is in the $(13)$-plane,
and see that $U_0 (\theta)$ can be written as 
\begin{equation}
\label{eq:U0Case1ssep}
U_0 (\theta) = U_0 (\theta, s=0) \, \mathrm{diag} ( e^{i \, \phi_s}, e^{-2 \, i \, \phi_s}, e^{i \, \phi_s})  \; ,
\end{equation}
where $U_0 (\theta, s=0)$ refers to the matrix $U_0 (\theta)$ for $s=0$. With this, it follows that the matrix $\eta$, see eq.~(\ref{eq:etaU0}), does not depend on the parameter $s$.
Thus, we can set $s$ without loss of generality to $s=1$. We have also numerically studied the results for different values of $s$ (both even and odd) and no dependence on this parameter is encountered.

\begin{figure}[t!]
    \centering
     \includegraphics[width=\textwidth]{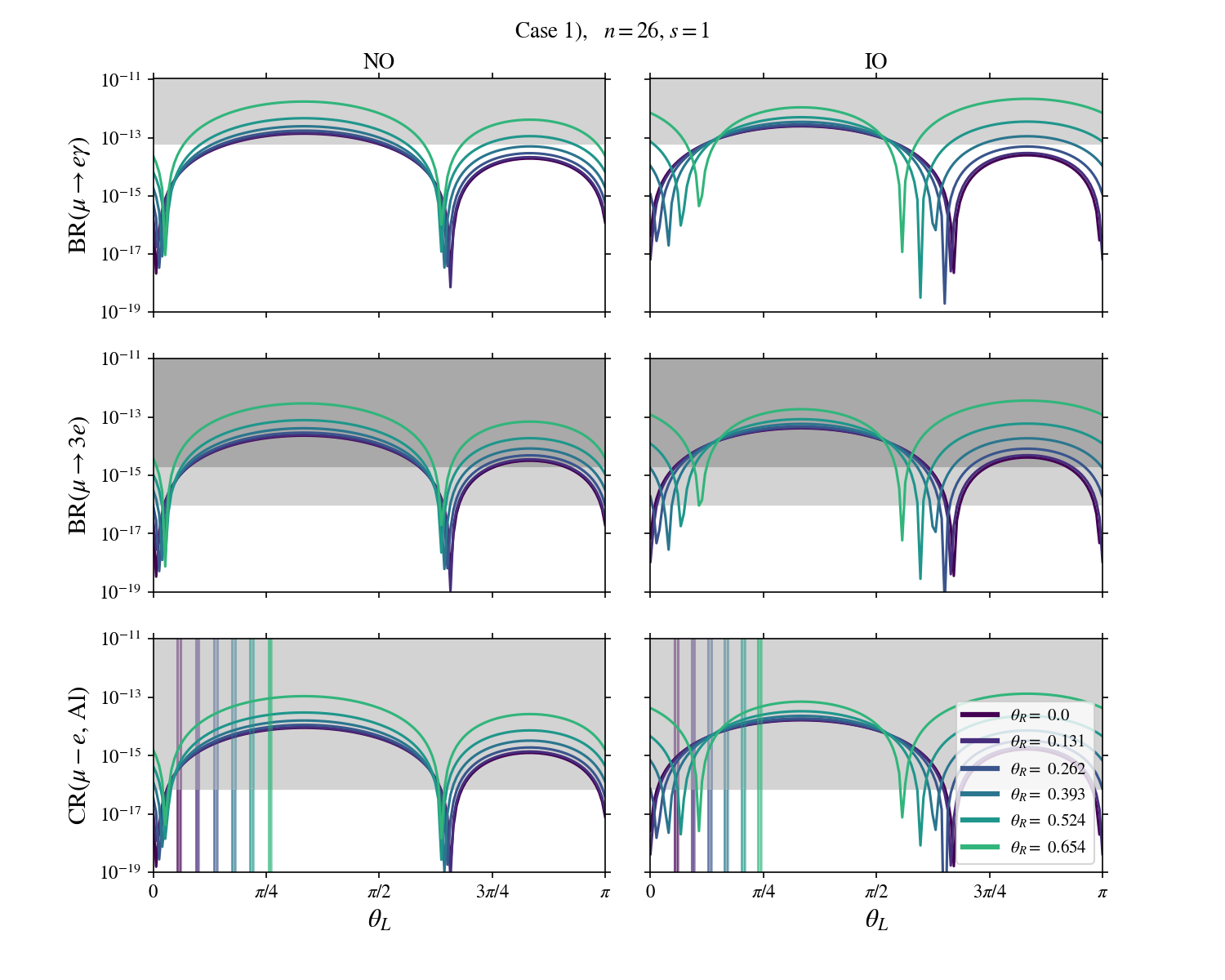}
    \caption{{\bf Case 1). Predictions for \mathversion{bold}$\mathrm{BR} (\mu\to e \gamma)$, $\mathrm{BR} (\mu\to 3 \, e)$ and $\mathrm{CR} (\mu-e, \mathrm{Al})$ as a function of the angle $\theta_L$\mathversion{normal}} in the upper, 
    middle and lower row.  This angle is strongly constrained by the measured lepton mixing angles, compare the coloured bands in the plots for $\mathrm{CR} (\mu-e, \mathrm{Al})$. 
    The index $n$ of the flavour symmetry is fixed to $n=26$ and the CP symmetry is chosen such that $s=1$. However, the obtained results do not depend on $s$. Left (right) plots are for light neutrino masses with NO (IO)
  and $m_0= 0.03 \, (0.015) \, \mathrm{eV}$. 
  Different values of the free angle $\theta_R$, see the different colours, 
    are explored. The scales $\mu_0$ and $M_0$ are set to $\mu_0=1 \, \mbox{keV}$ and $M_0=3 \, \mbox{TeV}$, respectively. Grey regions are excluded by the prospective bounds on the different BRs and CR with the darker (lighter) grey
    referring to Phase 1 (2) of Mu3E in the plots in the middle row. 
    }
    \label{fig:Case1_thetaL}
\end{figure}
A dependence of the size of $\mathrm{BR} (\mu\to e \gamma)$, $\mathrm{BR} (\mu\to 3 \, e)$ and $\mathrm{CR} (\mu-e, \mathrm{Al})$ 
on the angle $\theta_R$ is in general expected, since the matrix combination in square brackets in eq.~(\ref{eq:mnuform}) is not diagonal and the light neutrino mass spectrum is either 
given by eq.~(\ref{eq:mfyfthRCase1}),~(\ref{eq:mfyfthRCase1_strongNO}) or~(\ref{eq:mfyfthRCase1_strongIO}). We first explore the impact of this free angle by fixing the two scales $\mu_0$ and $M_0$ to $\mu_0=1 \, \mbox{keV}$ and $M_0=3 \, \mbox{TeV}$. Light neutrino masses either follow NO or IO with the lightest neutrino mass being set to $m_0=0.03 \, \mbox{eV}$ or $m_0=0.015 \, \mbox{eV}$, respectively. 
 The results for the BRs of $\mu\to e \gamma$ and
$\mu\to 3 \, e$ as well as the CR of $\mu-e$ conversion in aluminium are displayed in fig.~\ref{fig:Case1_thetaL} for different values of $\theta_R$, as indicated by the different colours, and for both light neutrino mass orderings 
in the left and right plots, respectively. 
The small vertical coloured bands in the plots for $\mathrm{CR} (\mu-e, \mathrm{Al})$ show the interval of $\theta_L$ which leads to an acceptable fit of the lepton mixing angles with $\chi^2 \leq 27$, as defined in eq.~(\ref{eq:chi2theta}) in appendix~\ref{app:numerics}. As mentioned, none of the current bounds on $\mu-e$ transitions can constrain the considered parameter space. Thus, we present in fig.~\ref{fig:Case1_thetaL} the exclusion due to the prospective limits as grey area, see eqs.~(\ref{eq:BRsmuefut}) and~(\ref{eq:CRmuefut}); in darker and lighter grey in the case of Phase 1 and Phase 2 of the experiment Mu3E in the plots in the middle. The constraining power of the future bound on $\mathrm{BR} (\mu\to e \gamma)$ is limited to the larger values of $\theta_R$ that are shown, while the bound on $\mathrm{BR} (\mu\to 3 \, e)$ expected from Phase 1 of Mu3E
can exclude all used values apart from vanishing $\theta_R$. Phase 2 of Mu3E and even more the prospective bound on $\mathrm{CR} (\mu-e, \mathrm{Al})$ have the potential to test all the displayed values of $\theta_R$ for $\mu_0 = 1 \, \mbox{keV}$ 
and $M_0=3 \, \mbox{TeV}$. We note that for these scales the obtained BRs and CR are for most of the studied values of $\theta_R$ slightly larger for light neutrino masses with IO than for a NO light neutrino mass spectrum.
 We can estimate their size easily for $\theta_R=0$, since for this choice the matrix combination in square brackets in eq.~(\ref{eq:mnuform}) is automatically diagonal. 
\begin{figure}[t!]
    \centering
    \includegraphics[width=\textwidth]{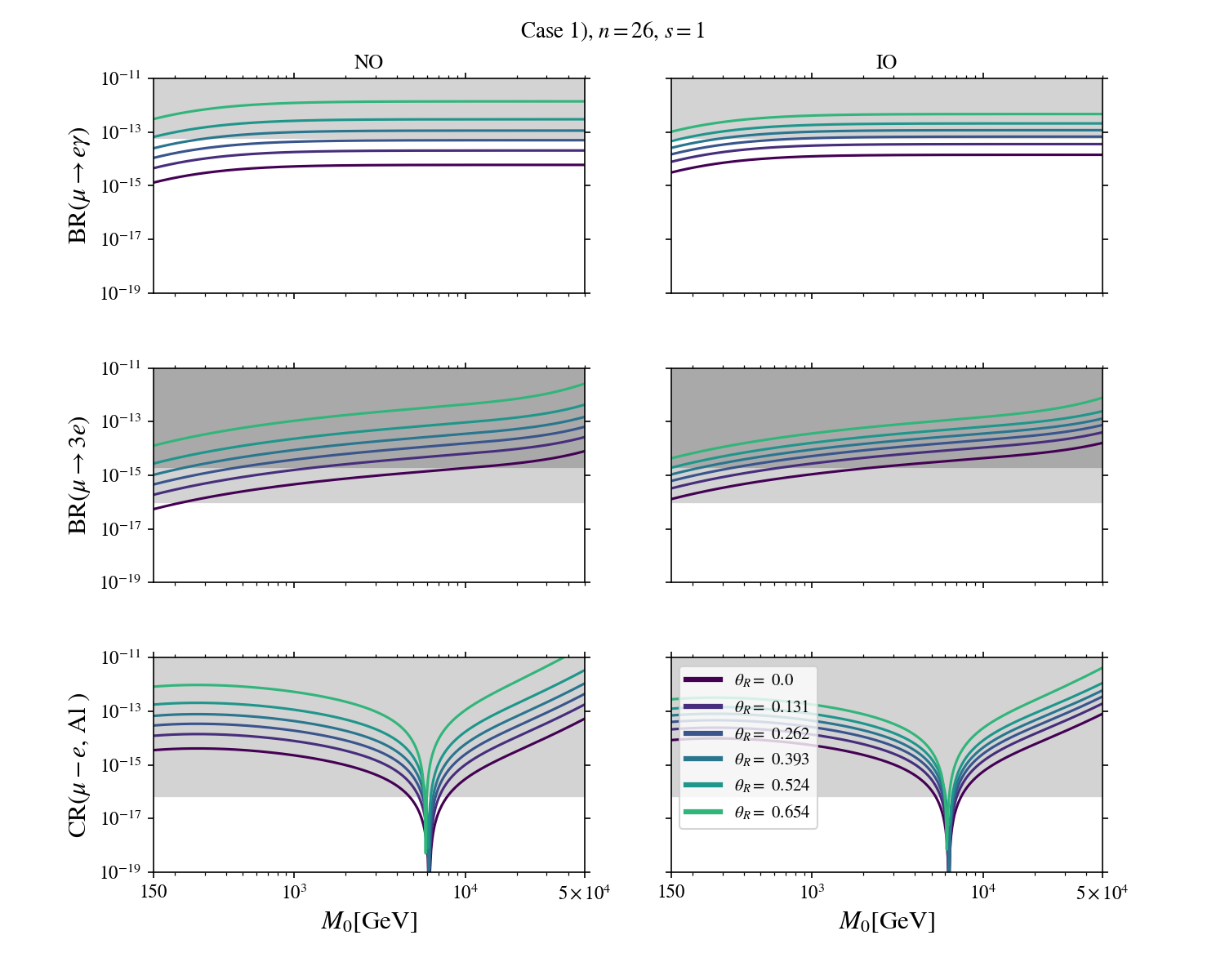}
    \caption{{\bf Case 1). Predictions for \mathversion{bold} $\mathrm{BR} (\mu\to e \gamma)$, $\mathrm{BR} (\mu\to 3 \, e)$ and $\mathrm{CR} (\mu-e, \mathrm{Al})$ as a function of $M_0$\mathversion{normal}}
    in the upper, middle and lower row. Note that the range of the scale $M_0$ is larger in these plots, $150 \, \mbox{GeV} \lesssim M_0 \lesssim 50 \, \mbox{TeV}$, than in the numerical scans.
    For the group theoretical parameters and light neutrino mass orderings compare fig.~\ref{fig:Case1_thetaL}.
    The parameter $\mu_0$ is set to $\mu_0=1 \, \mbox{keV}$ and $\theta_L$ to its best-fitting value for the lepton mixing angles. Six different values of $\theta_R$ between $0$ and $\frac{\pi}{4}$
    are explored, see different colours. The grey regions indicate exclusions of future experiments, see fig.~\ref{fig:Case1_thetaL}.
    }
    \label{fig:Case1_M0}
\end{figure}
  We use that the BRs and CR are mainly proportional to $|\eta_{e\mu}|^2$, as argued in section~\ref{sec:anacon}. The quantity $\eta_{e \mu}$ is calculated from eq.~(\ref{eq:etaU0}). It turns out to depend on $\theta_L$ (which can be written
  in terms of $\sin \theta_{13}$, see~\cite{Hagedorn:2014wha}) as well as on the differences of the squares of the Yukawa couplings $\Delta y_{ij}^2$, $\Delta y_{ij}^2=y_i^2-y_j^2$, 
   \begin{equation}
  \label{eq:etaexpressionCase1}
  \eta_{e \mu} = \frac{\eta_0^\prime}{6} \, \left( 2 \, \Delta y_{21}^2 - 3 \, (\sqrt{2-3 \, \sin^2 \theta_{13}} + \sin \theta_{13}) \, \sin \theta_{13} \, \Delta y_{31}^2 \right)  \; .
  \end{equation}
  The differences $\Delta y_{ij}^2$ can be expressed as the differences of the light neutrino masses, i.e.~$\Delta y_{ij}^2=\left(\frac{M_0^2}{\mu_0 \, \langle H \rangle^2}\right) \, (m_i - m_j)$.
   With this information, we find 
  \begin{equation}
  \label{eq:estimatesNOCase1}
\mathrm{BR} (\mu\to e \gamma) \approx  5.9 \times 10^{-15} \; , \;\; \mathrm{BR} (\mu\to 3 \, e) \approx 9.5 \times 10^{-16} \; , \;\; \mathrm{CR} (\mu-e, \mathrm{Al}) \approx 4.2 \times 10^{-16}
  \end{equation}
  for light neutrino masses with NO and $m_0=0.03 \, \mbox{eV}$ as well as  
    \begin{equation}
  \label{eq:estimatesIOCase1}
 \mathrm{BR} (\mu\to e \gamma) \approx 1.4 \times 10^{-14} \; , \;\; \mathrm{BR} (\mu\to 3 \, e) \approx 2.2 \times 10^{-15} \; , \;\; \mathrm{CR} (\mu-e, \mathrm{Al}) \approx 9.9 \times 10^{-16}
  \end{equation}
  for IO light neutrino masses with $m_0=0.015 \, \mbox{eV}$. 
 As one can see, these estimates agree well with the numerical results displayed in fig.~\ref{fig:Case1_thetaL}. 

\begin{figure}[t!]
    \centering
    \includegraphics[width=\textwidth]{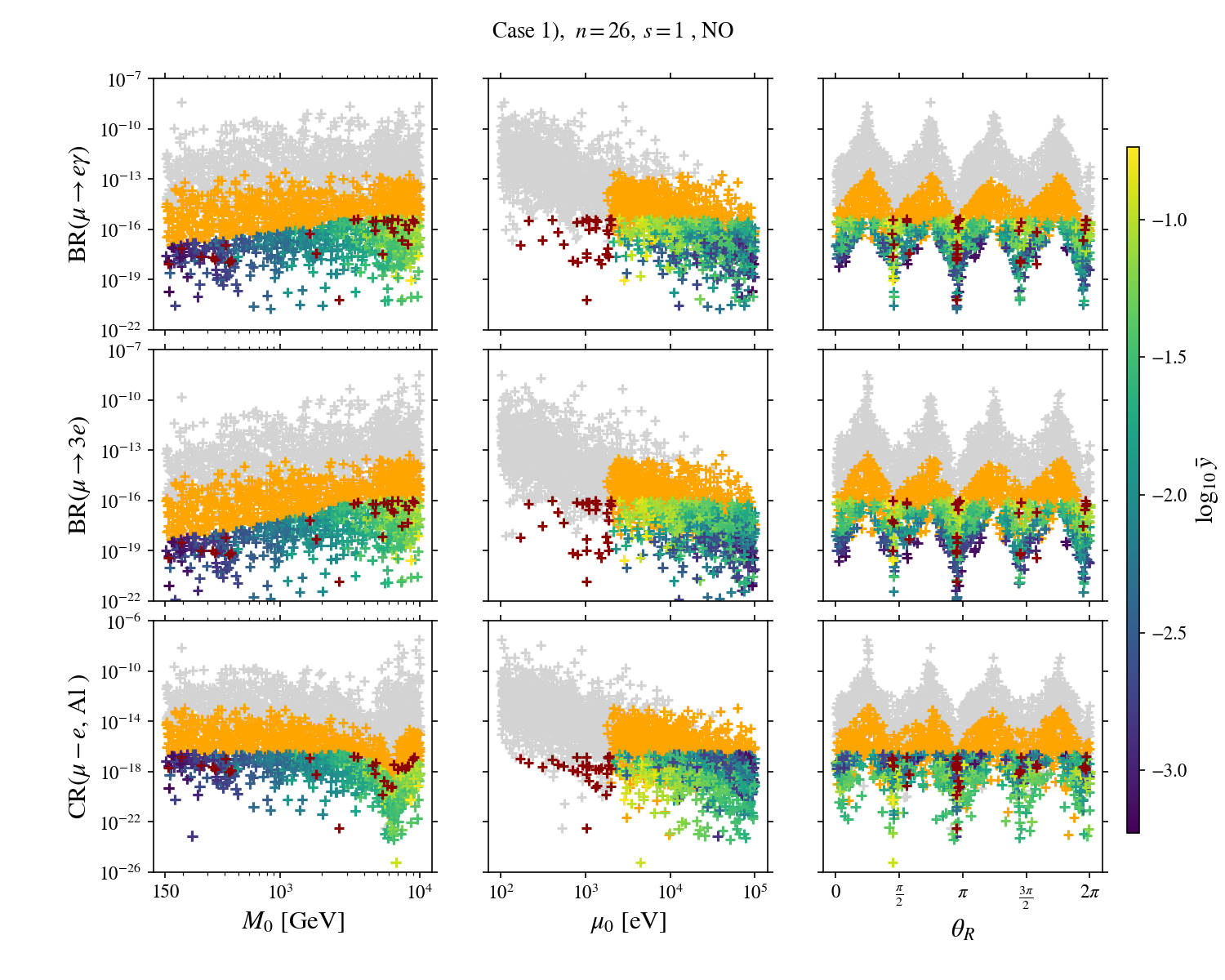}
    \caption{{\bf Case 1). Results of numerical scan for \mathversion{bold} $\mathrm{BR} (\mu\to e \gamma)$, $\mathrm{BR} (\mu\to 3 \, e)$ and $\mathrm{CR} (\mu-e, \mathrm{Al})$ varying $M_0$, $\mu_0$ and $\theta_R$
    \mathversion{normal}} in the ranges in eqs.~(\ref{eq:rangeM0}),~(\ref{eq:rangemu0}) and~(\ref{eq:rangethetaR}), respectively. The parameters $n$ and $s$ are the same as in figs.~\ref{fig:Case1_thetaL} and~\ref{fig:Case1_M0}.
    Lepton mixing angles are always accommodated well.
    The light neutrino mass ordering is fixed to NO and $m_0$ to $m_0=0.03 \, \mbox{eV}$. The average Yukawa coupling $\overline{y}$ is defined in eq.~(\ref{eq:ybar}). Points in grey, orange and red are excluded by different (current/future) 
    experimental bounds, see text for details. Points in other colours correspond to a certain value of $\overline{y}$, see colour bar, and pass all the imposed limits. 
    }
    \label{fig:Case1_scan}
\end{figure}
In fig.~\ref{fig:Case1_M0}, we vary in addition to $\theta_R$ also the scale $M_0$ (up to $50 \, \mbox{TeV}$), while still fixing $\mu_0=1 \, \mbox{keV}$. The value of $\theta_L$ is always fitted such that the lepton mixing angles are accommodated best. 
Furthermore, light neutrino masses are taken as in fig.~\ref{fig:Case1_thetaL}. We observe that the smaller values of $\theta_R$ that we show give rise to smaller BRs and CR and that the dependence on the angle $\theta_R$ appears 
less pronounced for light neutrino masses with IO and than with NO. Furthermore, we see that for the larger values of $\theta_R$ slightly smaller values of the BRs and CR are achieved for light neutrino masses with IO.
The dependence on $M_0$ is in general mild, in particular for $\mathrm{BR} (\mu\to e \gamma)$, if not for the cancellation occurring in the CR, see comments in section~\ref{sec:anacon}. As already shown in fig.~\ref{fig:Case1_thetaL},
the future bound on $\mathrm{BR} (\mu\to e \gamma)$ is only effective for the larger values of $\theta_R$, while the limit expected from Phase 1 of Mu3E reduces the available parameter space for all $\theta_R$ 
to $M_0 \lesssim 10 \, (3) \, \mbox{TeV}$ for light neutrino masses with NO (IO). Imposing the prospective bound of Mu3E Phase 2 on $\mathrm{BR} (\mu\to 3 \, e)$ leaves for $\mu_0= 1 \, \mbox{keV}$ 
only small values of $M_0$, $M_0 \lesssim 200 \, \mbox{GeV}$, for light neutrino masses 
with NO unconstrained, if $\theta_R=0$. In contrast to this, the forecasts from COMET and Mu2e clearly exclude this possibility and would only allow for viable parameter space close to $M_0 \approx 6.5 \, \mbox{TeV}$, where the cancellation
occurs.

\begin{figure}[t!]
    \centering
    \includegraphics[width=\textwidth]{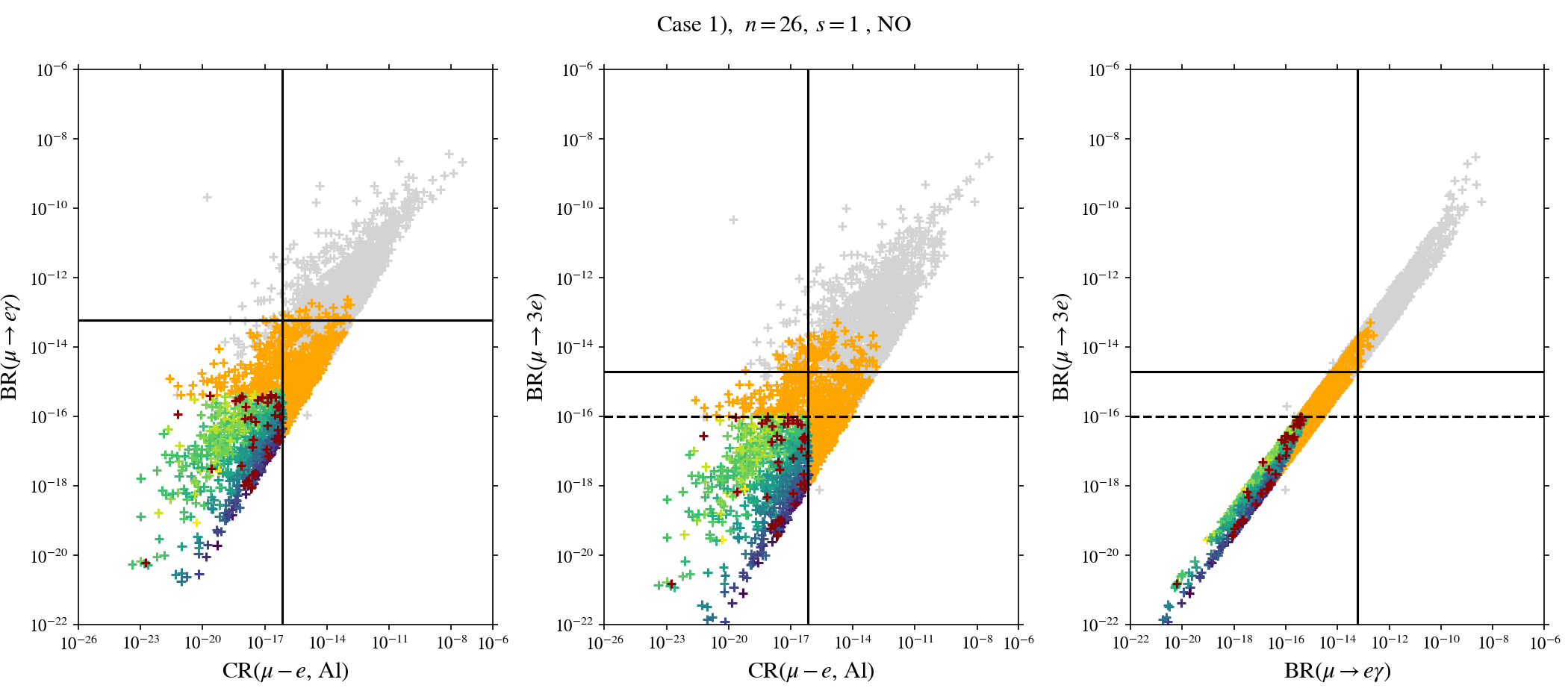}
    \caption{{\bf Case 1). Comparison of results for \mathversion{bold}$\mathrm{BR} (\mu\to e \gamma)$, $\mathrm{BR} (\mu\to 3 \, e)$ and $\mathrm{CR} (\mu-e, \mathrm{Al})$\mathversion{normal}} for $n=26$, $s=1$ and light neutrino masses
    with NO and $m_0=0.03 \,\mathrm{eV}$. The colour-coding is the same as in fig.~\ref{fig:Case1_scan}. For $\mathrm{BR} (\mu\to e \gamma)$ the black line indicates the future bound from MEG II, for $\mathrm{BR} (\mu\to 3 \, e)$ 
    the solid (dashed) black line corresponds to the expected limit from Mu3E Phase 1 (2) and for $\mathrm{CR} (\mu-e, \mathrm{Al})$ the black line represents the prospective bound from COMET.}
    \label{fig:Case1_correlationBRsCR}
\end{figure}
To comprehensively analyse the parameter space that we have scanned, we display in fig.~\ref{fig:Case1_scan} the results for the BRs and CR either as function of the scale $M_0$, $\mu_0$ or the free angle $\theta_R$.\footnote{Note that this plot can also be found in the proceedings~\cite{NOW2024}.} 
The different colour-coding indicates, on the one hand, which experimental constraint can/does exclude a data point, i.e.~for grey points at least one of the prospective bounds on $\mathrm{BR} (\mu\to e \gamma)$, 
$\mathrm{BR} (\mu\to 3 \, e)$ and $\mathrm{CR} (\mu-e, \mathrm{Al})$ is violated as well as the current limit on at least one of the elements of the matrix $\eta$, see eq.~(\ref{eq:etalimits}), while the orange points 
correspond to the situation in which at least one of the prospective bounds is violated, but all current constraints on the magnitude of the matrix elements $\eta_{\alpha\beta}$ are fulfilled and red points, eventually, represent data points 
that are excluded by
the current limits on $\eta$, but not by the future bounds on the three studied $\mu-e$ transitions.\footnote{We remark that not in all scans data points corresponding to red points are found.} 
On the other hand, the different colours which can be read off from the colour bar signal the average value of the Yukawa coupling, $\overline{y}$, which is defined as 
\begin{equation}
\label{eq:ybar}
\overline{y}= \frac 13 \, \left( y_1 + y_2 + y_3 \right) 
\end{equation}
for each viable data point that satisfies both the expected limits on the mentioned cLFV processes and the current bounds on $\eta_{\alpha\beta}$.\footnote{Note that first the grey points, then the orange ones, then the viable points 
which are coloured according to the colour bar and eventually the red points are plotted. We notice that the areas of the grey and the orange points can overlap, whereas the areas occupied by the viable points and by the grey 
and orange points usually hardly have an overlap.} 
We clearly see that a minimum value of $\mu_0$ around $2 \, \mathrm{keV}$ 
 is required in order to pass all bounds, while values of $M_0$ in the entire studied range, see eq.~(\ref{eq:rangeM0}), can lead to viable points (with corresponding size of $\overline{y}$). 
As mentioned, the behaviour of $\eta_{e\mu}$ and, thus, of the BRs and CR with respect to $\theta_R$ is determined by the behaviour of the Yukawa couplings. 
In particular, we observe enhanced values of the cLFV signals for $\cos 2 \, \theta_R$ being close to zero, since it entails at 
least one large Yukawa coupling, which can easiest be seen from eqs.~(\ref{eq:mfyfthRCase1_strongNO}) and~(\ref{eq:mfyfthRCase1_strongIO}). Smaller values of these signals are obtained, 
if the magnitude of $\cos 2 \, \theta_R$ is maximised, as this allows for smaller Yukawa couplings. As we see, the prospective bounds on $\mathrm{BR} (\mu\to 3 \, e)$ and $\mathrm{CR} (\mu-e, \mathrm{Al})$
are both effective, whereas all viable points correspond to $\mathrm{BR} (\mu\to e \gamma) \lesssim 6 \times 10^{-16}$, which is two orders of magnitude smaller than
the future limit, see eq.~(\ref{eq:BRsmuefut}). The results in fig.~\ref{fig:Case1_scan} are for light neutrino
masses with NO and $m_0=0.03 \, \mbox{eV}$. Those obtained for smaller $m_0$ (in particular, $m_0=0$) and the same light neutrino mass ordering look similar. Also light neutrino masses with IO and larger values of $m_0$, i.e.~the 
benchmark value $m_0=0.015 \, \mbox{eV}$, exhibit the same behaviour. However, if $m_0$ is sufficiently small (we have checked $m_0=0$ and $m_0=10^{-5} \, \mbox{eV}$ numerically), the dependence of the BRs and CR 
 on $\theta_R$ differs, since a cancellation among the terms of $\eta_{e\mu}$ becomes possible for certain values of $\theta_R$. Nevertheless, the expected size of the BRs and CR is the same as for larger $m_0$ and light neutrino masses with NO. 

Lastly, we display possible correlations among $\mathrm{BR} (\mu\to e \gamma)$, $\mathrm{BR} (\mu\to 3 \, e)$ and $\mathrm{CR} (\mu-e, \mathrm{Al})$ in fig.~\ref{fig:Case1_correlationBRsCR}.
Since all three observables mainly depend on $|\eta_{e \mu}|^2$, see section~\ref{sec:anacon}, we expect them to be proportional to each other. This is to a certain extent also visible in fig.~\ref{fig:Case1_correlationBRsCR}.
The shown results confirm the exclusion potential of the future experiments for the three different observables: limited for $\mathrm{BR} (\mu\to e \gamma)$ alone, considerable for $\mathrm{BR} (\mu\to 3 \, e)$
and even stronger for  $\mathrm{CR} (\mu-e, \mathrm{Al})$. 

%%%%%%%%%%%%%%%%%%%%%%%%%%%%%%%%%%%%%%%%%%%%%%%%%%%%%%%
\subsection{Results for Case 2)}
\label{sec:rescase2}
%%%%%%%%%%%%%%%%%%%%%%%%%%%%%%%%%%%%%%%%%%%%%%%%%%%%%%%

In the discussion of the results for Case 2), it is decisive whether $t$ is odd (equivalent to $u$ odd, see eq.~(\ref{eq:defuv}))
or even (meaning $u$ even), since for $t$ odd the matrix combination in square brackets in eq.~(\ref{eq:mnuform}) is not automatically diagonal 
 and the light neutrino masses depend on the angle $\theta_R$. Indeed, the formulae for them are the same as those in eqs.~(\ref{eq:mfyfthRCase1})-(\ref{eq:mfyfthRCase1_strongIO}) for Case 1) with $\cos 2 \, \theta_R$
 being replaced by $\sin 2 \, \theta_R$.  
Consequently, we in general observe a dependence of the flavour observables on this angle if $t$ is odd, while this does not happen
for $t$ even, because the mentioned matrix combination is then always diagonal and the light neutrino masses do not depend on $\theta_R$, see eq.~(\ref{eq:mfyf}).
As expected and observed in the numerical analysis, the results do not depend on whether $s$ is even or odd.

Before focussing on values of $u$ (or equivalently of the combination $2 \, s - t$) that allow for a good fit of the experimental data on the lepton mixing angles, see comment below eq.~(\ref{eq:defuv}), we study the general dependence
of the results for the three different $\mu - e$ transitions on the parameters of this scenario. We first remark that, similar to the parameter $s$ for Case 1), see eq.~(\ref{eq:U0Case1ssep}), the parameter $v$ for Case 2)
does not appear in the expression for $\eta$ as given in eq.~(\ref{eq:etaU0}), since the matrix $U_0 (\theta)$ can be written as
\begin{equation}
U_0 (\theta) = U_0 (\theta, v=0) \,  \mathrm{diag} ( e^{i \, \phi_v/6}, e^{-i \, \phi_v/3}, e^{i \, \phi_v/6}) \; ,
\end{equation}
where $U_0 (\theta, v=0)$ refers to the matrix $U_0 (\theta)$ with $v=0$. Here we have used the form of $\Omega ({\bf 3})$ shown in eq.~(\ref{eq:Omega3_Case2}) in appendix~\ref{app:matrices} 
and that the relevant rotation occurs in the (13)-plane. 

\begin{figure}[t!]
    \centering
    \includegraphics[width=\textwidth]{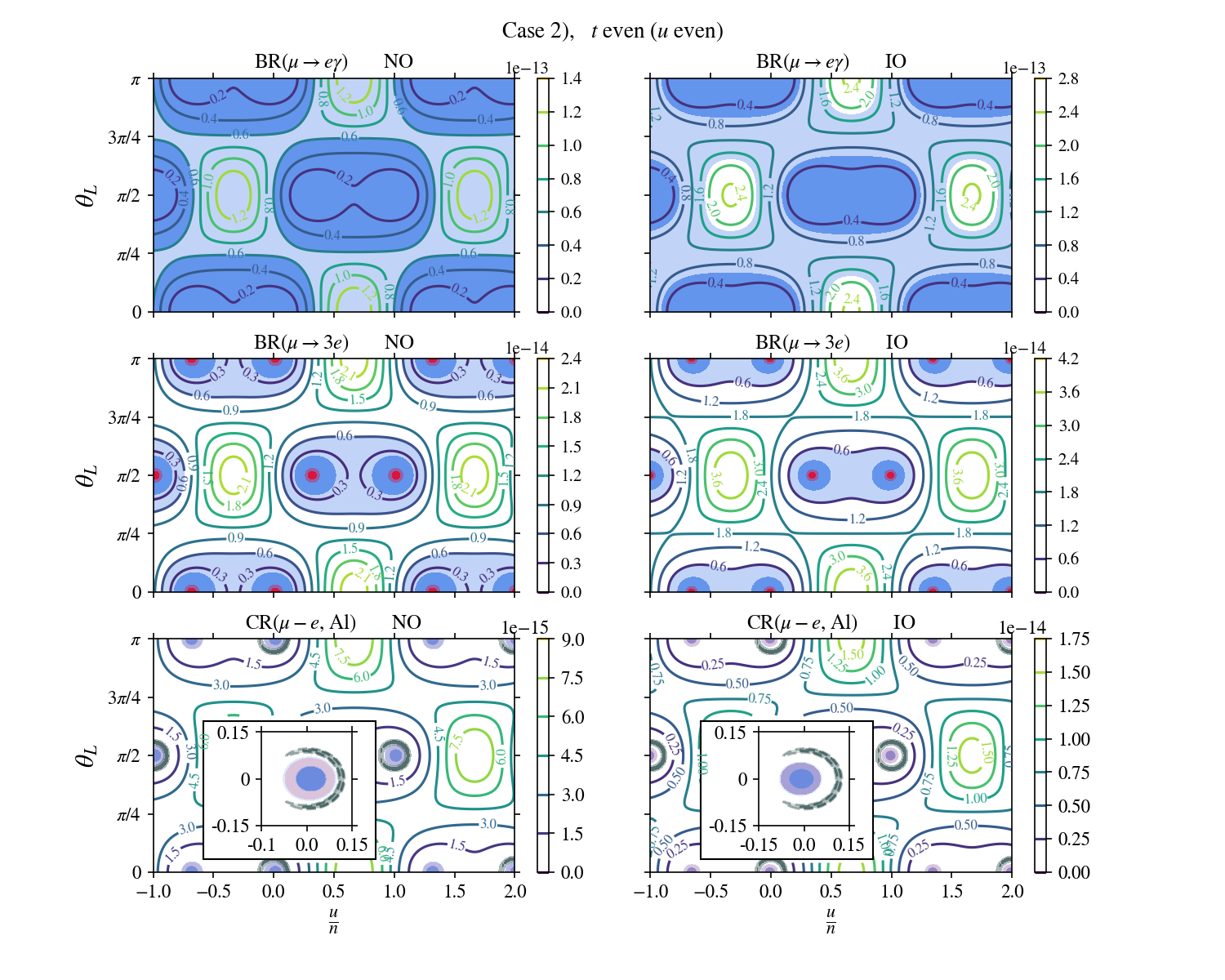}
    \caption{{\bf Case 2), \mathversion{bold}$t$ even ($u$ even). Predictions for $\mathrm{BR} (\mu\to e \gamma)$, $\mathrm{BR} (\mu\to 3 \, e)$ and $\mathrm{CR} (\mu-e, \mathrm{Al})$ in the $\frac un-\theta_L$-plane} in the upper, 
    middle and lower row.\mathversion{normal}  Left (right) plots are for light neutrino masses with NO (IO)
  and $m_0= 0.03 \, (0.015) \, \mathrm{eV}$. The scales $\mu_0$ and $M_0$ are set to $\mu_0=1 \, \mbox{keV}$ and $M_0=3 \, \mbox{TeV}$, respectively. Dark (light) blue and red regions indicate the 
  regions of parameter space compatible at the $1 \, \sigma$ ($3 \, \sigma$) level with different prospective bounds on the BRs and CR, see text for details. In the plots for $\mathrm{CR} (\mu-e, \mathrm{Al})$ grey regions
    indicate the parameter space in which the lepton mixing angles can be accommodated, i.e.~dark (light) grey for $\chi^2 \leq 100 \, (300)$ and in the insets $\chi^2 \leq 27$.
    Simplified versions of these plots, with only contour lines or only experimentally preferred regions, can be found in appendix~\ref{app:Case2Case3b1}, see figs.~\ref{fig:Case2_un_thetaL_teven_simplified_contours} and~\ref{fig:Case2_un_thetaL_teven_simplified_limits}.
    }
    \label{fig:Case2_un_thetaL_teven}
\end{figure}

We, then, display for the choice $t$ even ($u$ even) the results for $\mathrm{BR} (\mu\to e \gamma)$, $\mathrm{BR} (\mu\to 3 \, e)$ and $\mathrm{CR} (\mu-e, \mathrm{Al})$ in 
the $\frac{u}{n}-\theta_L$-plane\footnote{According to the definition of $u$ in eq.~(\ref{eq:defuv}) it varies between $- (n-1)$ and $2 \, (n-1)$ and should be an integer. In this analysis, we treat $\frac un$ as
continuous parameter in the interval $-1 \leq \frac un \leq 2$.} for a fixed value of $\mu_0$ and $M_0$, 
$\mu_0=1 \, \mbox{keV}$ and $M_0=3 \, \mbox{TeV}$, and light neutrino masses with NO (IO) and non-zero $m_0$, see eq.~(\ref{eq:m0bm}), in the left (right) of fig.~\ref{fig:Case2_un_thetaL_teven}.
As mentioned, the angle $\theta_R$ is irrelevant and thus not specified. For $\mathrm{BR} (\mu\to e \gamma)$, shown in the upper row, the dark (light) blue regions indicate the parameter space in which
the prospective bound from the experiment MEG II is passed at the $1 \, \sigma$ ($3 \, \sigma$) level, see eq.~(\ref{eq:BRsmuefut}). As one can see, the impact of this future limit is mild on the considered parameter space, while being 
slightly stronger for light neutrino masses with IO. Similarly, for $\mathrm{BR} (\mu\to 3 \, e)$, see plots in the middle row, the dark (light) blue regions indicate the parameter space
compatible with the expected bound from Mu3E Phase 1 at the $1 \, \sigma$ ($3 \, \sigma$) level, see eq.~(\ref{eq:BRsmuefut}). We observe that about half of the parameter space in the shown plane can be disfavoured; again, for light
neutrino masses with IO the exclusion reach is slightly larger. The impact of the prospective limit from Mu3E Phase 2 at the $1 \, \sigma$ ($3 \, \sigma$) level on the parameter space is
displayed with the dark (light) red regions. We clearly see that the viable parameter space is considerably reduced. The lower row of fig.~\ref{fig:Case2_un_thetaL_teven} contains the corresponding plots
 for $\mathrm{CR} (\mu-e, \mathrm{Al})$. As expected, the future bounds from COMET and Mu2e, respectively, see eq.~(\ref{eq:CRmuefut}), can only be passed in small regions of the parameter space.
These are shown in dark (light) red for the bound from COMET at the $1 \, \sigma$ ($3 \, \sigma$) level and in dark (light) blue for Mu2e at the same $\sigma$ levels. Since these bounds are very similar, the
regions are largely super-imposed. As explained in section~\ref{sec:leptonmixing}, the ratio $\frac un$  and the free angle $\theta_L$ are also constrained by the requirement to accommodate the lepton mixing angles well. For this reason, 
we show in the plots for $\mathrm{CR} (\mu-e, \mathrm{Al})$ in dark (light) grey the parameter space which leads to $\chi^2 \leq 100 \, (300)$, when using the $\chi^2$-function as defined in eq.~(\ref{eq:chi2theta}) in 
appendix~\ref{app:numerics}. These large values are chosen in order to ensure the visibility of the corresponding parameter space in the $\frac un-\theta_L$-plane. In the insets the regions with $\chi^2 \leq 27$
are displayed in grey. While compatible with the future bound from MEG II and mostly with the one from Mu3E Phase 1,
this part of the parameter space will be excluded by Mu3E Phase 2 and the upcoming searches for $\mu-e$ conversion in aluminium for the chosen values of the two scales $\mu_0$ and $M_0$.

\begin{figure}[t!]
    \centering
    \includegraphics[width=\textwidth]{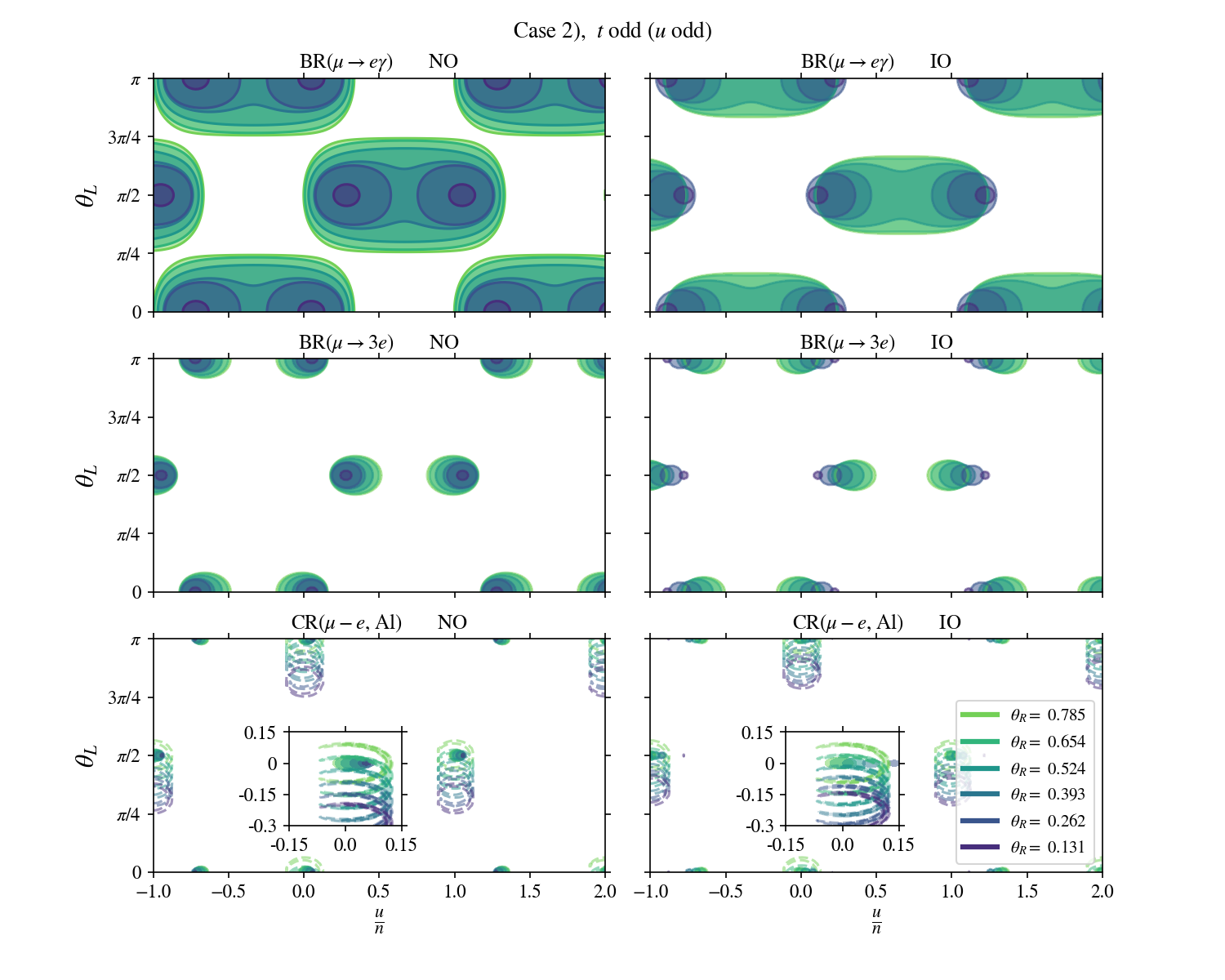}
    \caption{{\bf Case 2), \mathversion{bold}$t$ odd ($u$ odd). Predictions for $\mathrm{BR} (\mu\to e \gamma)$, $\mathrm{BR} (\mu\to 3 \, e)$ and $\mathrm{CR} (\mu-e, \mathrm{Al})$ in the $\frac un-\theta_L$-plane} in the upper, 
    middle and lower row.\mathversion{normal} Light neutrino masses, $\mu_0$ and $M_0$ are chosen as in fig.~\ref{fig:Case2_un_thetaL_teven}. For six different values of $\theta_R$, the parameter space 
  that is compatible at the $1 \, \sigma$ level
  with the future bound on $\mathrm{BR} (\mu\to e \gamma)$ from MEG II, on $\mathrm{BR} (\mu\to 3 \, e)$ from Mu3E Phase 1 and on $\mathrm{CR} (\mu-e, \mathrm{Al})$ from COMET, respectively, is displayed. 
   In the plots for $\mathrm{CR} (\mu-e, \mathrm{Al})$, different light-coloured regions (with dashed contours)
    indicate the parameter space in which the lepton mixing angles can be accommodated (corresponding to $\chi^2 \leq 100$ and in the insets to $\chi^2 \leq 27$).}
    \label{fig:Case2_un_thetaL_todd}
\end{figure}
In fig.~\ref{fig:Case2_un_thetaL_todd} the corresponding results for $t$ odd ($u$ odd) are displayed for several fixed values of $\theta_R$. Again, $\mu_0$ and $M_0$ are set to $\mu_0=1 \, \mbox{keV}$ and $M_0=3 \, \mbox{TeV}$.
 As for $t$ even, we study the three different $\mu-e$ transitions for both light neutrino mass orderings, see left and right plots, respectively. In the upper row the parameter space which is compatible with the prospective bound 
on $\mathrm{BR} (\mu\to e \gamma)$ from MEG II at the $1 \, \sigma$ level is shown in different colours corresponding to the six different values of $\theta_R$. As we can see, the smaller the shown value of $\theta_R$ is, the smaller the allowed parameter space in the $\frac un-\theta_L$-plane is.
Furthermore, the viable parameter space for light neutrino masses with IO is slightly reduced compared to the one for NO. The same observations can be made when analysing the parameter space allowed by the future limit on $\mathrm{BR} (\mu\to 3 \, e)$ from Mu3E 
Phase 1 at the $1 \, \sigma$ level. We note that, as expected, the resulting allowed portion of parameter space is considerably smaller than in the case of $\mathrm{BR} (\mu\to e \gamma)$. Applying the prospective bound on $\mu-e$ conversion in aluminium from COMET 
at the $1 \, \sigma$ level  to the parameter space, we find that only tiny regions remain allowed in the $\frac un-\theta_L$-plane, see plots in the lower row in fig.~\ref{fig:Case2_un_thetaL_todd}. In these plots, we also display the 
areas in which the lepton mixing angles are accommodated to a certain degree, i.e.~$\chi^2 \leq 100$, see eq.~(\ref{eq:chi2theta}) in appendix~\ref{app:numerics}, in different light colours (and with dashed contours) depending on the value 
of $\theta_R$.\footnote{Again, a larger value of $\chi^2$ is used in order to increase visibility.} In the insets, instead the parameter space with $\chi^2 \leq 27$ is displayed.
 Overall, we see that it is very challenging, if not impossible, for all shown values of $\theta_R$ to reconcile the requirement to fit the lepton mixing
angles well with passing the expected strong bounds on $\mu-e$ conversion in aluminium.
We remark that for $\theta_R \approx \frac{\pi}{4}$ we obtain areas very similar to those for $t$ even. This is consistent, since in this case the matrix combination in square brackets in eq.~(\ref{eq:mnuform}) is (nearly) diagonal and the light neutrino masses are (almost) proportional to the square of the Yukawa couplings, see eq.~(\ref{eq:mfyf}), as it always happens for $t$ even.

\begin{figure}[t!]
    \centering
    \includegraphics[width=\textwidth]{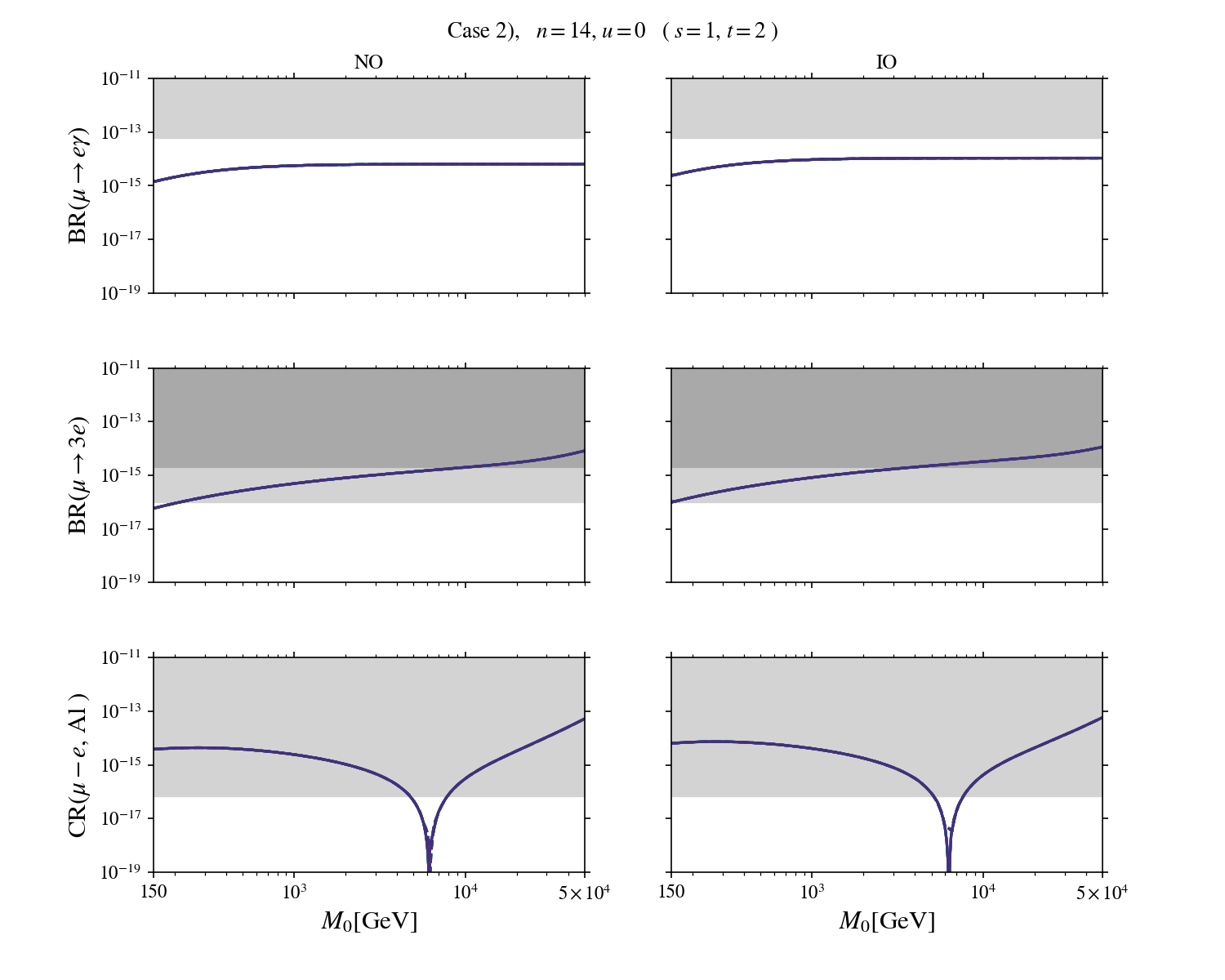}
    \caption{{\bf Case 2), \mathversion{bold}$t$ even ($u$ even). Predictions for $\mathrm{BR} (\mu\to e \gamma)$, $\mathrm{BR} (\mu\to 3 \, e)$ and $\mathrm{CR} (\mu-e, \mathrm{Al})$ as a function of $M_0$\mathversion{normal}}
    in the upper, middle and lower row. The group theory parameters are chosen as $n=14$, $s=1$ and $t=2$, leading to $u=0$. 
    Light neutrino masses with NO (IO) correspond to the plots in the left (right). Solid and dashed lines (almost always super-imposed) refer to the two different values of the angle $\theta_L$
    that allow for best-fitting the lepton mixing angles.
    For the range of $M_0$, the value of $\mu_0$ and the grey regions, see fig.~\ref{fig:Case1_M0} of Case 1).}
    \label{fig:Case2_M0_teven}
\end{figure}
In order to further study the results for Case 2), we specify the values of $n$ and $u$, corresponding to certain combinations of $s$ and $t$, such that the lepton mixing angles can be accommodated well. 
As has been shown in~\cite{Hagedorn:2014wha,Hagedorn:2021ldq,Drewes:2022kap}, a viable choice is
\begin{equation}
n=14 \;\;\; \mbox{together with} \;\;\; u=-1,0,1 \; ,
\end{equation}
since then $\frac{u}{n}$ is small enough. 
In terms of the parameters $s$ and $t$, see eq.~(\ref{eq:defuv}), the mentioned values of $u$ correspond to e.g.~$s=0$ and $t=0$ as well as $s=1$ and $t=2$ for $u=0$, $s=0$ and $t=1$ for $u=-1$ 
and $s=1$ and $t=1$ for $u=1$.\footnote{The choice $s=0$ and $t=0$ leads to trivial Majorana phases, since $I_1$ and $I_2$ are zero, see e.g.~\cite{Hagedorn:2014wha}.} 
In the following, we numerically explore these combinations of $s$ and $t$.
As is known~\cite{Hagedorn:2014wha}, two different values of the free angle lead to a good fit of the lepton mixing parameters and we consider both of these.
 Like for Case 1), we first analyse the behaviour of the BRs and CR for varying $M_0$, while fixing $\mu_0$ to $\mu_0=1 \, \mbox{keV}$, 
and either assuming light neutrino masses with NO or IO. 

For $t$ even, we take $s=1$ and $t=2$ (leading to $u=0$), as shown in fig.~\ref{fig:Case2_M0_teven}.
This leads to $\sin^2 \theta_{13} \approx 0.022$, $\sin^2 \theta_{12} \approx 0.34$ and $\sin^2 \theta_{23}=0.5$, in agreement with~\cite{Hagedorn:2014wha,Hagedorn:2021ldq,Drewes:2022kap}. 
Clearly, there is no dependence on the angle $\theta_R$. Also the two different values of $\theta_L$ that permit a good fit to lepton mixing data lead to nearly identical results,
since the solid and dashed curves are almost always super-imposed. Here the solid curve represents the results for the smaller value of the angle $\theta_L$ (which equals $\theta_L\approx 0.184$ for light neutrino masses of both orderings), 
while the dashed curve corresponds to those obtained for 
the larger value of $\theta_L$ (which sums up to $\pi$ together with the smaller value of $\theta_L$), compare~\cite{Hagedorn:2014wha,Drewes:2022kap}.
In fig.~\ref{fig:Case2_M0_teven} we display the same experimental constraints for the BRs of $\mu \to e \, \gamma$ and $\mu \to 3 \, e$
and $\mathrm{CR} (\mu-e, \mathrm{Al})$ as in fig.~\ref{fig:Case1_M0} for Case 1). The conclusions which can be drawn regarding the parameter space compatible with these future limits are very similar
to those found for Case 1) and $\theta_R=0$, namely the future bound on $\mathrm{BR} (\mu\to e \gamma)$ is passed for the shown interval of $M_0$, while the prospective limit from Mu3E Phase 1 can only constrain
larger masses, i.e.~$M_0 \gtrsim 10 \, (3) \, \mbox{TeV}$ for light neutrino masses with NO (IO). In contrast to this, Mu3E Phase 2 is expected to reduce the allowed parameter space to $M_0 \lesssim 200 \, \mbox{GeV}$ for light neutrino masses with NO
and the future bound on $\mathrm{CR} (\mu-e, \mathrm{Al})$ forces $M_0$ to lie in the interval $5 \, \mbox{TeV} \lesssim M_0 \lesssim 8 \, \mbox{TeV}$ for both light neutrino mass orderings -- close to the value of $M_0$ which leads to a (complete) cancellation
in the CR, see eq.~(\ref{eq:M0cancelCR}).

\begin{figure}[t!]
    \centering
    \includegraphics[width=\textwidth]{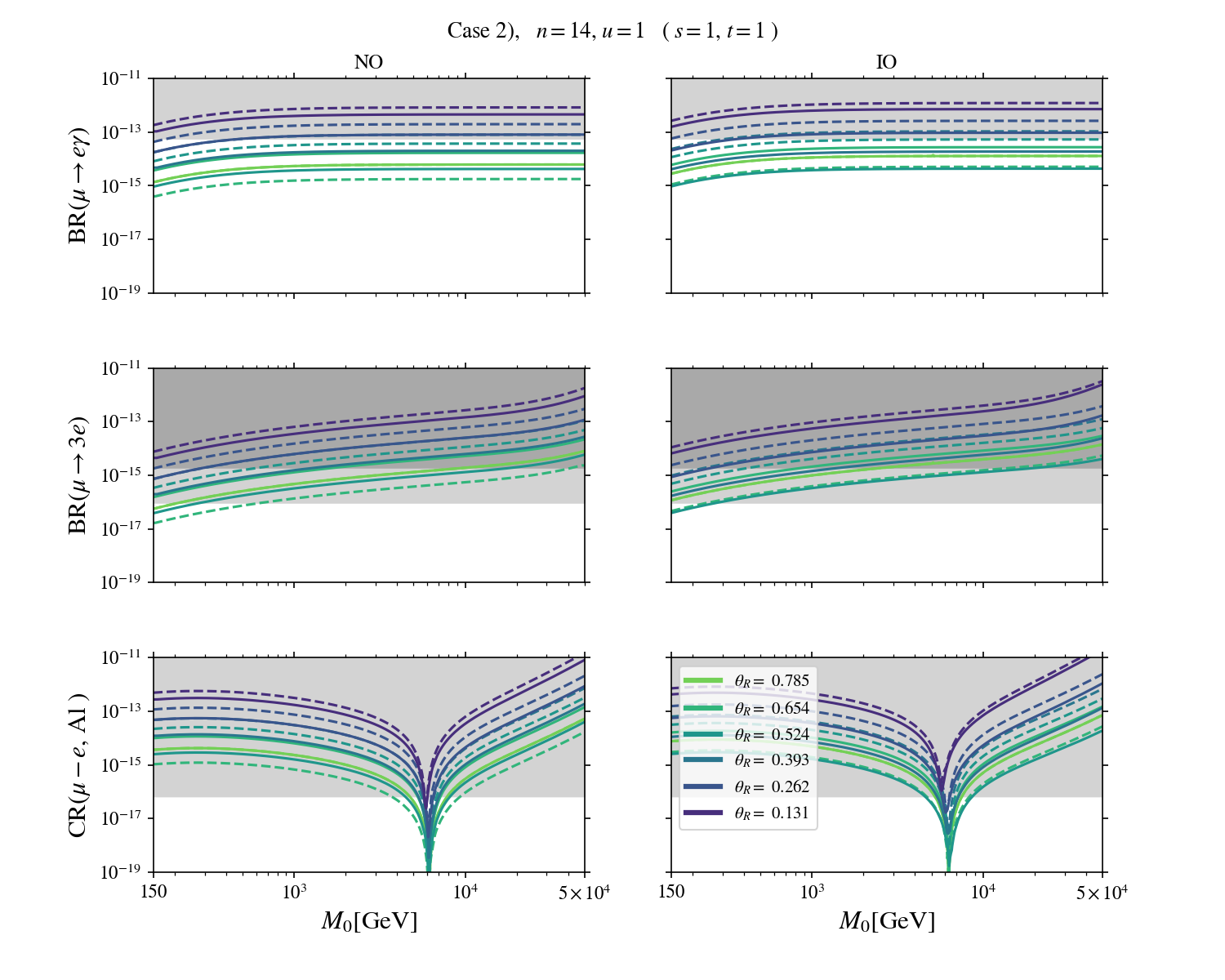}
    \caption{{\bf Case 2),  \mathversion{bold}$t$ odd ($u$ odd). Predictions for $\mathrm{BR} (\mu\to e \gamma)$, $\mathrm{BR} (\mu\to 3 \, e)$ and $\mathrm{CR} (\mu-e, \mathrm{Al})$ as a function of $M_0$\mathversion{normal}}
    in the upper, middle and lower row. The group theory parameters are chosen as $n=14$, $s=1$ and $t=1$, corresponding to $u=1$. The conventions in these plots are the same as in figs.~\ref{fig:Case1_M0} and~\ref{fig:Case2_M0_teven}.}
    \label{fig:Case2_M0_todd}
\end{figure}
 \begin{figure}[t!]
    \centering
    \includegraphics[width=\textwidth]{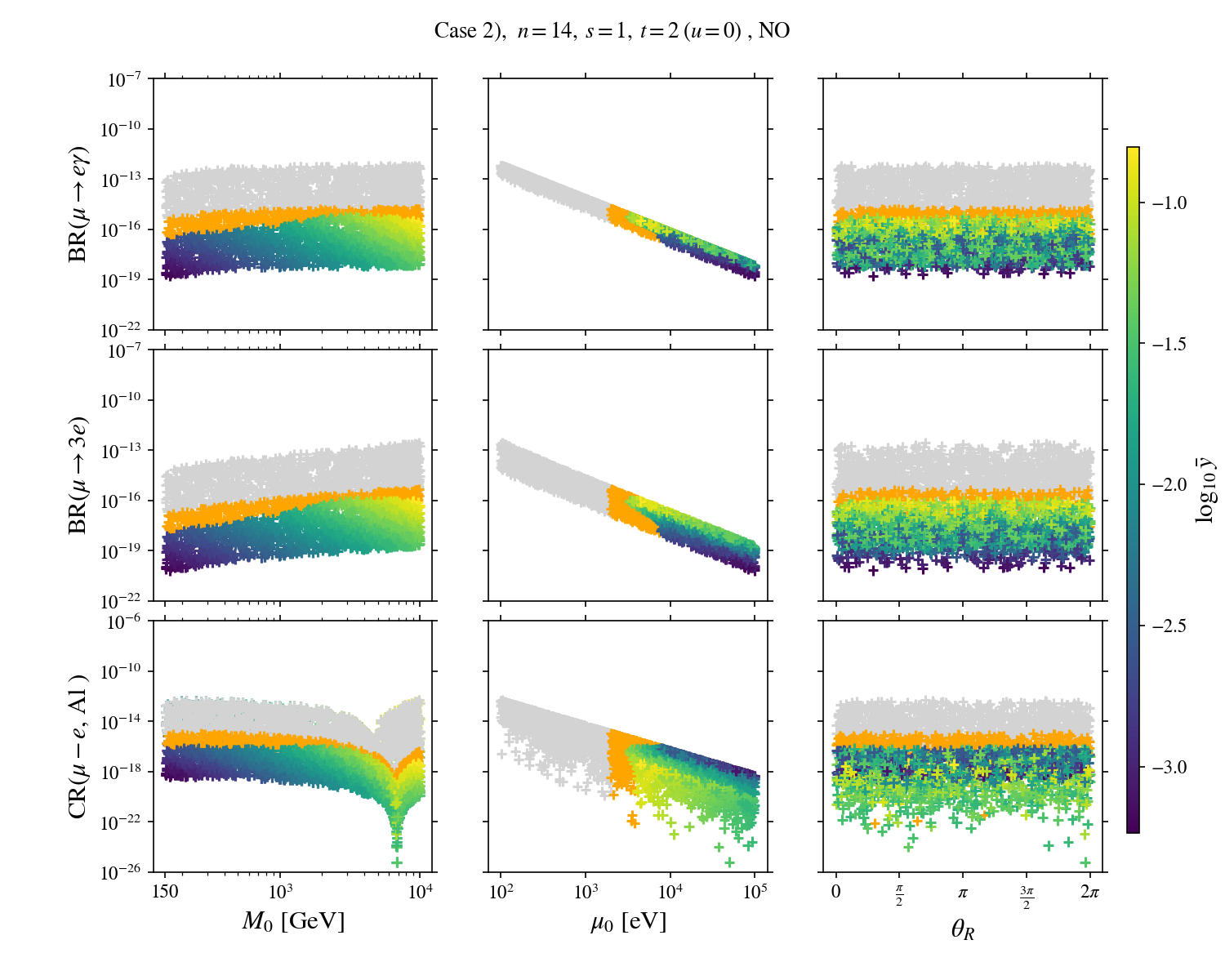}
    \caption{{\bf Case 2), \mathversion{bold}$t$ even ($u$ even). Results of numerical scan for  $\mathrm{BR} (\mu\to e \gamma)$, $\mathrm{BR} (\mu\to 3 \, e)$ and $\mathrm{CR} (\mu-e, \mathrm{Al})$ varying $M_0$, $\mu_0$ and $\theta_R$
    \mathversion{normal}} in the ranges in eqs.~(\ref{eq:rangeM0}),~(\ref{eq:rangemu0}) and~(\ref{eq:rangethetaR}), respectively. As example of the group theory parameters $n=14$, $s=1$ and $t=2$, corresponding to $u=0$, is chosen.     
    For conventions, see fig.~\ref{fig:Case1_scan}.}
    \label{fig:Case2_scan_teven}
\end{figure}
For $t$ odd, we employ as example $s=1$ and $t=1$ which give rise to $u=1$. For this value of $u$ and $n=14$, the lepton mixing angles read
$\sin^2 \theta_{13} \approx 0.022$, $\sin^2 \theta_{12} \approx 0.34$ and $\sin^2 \theta_{23} \approx 0.56$, see also~\cite{Hagedorn:2014wha,Hagedorn:2021ldq,Drewes:2022kap}. 
Using the same conventions as in figs.~\ref{fig:Case1_M0} and~\ref{fig:Case2_M0_teven}, we display the results in fig.~\ref{fig:Case2_M0_todd} for six different values 
of the angle $\theta_R$ in different colours. While for $t$ even, we hardly see any difference in the results for the two different values of $\theta_L$ that lead to a good fit of the lepton mixing angles, the 
solid and dashed curves can be clearly distinguished in fig.~\ref{fig:Case2_M0_todd}.\footnote{The smaller value of $\overline{\theta}_L$, the angle appearing in the matrix $\tilde U_\nu$, see eq.~(\ref{eq:UnutildethetaLbar}), 
is given by $\overline{\theta}_L\approx0.146$ for light neutrino masses with NO and
$\overline{\theta}_L\approx0.148$ for IO light neutrino masses, while the larger value always sums up to $\pi$ together with the smaller one. For completeness, we mention that for the choice $u=-1$ the same values of $\overline{\theta}_L$
are obtained. For further details, we refer to e.g.~\cite{Hagedorn:2014wha,Drewes:2022kap}.} 
We also observe that the larger of the shown values of $\theta_R$ result in smaller values of the BRs and CR, except for
$\theta_R=\frac{\pi}{4}$. For $\theta_R=\frac{\pi}{4}$ we instead find very similar results as for the combination $s=1$, $t=2$ ($u=0$), see fig.~\ref{fig:Case2_M0_teven}. This is expected, since for $\theta_R=\frac{\pi}{4}$
the matrix combination in square brackets in eq.~(\ref{eq:mnuform}) is automatically diagonal and the light neutrino masses are proportional to the squares of the Yukawa couplings even for $t$ odd.
Comparing the results for light neutrino masses with NO and IO, they appear to be very similar, compare plots in the left and the right in fig.~\ref{fig:Case2_M0_todd}. For the larger values of $\theta_R$
that we show the prospective bound on $\mathrm{BR} (\mu\to e \gamma)$ is passed, while the expected limit from Mu3E Phase 1 is more constraining, especially for larger values of $M_0$.
The future bound from Mu3E Phase 2 can exclude all parameter space for $\mu_0=1 \, \mbox{keV}$, unless $M_0 \lesssim 600 \, (300) \, \mbox{GeV}$ for light neutrino masses with NO (IO) and certain larger values of $\theta_R$ are chosen. 
The limit expected from COMET (and also Mu2e), on the other hand, restricts the allowed parameter space to values of $M_0$ close to $M_0 \approx 6.5 \, \mbox{TeV}$, for which a cancellation occurs in the CR.
 Furthermore, we note that the choice $s=0$ and $t=1$, corresponding to $u=-1$, leads to similar results as shown in fig.~\ref{fig:Case2_M0_todd}.\footnote{For this value of $u$ and $n=14$, the atmospheric mixing angle is best-fitted to $\sin^2 \theta_{23} \approx 0.44$,
while the reactor and the solar mixing angles are like for $u=0$ and $u=1$.}  

\begin{figure}[t!]
    \centering
    \includegraphics[width=\textwidth]{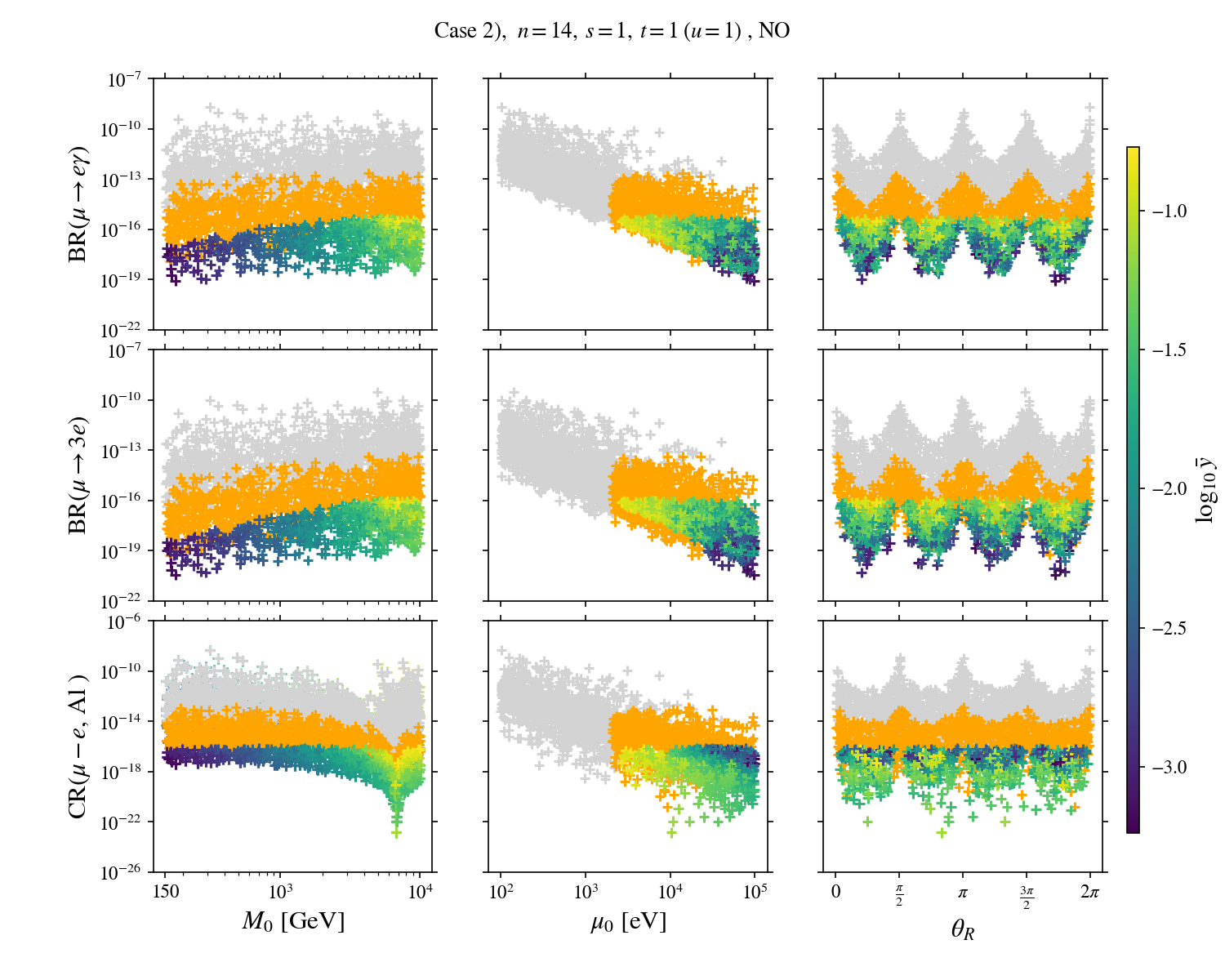}
    \caption{{\bf Case 2), \mathversion{bold}$t$ odd ($u$ odd). Results of numerical scan for  $\mathrm{BR} (\mu\to e \gamma)$, $\mathrm{BR} (\mu\to 3 \, e)$ and $\mathrm{CR} (\mu-e, \mathrm{Al})$ varying $M_0$, $\mu_0$ and $\theta_R$
    \mathversion{normal}} in the ranges in eqs.~(\ref{eq:rangeM0}),~(\ref{eq:rangemu0}) and~(\ref{eq:rangethetaR}), respectively. As example of the group theory parameters $n=14$, $s=1$ and $t=1$, meaning $u=1$, is chosen.     
    For conventions, see fig.~\ref{fig:Case1_scan}.}
    \label{fig:Case2_scan_todd}
\end{figure}
Before moving on to the results of the numerical scans, we estimate the size of the BRs and CR with the help of the analytic approximations, found in section~\ref{sec:anacon}. Similar to Case 1), see eq.~(\ref{eq:etaexpressionCase1}), 
we first express the most relevant element
of the matrix $\eta$, $\eta_{e \mu}$, in terms of $\Delta y_{ij}^2$ and lepton mixing observables which replace the combinations of $\theta_L$ and $\frac{u}{n}$ that show up 
\begin{equation}
\label{eq:etaexpressionCase2}
\eta_{e \mu} = \frac{\eta_0^\prime}{6} \, \left( 2 \, \Delta y_{21}^2 + 3 \, (1 - 2 \, (\sin^2 \theta_{13} + \sin^2 \theta_{23} - \sin^2 \theta_{13} \, \sin^2 \theta_{23}) - 6 \, i \, J_{\mathrm{CP}}) \, \Delta y_{31}^2 \right) \; .
\end{equation}
This formula holds for $t$ even as well as $t$ odd and $\theta_R=\frac{\pi}{4}$. Then, fixing $\mu_0=1 \, \mbox{keV}$ and $M_0=3 \, \mbox{TeV}$, we find for $u=0$ ($u=1$) and light neutrino masses with NO and $m_0=0.03 \, \mbox{eV}$
\begin{equation}
\label{eq:BRsCRestimateCase2NO}
\mathrm{BR} (\mu\to e \gamma) \approx  6.2 \, (5.7) \times 10^{-15} \; , \;\; \mathrm{BR} (\mu\to 3 \, e) \approx 9.9 \, (9.1) \times 10^{-16} \; , \;\; \mathrm{CR} (\mu-e, \mathrm{Al}) \approx 4.3 \, (4.0) \times 10^{-16}
\end{equation}
as well as for IO light neutrino masses and $m_0=0.015 \, \mbox{eV}$
 \begin{equation}
\label{eq:BRsCRestimateCase2IO}
\mathrm{BR} (\mu\to e \gamma) \approx  1.1 \, (1.2) \times 10^{-14} \; , \;\; \mathrm{BR} (\mu\to 3 \, e) \approx 1.7 \, (1.9) \times 10^{-15} \; , \;\; \mathrm{CR} (\mu-e, \mathrm{Al}) \approx 7.4 \, (8.4) \times 10^{-16} \; .
\end{equation}
These estimates are consistent with the results displayed in figs.~\ref{fig:Case2_M0_teven} and~\ref{fig:Case2_M0_todd}.

\begin{figure}[t!]
    \centering
    \includegraphics[width=\textwidth]{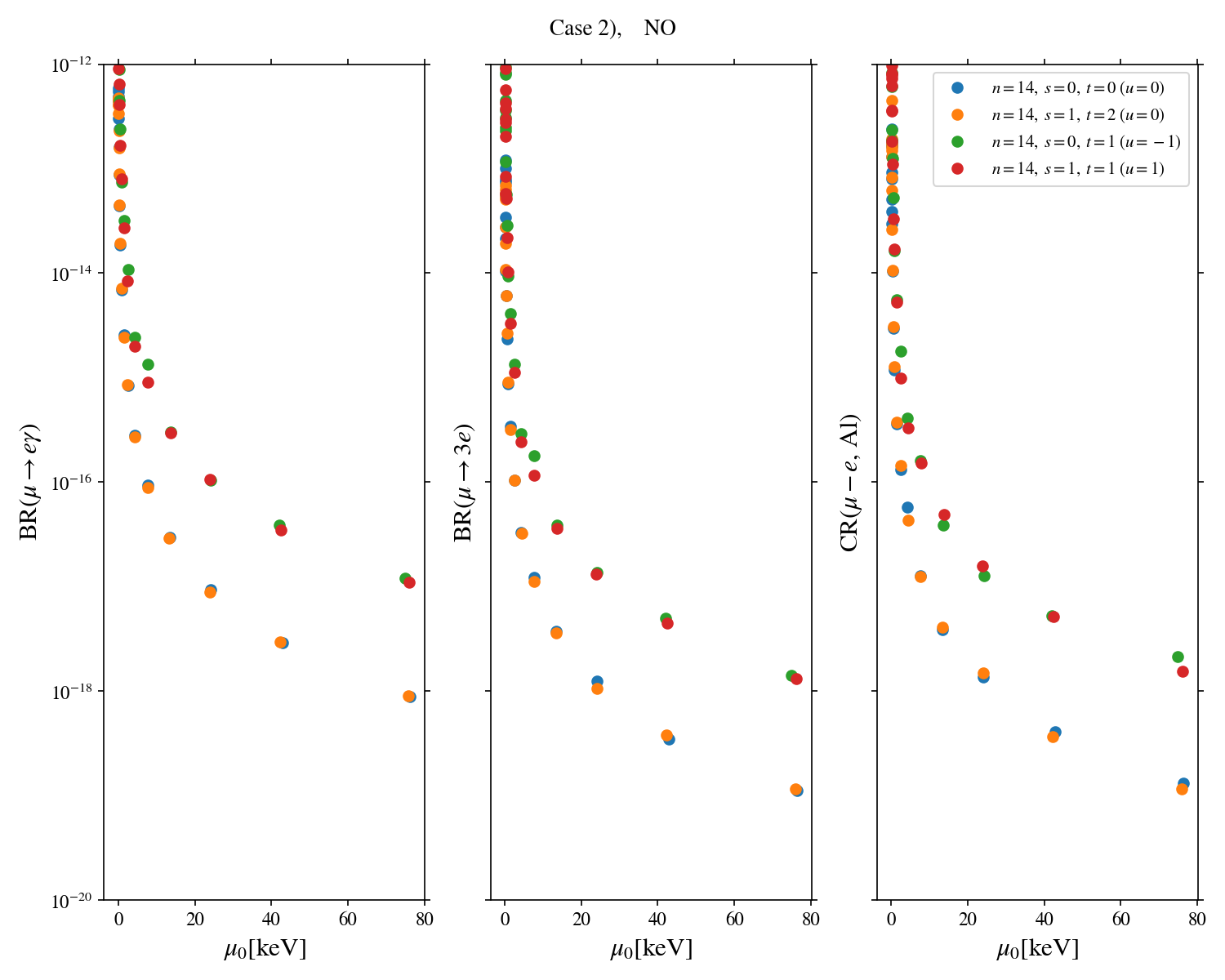}
    \caption{{\bf Case 2). Comparison of results for \mathversion{bold}$\mathrm{BR} (\mu\to e \gamma)$, $\mathrm{BR} (\mu\to 3 \, e)$ and $\mathrm{CR} (\mu-e, \mathrm{Al})$ for different combinations of $s$ and $t$ (choices of $u$)}\mathversion{normal}. 
   The group index $n$ is fixed to $n=14$. Light neutrino masses
    follow NO and $m_0=0.03 \,\mathrm{eV}$. The shown averages are made over the results of the scans in logarithmically uniform bins of the scale $\mu_0$. For details see text.}
    \label{fig:Case2_alltogether}
\end{figure}
 In figs.~\ref{fig:Case2_scan_teven} and~\ref{fig:Case2_scan_todd} we provide results of the numerical scans for $t$ even corresponding to $u$ even ($s=1$, $t=2$ and thus $u=0$) and for $t$ odd/$u$ odd ($s=1$, $t=1$, meaning $u=1$), respectively. 
 The colour-coding is the same as in fig.~\ref{fig:Case1_scan} and the definition of the quantity $\overline{y}$ can be found in eq.~(\ref{eq:ybar}). We see that the same orders of magnitude for the BRs and CR are
 predicted as in Case 1) and that also for Case 2) the prospective limits for both processes $\mu\to 3 \, e$ and $\mu-e$ conversion in aluminium can be reached, while $\mathrm{BR} (\mu\to e \gamma)$ is 
 expected to be smaller than $ 6 \times 10^{-16}$ in the portion of parameter space that is allowed. Furthermore, this part of the parameter space is characterised by $\mu_0$ being larger than
 $2 \, \mbox{keV}$, whereas the scanned intervals of $M_0$ and $\theta_R$ are not constrained. Nevertheless, we also observe certain differences compared to the results for Case 1). In particular, both BRs reveal lower limits,
 $\mathrm{BR} (\mu\to e \gamma) \gtrsim 10^{-19}$ and  $\mathrm{BR} (\mu\to 3 \, e)  \gtrsim 10^{-21}$, and the CR is predicted to be larger than $5 \times 10^{-22}$ for values of $M_0$ which are not too close to $M_0 \approx 6.5 \, \mbox{TeV}$,
 where the cancellation occurs.\footnote{\label{ft:cancel} We actually exclude all values of the CR which are obtained for $4.5 \, \mbox{TeV} \lesssim M_0 \lesssim 7.5 \, \mbox{TeV}$.} As expected, for $t$ even, see fig.~\ref{fig:Case2_scan_teven}, no dependence of the flavour observables on the angle $\theta_R$ is seen. 
  Consequently, we observe less spread in the data points. 
 For $t$ odd, see fig.~\ref{fig:Case2_scan_todd}, a clear dependence on $\theta_R$ is observed. In contrast to the results for Case 1), the BRs and CR are now enhanced for $\sin 2 \, \theta_R$ close to zero, since then at least one Yukawa coupling
 becomes large (compare the results found in~\cite{Drewes:2022kap}), while for $|\sin 2 \, \theta_R|$ being maximal the Yukawa couplings and, hence, the size of the cLFV signals are suppressed. This behaviour is very similar for light neutrino
 masses with IO and $m_0=0.015 \, \mbox{eV}$ as well as in situations in which the lightest neutrino mass $m_0$ is set to zero.
 
Eventually, we compare the results of the numerical scans for the four different combinations $\{s=0, t=0\}$, $\{s=1, t=2\}$ (corresponding to $u=0$), $\{s=0, t=1\}$ ($u=-1$) and $\{s=1, t=1\}$ ($u=1$)
in fig.~\ref{fig:Case2_alltogether}. In order to do so, we divide the scanned range of $\mu_0$ in 30 logarithmically uniform bins and also the data points accordingly. Then, we take the average of the 
results for the two BRs and the CR in each bin and plot it with respect to $\mu_0$.\footnote{In fig.~\ref{fig:Case2_alltogether}, we take as value representing a bin the central value of $\mu_0$ in this bin.} We see that the two
combinations corresponding to $u=0$ lead to (nearly) the same averages. This is supported by the observation that the matrix element $\eta_{e \mu}$ in eq.~(\ref{eq:etaexpressionCase2}) can be written
in terms of the lepton mixing angles and $J_{\mathrm{CP}}$ that all depend on $s$ and $t$ only through $u$, compare~\cite{Hagedorn:2014wha}. Furthermore, it can be observed that the resulting
averages are smaller by up to a factor 10 to 25, if $u=0$. We note that also the two combinations corresponding to $u=-1$ and $u=1$, respectively,
lead to very similar results for the cLFV signals.
The reason why the results for $u=-1$ and $u=1$ are larger than those for $u=0$ (mainly) lies in the fact that the displayed averages also take into account the
part of the parameter space in which $\theta_R$ fulfils $\sin 2 \, \theta_R \approx 0$, entailing at least one large Yukawa coupling and, thus, larger BRs and CR. This can also explain 
 the similarity of the results for $u=-1$ and $u=1$. Moreover,
  we note that the analogue of fig.~\ref{fig:Case2_alltogether} looks very similar for light neutrino masses with IO and $m_0=0.015 \, \mbox{eV}$.

%%%%%%%%%%%%%%%%%%%%%%%%%%%%%%%%%%%%%%%%%%%%%%%%%%%%%%%
\subsection{Results for Case 3 a)}
\label{sec:rescase3a}
%%%%%%%%%%%%%%%%%%%%%%%%%%%%%%%%%%%%%%%%%%%%%%%%%%%%%%%

Like for Case 2), also for Case 3 a) it is relevant which residual symmetries, encoded in the parameters $m$ and $s$, are preserved in the neutral lepton sector
in order to determine whether the results depend in general on the angle $\theta_R$ or not. Concretely, for $m$ and $s$ being both even or being both odd
no dependence on $\theta_R$ is expected, while for the combination $m$ even and $s$ odd or vice versa such a dependence is anticipated, because the matrix combination 
in square brackets in eq.~(\ref{eq:mnuform}) is not diagonal and the light neutrino masses depend on $\theta_R$, see section~\ref{sec2}. It is, thus, important 
to consider numerical examples which represent these different possibilities. For this reason, we choose, as in~\cite{Drewes:2022kap}, one example with $m$ odd,
\begin{equation}
\label{eq:modd_Case3a}
n=16 \;\;\; \mbox{and} \;\;\; m=1 \; ,
\end{equation}
leading to $\sin^2 \theta_{13} \approx 0.025$ and $\sin^2 \theta_{23} \approx 0.61$,\footnote{We note that the value of the reactor mixing angle is slightly outside the experimentally preferred $3 \, \sigma$ range~\cite{Esteban:2020cvm}.}
and another one with $m$ even,
 \begin{equation}
\label{eq:meven_Case3a}
n=34 \;\;\; \mbox{and} \;\;\; m=2 \; ,
\end{equation}
giving rise to $\sin^2 \theta_{13} \approx 0.022$ and $\sin^2 \theta_{23} \approx 0.61$. The solar mixing angle is fixed by the choice of the CP symmetry, i.e.~the parameter $s$,
and the free angle. All values of $s$, $0 \leq s \leq n-1$, allow for a good fit of the lepton mixing angles, usually for two different values of the free angle.
For examples of viable fits, see~\cite{Hagedorn:2014wha} and more recently~\cite{Drewes:2022kap}. 

\begin{figure}[t!]
    \centering
    \includegraphics[width=\textwidth]{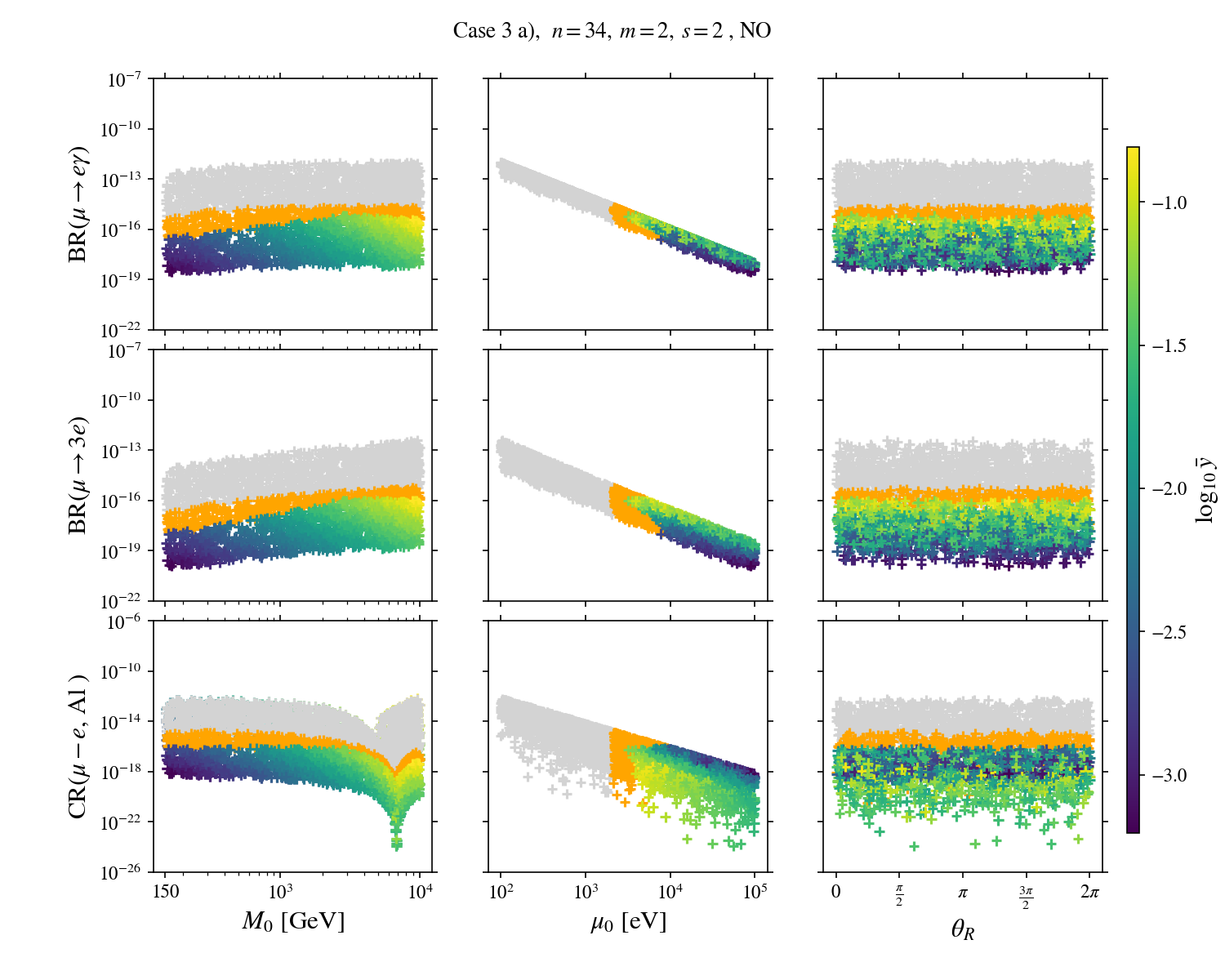}
    \caption{{\bf Case 3 a), \mathversion{bold}$m$ and $s$ both even (or both odd). Results of numerical scan for $\mathrm{BR} (\mu\to e \gamma)$, $\mathrm{BR} (\mu\to 3 \, e)$ and $\mathrm{CR} (\mu-e, \mathrm{Al})$ 
    varying $M_0$, $\mu_0$ and $\theta_R$
    \mathversion{normal}} in the ranges in eqs.~(\ref{eq:rangeM0}),~(\ref{eq:rangemu0}) and~(\ref{eq:rangethetaR}), respectively. As example of the group theory parameters $n=34$, $m=2$ and $s=2$ is chosen.     
    For conventions, see fig.~\ref{fig:Case1_scan}.}
    \label{fig:Case3a_scan_mseven_msodd}
\end{figure}
With this information, we can study the signal strength of the BRs and CR for $n$ and $m$ given in the $\frac{s}{n}-\theta_L$-plane (treating $\frac sn$ as continuous parameter between $0$ and $1$, but still keeping track of whether $s$ is even or odd).   We fix the values of $\mu_0$ and $M_0$, e.g.~$\mu_0= 1 \, \mbox{keV}$ and $M_0=3 \, \mbox{TeV}$, 
and the light neutrino mass spectrum. Then, we superimpose the results of the fit to the solar mixing angle, similar to the analysis for Case 2) found in figs.~\ref{fig:Case2_un_thetaL_teven} and~\ref{fig:Case2_un_thetaL_todd}. 
We find for all possible combinations of $m$ and $s$ being even and odd as well as for light neutrino masses with NO (IO) and $m_0=0.03 \, (0.015) \, \mbox{eV}$ 
that no constraint on the parameter space arises from the prospective bound on $\mu\to e \gamma$ from MEG II and on $\mu\to 3 \, e$ from Mu3E Phase 1 at the $3 \, \sigma$ level, while 
the future limit of Mu3E Phase 2 and the expected reach of the experiments COMET and Mu2e exclude the entire parameter space. Thus, we do not display
any corresponding plots. 

\begin{figure}[t!]
    \centering
    \includegraphics[width=\textwidth]{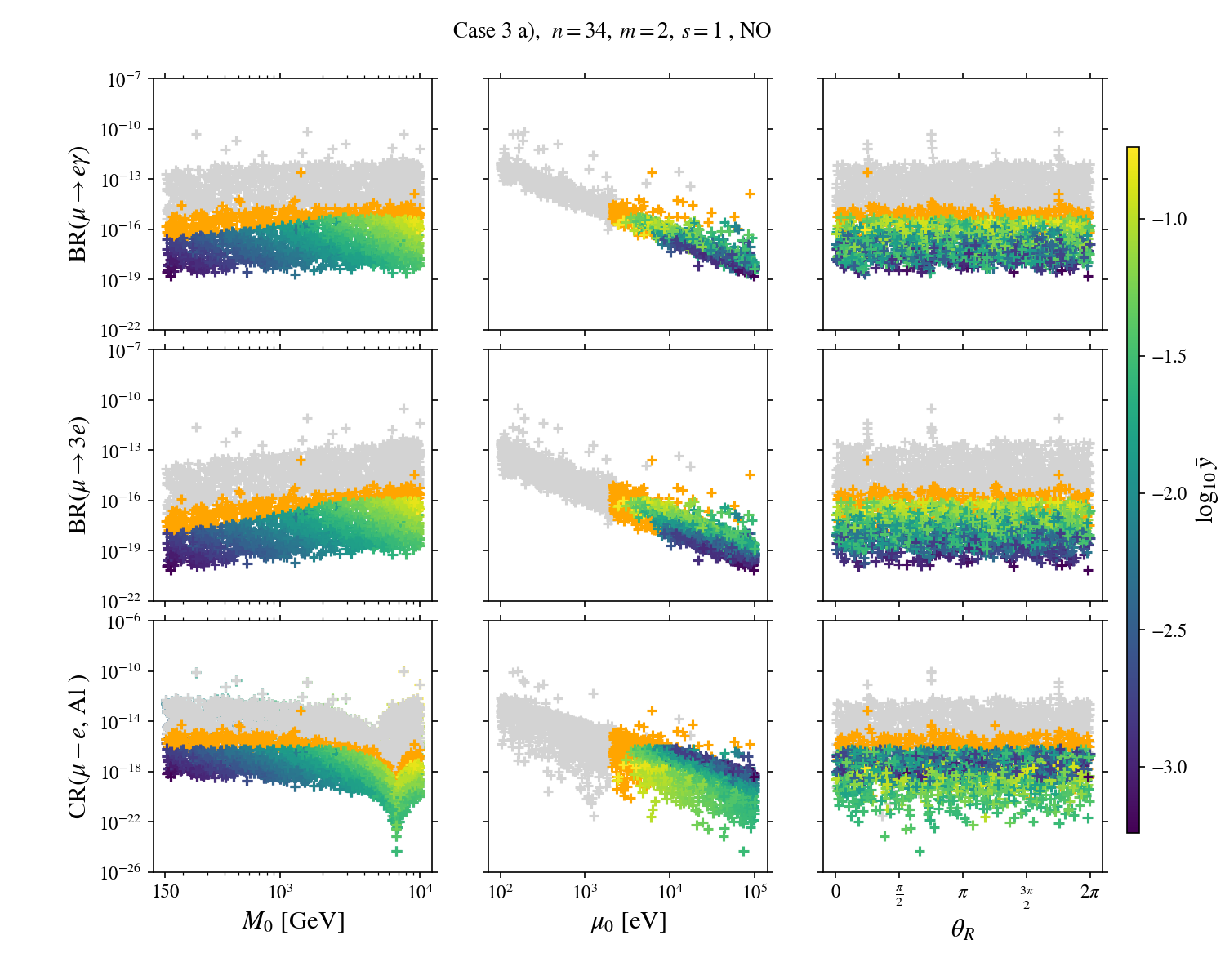}
    \caption{{\bf Case 3 a), \mathversion{bold}$m$ even and $s$ odd (or vice versa). Results of numerical scan for $\mathrm{BR} (\mu\to e \gamma)$, $\mathrm{BR} (\mu\to 3 \, e)$ and $\mathrm{CR} (\mu-e, \mathrm{Al})$ 
    varying $M_0$, $\mu_0$ and $\theta_R$
    \mathversion{normal}} in the ranges in eqs.~(\ref{eq:rangeM0}),~(\ref{eq:rangemu0}) and~(\ref{eq:rangethetaR}), respectively. As example of the group theory parameters $n=34$, $m=2$ and $s=1$ is chosen.     
    For conventions, see fig.~\ref{fig:Case1_scan}.}
    \label{fig:Case3a_scan_msmixed}
\end{figure}
We instead continue with giving an estimate for $m$ and $s$ both even or both odd as well as for $m$ even and $s$ odd or vice versa for $\theta_R=0$ (since then the matrix combination in square brackets
in eq.~(\ref{eq:mnuform}) is diagonal) that is based on the analytic form of $\eta_{e \mu}$. For this, we replace the terms depending on $\frac{m}{n}$ with the sine of the reactor mixing angle  
and take the limit $\theta_L$ being zero (using $\theta_L \approx \pi$ leads to the same result), compare section~\ref{sec:leptonmixing}
and~\cite{Hagedorn:2014wha}. Then, we have
\begin{equation}
\label{eq:etaemu_Case3a}
\eta_{e \mu} \approx \frac{\eta_0^\prime}{6} \, e^{\frac{4 \, \pi \, i}{3}} \, \left( 2 \, \Delta y_{21}^2 - 3 \, ( \sqrt{2-3 \, \sin^2 \theta_{13}} + \sin \theta_{13} ) \, \sin \theta_{13} \, \Delta y_{31}^2 \right) \; .
\end{equation}
This expression coincides with the one obtained for Case 1), see eq.~(\ref{eq:etaexpressionCase1}), up to the overall phase. Consequently, we refer to the estimates of the BRs and CR found
in eqs.~(\ref{eq:estimatesNOCase1}) and~(\ref{eq:estimatesIOCase1}) for light neutrino masses with NO and IO and non-vanishing $m_0$, respectively.

\begin{figure}[t!]
    \centering
    \includegraphics[width=\textwidth]{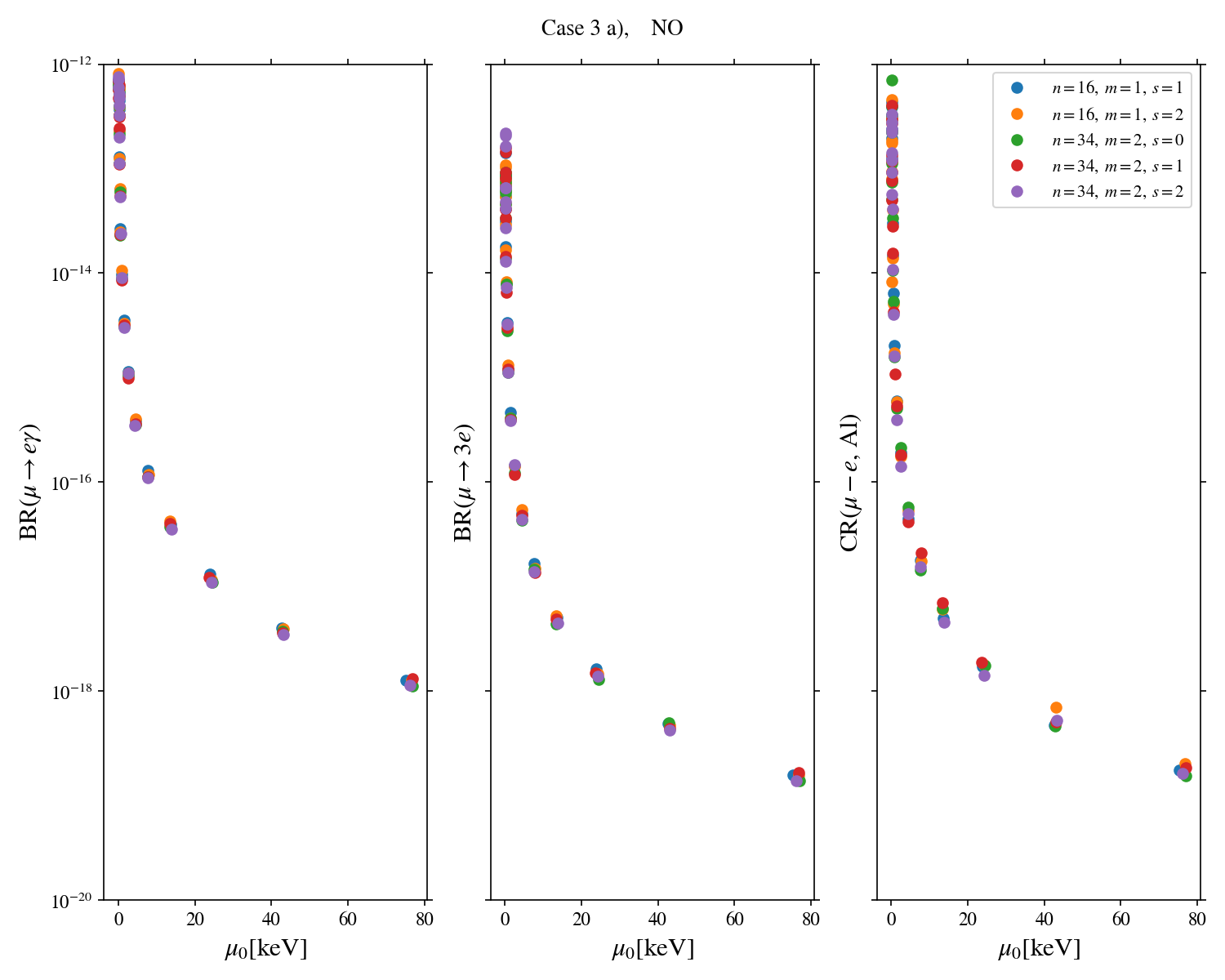}
    \caption{{\bf Case 3 a). Comparison of results for \mathversion{bold}$\mathrm{BR} (\mu\to e \gamma)$, $\mathrm{BR} (\mu\to 3 \, e)$ and $\mathrm{CR} (\mu-e, \mathrm{Al})$ for different combinations of $m$ and $s$}\mathversion{normal}. 
   The group index $n$ is either fixed to $n=16$ or to $n=34$. For further conventions and details see fig.~\ref{fig:Case2_alltogether} for Case 2).}
    \label{fig:Case3a_alltogether}
\end{figure}

We perform numerical scans for the concrete choices $n=16$, $m=1$ and $s=1$ as well as $s=2$ and for $n=34$, $m=2$ and $0 \leq s \leq 2$.
The results are presented in figs.~\ref{fig:Case3a_scan_mseven_msodd}-\ref{fig:Case3a_alltogether} for light neutrino masses with NO and non-zero $m_0$. 

As expected, in fig.~\ref{fig:Case3a_scan_mseven_msodd} no dependence on the free angle $\theta_R$ is observed. Although these plots assume $m$ and $s$ both even (since we use $n=34$),
the results for $m$ and $s$ both odd (for $n=16$) look very similar.  For light neutrino masses with IO and $m_0=0.015 \, \mbox{eV}$ we also find comparable results. We have verified that 
taking $m_0=0$ leads to very similar plots as well. Like for Case 2), $t$ even ($u$ even), the viable data points can be characterised by $\mu_0 \gtrsim 3 \, \mbox{keV}$, while $M_0$ and $\theta_R$
are not restricted. Furthermore, the future limit on $\mathrm{BR} (\mu\to 3 \, e)$ as well as on $\mathrm{CR} (\mu-e, \mathrm{Al})$ can be saturated for the viable points, whereas the maximum value of
$\mathrm{BR} (\mu\to e \gamma)$ is around $6 \times 10^{-16}$ and thus much lower than the prospective bound from MEG II. We also remark that both BRs appear to be bounded from below,
$\mathrm{BR} (\mu\to e \gamma) \gtrsim 10^{-19}$ and $\mathrm{BR} (\mu\to 3 \, e) \gtrsim 10^{-20}$. The CR of $\mu-e$ conversion in aluminium reveals -- for $M_0 \lesssim 4.5 \, \mbox{TeV}$ or $M_0 \gtrsim 7.5 \, \mbox{TeV}$ --
also a lower limit, $\mathrm{CR} (\mu-e, \mathrm{Al}) \gtrsim 10^{-21}$.

For the choice $m$ even and $s$ odd (for $n=34$) we expect the results to depend on $\theta_R$ and this dependence should be similar to the one found for Case 1), compare~\cite{Drewes:2022kap}. 
As can be observed in fig.~\ref{fig:Case3a_scan_msmixed}, the dependence is much milder (for light neutrino masses with NO and $m_0=0.03 \, \mbox{eV}$) than
 in the other studied cases, Case 1) (see fig.~\ref{fig:Case1_scan}), Case 2) for $t$ odd (see fig.~\ref{fig:Case2_scan_todd}) and Case 3 b.1) for $m$ even and $s$ odd or vice versa (see fig.~\ref{fig:Case3b1_scan_msmixed}).
We have confirmed that for $m$ odd and $s$ even (employing $n=16$) a very similar result is obtained. Assuming light neutrino masses with IO and $m_0=0.015 \, \mbox{eV}$ also leads to plots
similar to those shown in fig.~\ref{fig:Case3a_scan_msmixed}. For NO light neutrino masses and $m_0$ being very small ($m_0=0$ or $m_0=10^{-5} \, \mbox{eV}$) we find instead
 a larger dependence on $\theta_R$ like for Case 1) in fig.~\ref{fig:Case1_scan}, when considering e.g.~$n=16$, $m=1$ and $s=2$. Regarding the experimental prospects to detect cLFV signals involving
 the muon and the electron for Case 3 a) and $m$ even and $s$ odd or vice versa, these are similar to those for $m$ and $s$ both even or both odd, namely for the studied examples we obtain
 $4 \times 10^{-21} \lesssim \mathrm{BR} (\mu\to e \gamma) \lesssim 6 \times 10^{-16}$, $\mathrm{BR} (\mu\to 3 \, e) \gtrsim 10^{-22}$ and $\mathrm{CR} (\mu-e, \mathrm{Al}) \gtrsim 10^{-21}$, if the cancellation region, see footnote~\ref{ft:cancel},
 is avoided.

Lastly, we would like to compare the different choices of $n$, $m$ and $s$ that we have analysed numerically. Like for Case 2), see fig.~\ref{fig:Case2_alltogether}, we make use of averages corresponding to a certain binning of the data
in the parameter $\mu_0$. We find that the results of the BRs and CR are nearly independent of the choice of $n$, $m$ and $s$, as can be clearly seen in fig.~\ref{fig:Case3a_alltogether}, where we show the 
results for light neutrino masses with NO and non-zero $m_0$. This behaviour can be explained with the fact that (almost) no enhancement of the BRs and CR is observed for certain values of the angle $\theta_R$, compare fig.~\ref{fig:Case3a_scan_msmixed}.
 The corresponding plots for light neutrino masses with IO and $m_0=0.015 \, \mbox{eV}$ reveal the same features, as expected from the above discussion.

%%%%%%%%%%%%%%%%%%%%%%%%%%%%%%%%%%%%%%%%%%%%%%%%%%%%%%%
\subsection{Results for Case 3 b.1)}
\label{sec:rescase3b1}
%%%%%%%%%%%%%%%%%%%%%%%%%%%%%%%%%%%%%%%%%%%%%%%%%%%%%%%

 %
 \begin{figure}[h!]
    \centering
    \includegraphics[width=\textwidth]{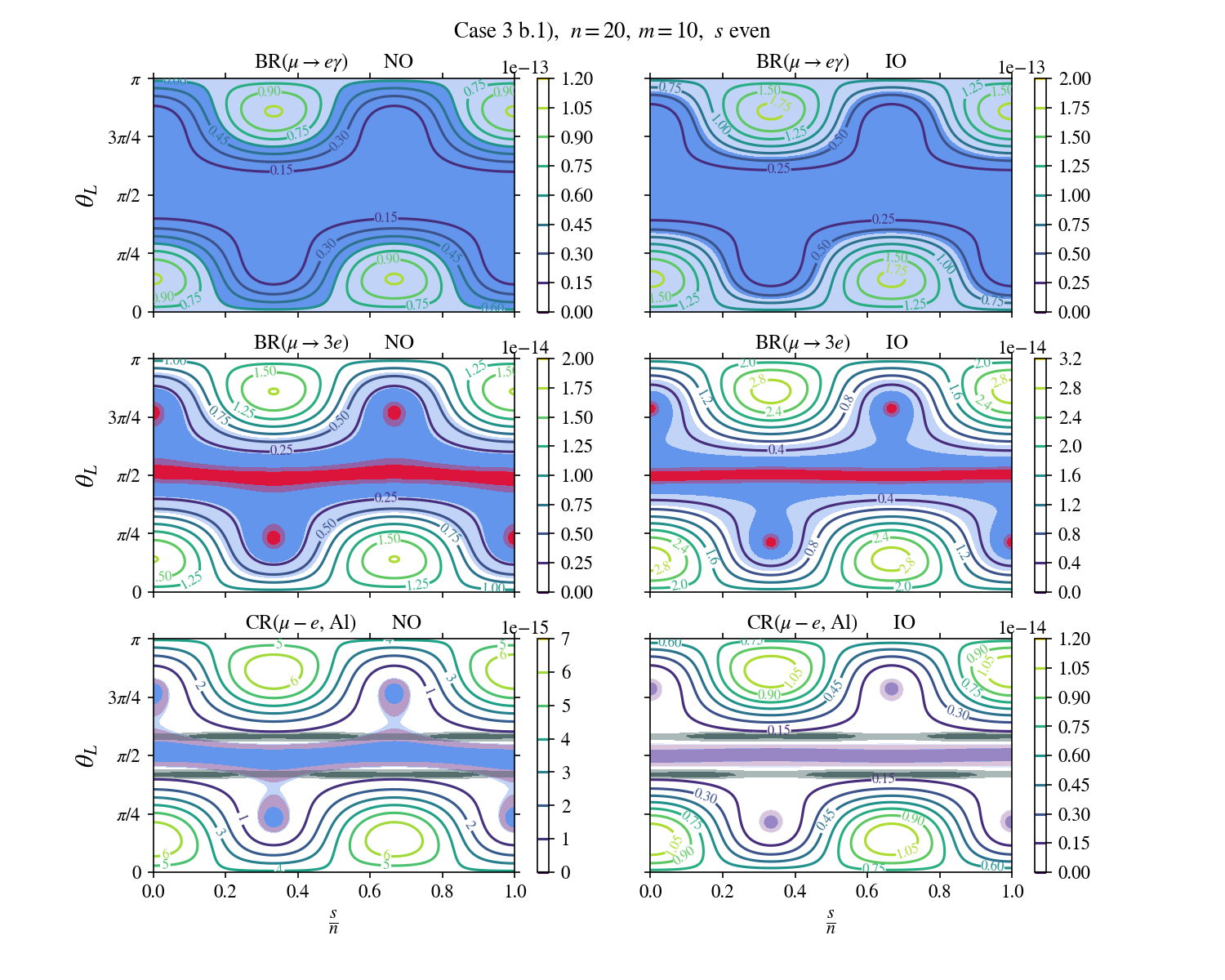}
    \caption{{\bf Case 3 b.1), \mathversion{bold}$m$ and $s$ both even. Predictions for $\mathrm{BR} (\mu\to e \gamma)$, $\mathrm{BR} (\mu\to 3 \, e)$ and $\mathrm{CR} (\mu-e, \mathrm{Al})$ in the $\frac sn-\theta_L$-plane} in the upper, 
    middle and lower row.\mathversion{normal}  The group theory parameters are chosen as $n=20$ and $m=10$. The conventions are like in fig.~\ref{fig:Case2_un_thetaL_teven} of Case 2) and $t$ even ($u$ even). 
    Simplified versions of these plots, with only contour lines or only experimentally preferred regions, can be found in appendix~\ref{app:Case2Case3b1}, see figs.~\ref{fig:Case3b1_sn_thetaL_m10seven_simplified_contours} and~\ref{fig:Case3b1_sn_thetaL_m10seven_simplified_limits}.
        Plots reflecting the features of examples with $m$ and $s$ both odd
    can be found in figs.~\ref{fig:Case3b1_sn_thetaL_m9seven} and~\ref{fig:Case3b1_sn_thetaL_m11seven}, since these are displayed for the choice $\theta_R=0$.}
\label{fig:Case3b1_sn_thetaL_m10seven}
\end{figure}
\noindent The relevant parameters for Case 3 b.1) are the same as for Case 3 a). Given the different assignment of the light neutrino masses, see e.g.~eq.~(\ref{eq:mfyf_Case3b1}), the values of the parameters $n$, $m$ and $s$ 
that can lead to a good agreement with the experimental data on lepton mixing angles
differ for Case 3 b.1) from those for Case 3 a). 
As mentioned in section~\ref{sec:leptonmixing}, see also~\cite{Hagedorn:2014wha}, the parameter $m$ is preferred to 
be close to $\frac n2$ in order to correctly accommodate the measured value of the solar mixing angle. Unlike for Case 3 a), the smallest value of the index $n$ that permits a good fit can be small, e.g.~even $n=2$. 
Nevertheless, we study numerically the example 
\begin{equation}
n=20 \;\;\; \mbox{and} \;\;\; m=9,10,11 \; ,
\end{equation}
 \begin{figure}[t!]
    \centering
    \includegraphics[width=\textwidth]{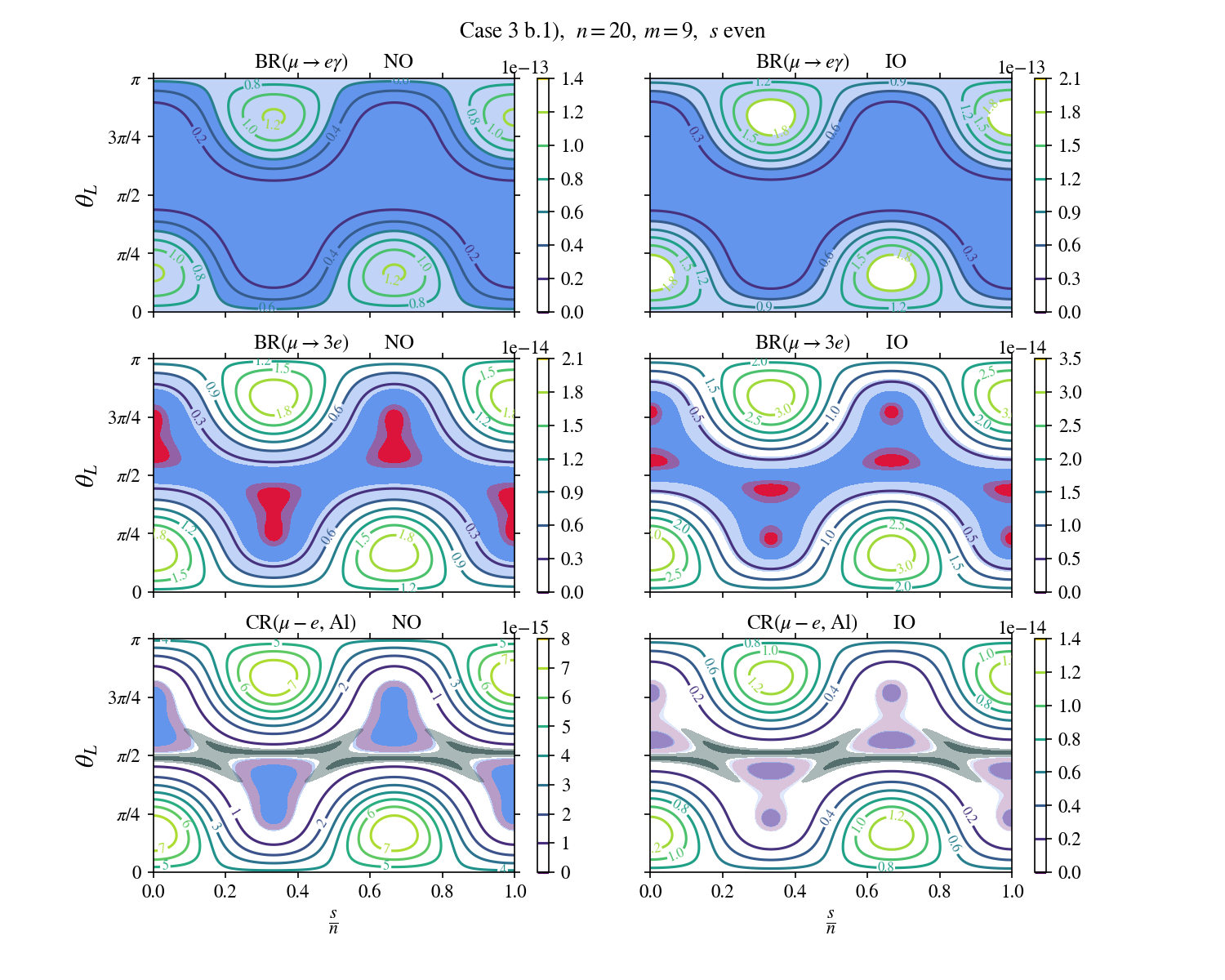}
    \caption{{\bf Case 3 b.1), \mathversion{bold}$m$ odd, $s$ even and $\theta_R=0$. Predictions for $\mathrm{BR} (\mu\to e \gamma)$, $\mathrm{BR} (\mu\to 3 \, e)$ and $\mathrm{CR} (\mu-e, \mathrm{Al})$ in the $\frac sn-\theta_L$-plane} in the upper, 
    middle and lower row.\mathversion{normal}  The group theory parameters are chosen as $n=20$ and $m=9$. For the remaining conventions see fig.~\ref{fig:Case3b1_sn_thetaL_m10seven}. 
        Simplified versions of these plots, with only contour lines or only experimentally preferred regions, can be found in appendix~\ref{app:Case2Case3b1}, see figs.~\ref{fig:Case3b1_sn_thetaL_m9seven_simplified_contours} and~\ref{fig:Case3b1_sn_thetaL_m9seven_simplified_limits}.
    The dependence on $\theta_R$ is illustrated with six different values of $\theta_R$ in different colours in fig.~\ref{fig:Case3b1_sn_thetaL_m9seven_thR} in the same appendix. Results for an example of the combination $m$ even and $s$ odd and different values of $\theta_R$ are found in fig.~\ref{fig:Case3b1_sn_thetaL_m10sodd_thR} also in this appendix.}
\label{fig:Case3b1_sn_thetaL_m9seven}
\end{figure}
since a larger $n$ allows to analyse, on the one hand, values of $m$ which are not equal to $\frac n2$ and, on the other hand, several different values of the parameter $s$ (corresponding to different CP symmetries), 
 because $s$ varies between $0$ and $n-1$. Results for the lepton mixing angles can be found in~\cite{Hagedorn:2014wha,Drewes:2022kap}.
Like for Case 3 a), we expect that the results for $m$ and $s$ both even or both odd are independent of the angle $\theta_R$, whereas those for $m$ even and $s$ odd or vice versa reveal a dependence on this parameter. 
This expectation is confirmed in the following.

 \begin{figure}[t!]
    \centering
    \includegraphics[width=\textwidth]{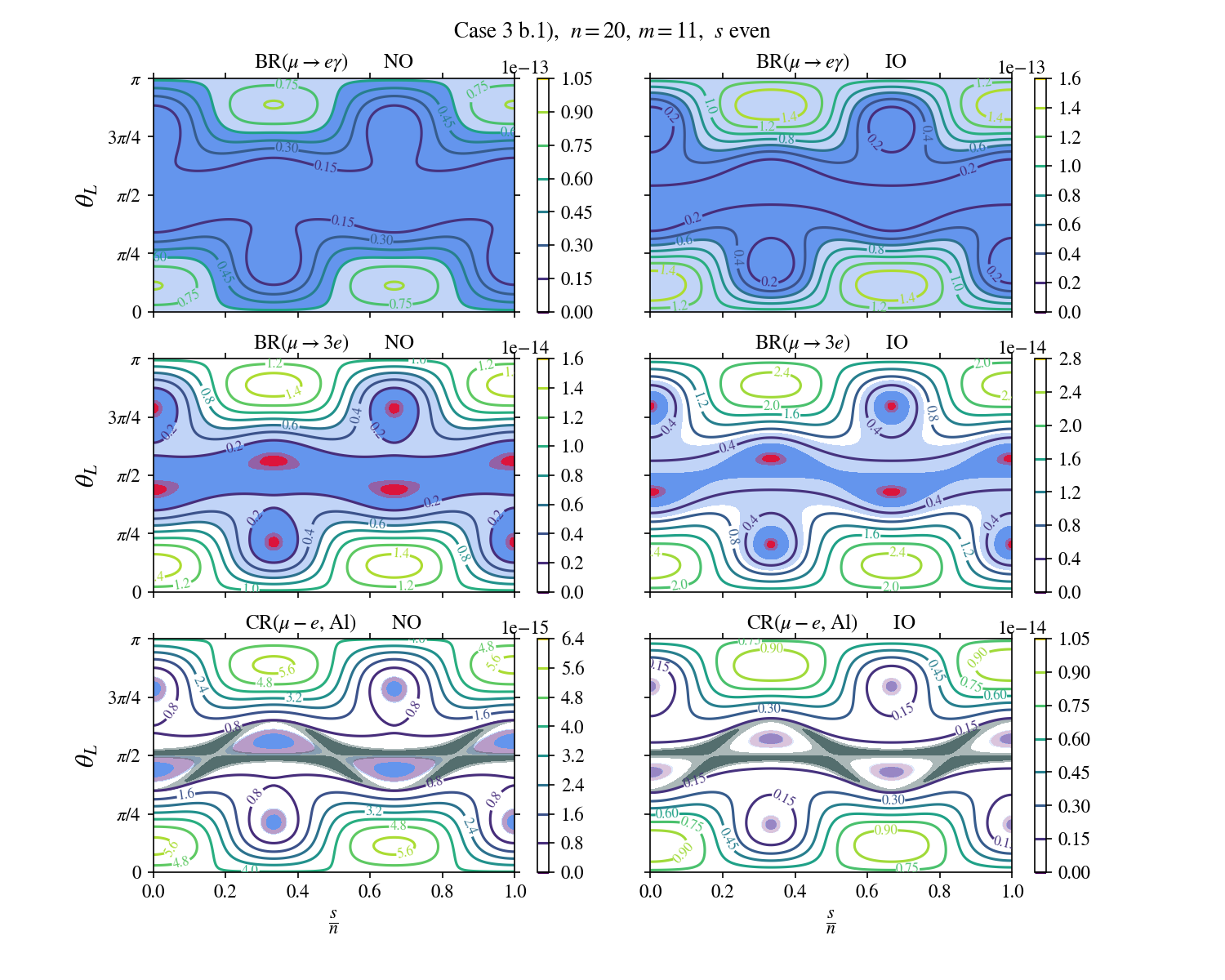}
    \caption{{\bf Case 3 b.1), \mathversion{bold}$m$ odd, $s$ even and $\theta_R=0$. Predictions for $\mathrm{BR} (\mu\to e \gamma)$, $\mathrm{BR} (\mu\to 3 \, e)$ and $\mathrm{CR} (\mu-e, \mathrm{Al})$ in the $\frac sn-\theta_L$-plane} in the upper, 
    middle and lower row.\mathversion{normal}  The group theory parameters are chosen as $n=20$ and $m=11$. For the remaining conventions see fig.~\ref{fig:Case3b1_sn_thetaL_m10seven}. 
        Simplified versions of these plots, with only contour lines or only experimentally preferred regions, can be found in appendix~\ref{app:Case2Case3b1}, see figs.~\ref{fig:Case3b1_sn_thetaL_m11seven_simplified_contours} and~\ref{fig:Case3b1_sn_thetaL_m11seven_simplified_limits}.    
    The dependence on $\theta_R$ is illustrated with six different values of $\theta_R$ in different colours in fig.~\ref{fig:Case3b1_sn_thetaL_m11seven_thR} in the same appendix. Results for an example of the combination $m$ even and $s$ odd and different values of $\theta_R$ are found in fig.~\ref{fig:Case3b1_sn_thetaL_m10sodd_thR} also in this appendix.}
\label{fig:Case3b1_sn_thetaL_m11seven}
\end{figure}
We begin the analysis of the BRs and CR by evaluating their sizes in the $\frac sn-\theta_L$-plane\footnote{The ratio $\frac sn$ is taken to be continuous, $0 \leq \frac sn \leq 1$, and $s$ is either assumed to be even or odd.} 
for a fixed ratio of $\frac mn$, $\frac mn=\frac{9}{20}$, $\frac mn=\frac{10}{20}$ and $\frac mn=\frac{11}{20}$, as well as for fixed values of the 
two scales $\mu_0$ and $M_0$, $\mu_0=1 \, \mbox{keV}$ and $M_0= 3 \, \mbox{TeV}$, see figs.~\ref{fig:Case3b1_sn_thetaL_m10seven}-\ref{fig:Case3b1_sn_thetaL_m11seven}. The colour-coding in these figures is the 
same as in fig.~\ref{fig:Case2_un_thetaL_teven} for Case 2). In particular, the dark (light) grey regions in the plots for $\mathrm{CR} (\mu-e, \mathrm{Al})$ indicate the parameter space in which the lepton mixing angles are
accommodated to a certain degree, i.e.~$\chi^2 \leq 100 \, (300)$. These regions are found to be consistent with those shown in corresponding plots in~\cite{Hagedorn:2014wha,Drewes:2022kap}. 
The plots in the left in figs.~\ref{fig:Case3b1_sn_thetaL_m10seven}-\ref{fig:Case3b1_sn_thetaL_m11seven} assume that light neutrino masses follow NO 
with $m_0=0.03 \, \mbox{eV}$, while the plots in the right in these figures are based on an IO light neutrino mass spectrum with $m_0=0.015 \, \mbox{eV}$. Fig.~\ref{fig:Case3b1_sn_thetaL_m10seven} represents the combination 
$m$ and $s$ both even and, thus, no dependence on $\theta_R$ is expected, and also observed, while in figs.~\ref{fig:Case3b1_sn_thetaL_m9seven} 
and~\ref{fig:Case3b1_sn_thetaL_m11seven} $m$ is odd and $s$ even such that the results depend on $\theta_R$. Here, they are shown 
for $\theta_R=0$, for which the matrix combination in square brackets in eq.~(\ref{eq:mnuform}) is diagonal. We complete the set of plots by figs.~\ref{fig:Case3b1_sn_thetaL_m9seven_thR}-\ref{fig:Case3b1_sn_thetaL_m10sodd_thR} in appendix~\ref{app:Case2Case3b1}
in which we display for $\frac mn=\frac{9}{20}$ and $\frac mn=\frac{11}{20}$ and $s$ even as well as for  $\frac mn=\frac{10}{20}$ and $s $ odd 
the results for six different values of the angle $\theta_R$ in different colours, similar to e.g.~fig.~\ref{fig:Case2_un_thetaL_todd} for Case 2) and $t$ odd.
We note that the results for $\frac mn=\frac{9}{20}$ ($\frac mn=\frac{11}{20}$) and $s$ even agree for the choice $\theta_R=0$ with those for $\frac mn=\frac{9}{20}$ ($\frac mn=\frac{11}{20}$) and $s$ odd
and, similarly, the choice $\frac mn=\frac{10}{20}$ and $s$ odd with $\theta_R=0$ reproduces the results of $\frac mn=\frac{10}{20}$ and $s$ even. Consequently, the results for all possible combinations of $m$ and $s$
even and odd are covered with the six figures, figs.~\ref{fig:Case3b1_sn_thetaL_m10seven}-\ref{fig:Case3b1_sn_thetaL_m11seven} and 
figs.~\ref{fig:Case3b1_sn_thetaL_m9seven_thR}-\ref{fig:Case3b1_sn_thetaL_m10sodd_thR}. We see for $m$ and $s$ both even or both odd as well as for $m$ even and $s$ odd or vice versa
and $\theta_R=0$ that in most situations the prospective bound on $\mathrm{BR} (\mu\to e \gamma)$ is passed  in the entire parameter space -- at least at the $3 \, \sigma$ level, whereas the limit expected from Mu3E Phase 1
excludes about half of the parameter space at the $3 \, \sigma$ level and the future bound from Mu3E Phase 2 considerably reduces the viable regions of parameter space. Employing the 
expected limits from COMET and Mu2e only leaves a rather small portion of parameter space in the $\frac sn-\theta_L$-plane allowed, which in general is hardly compatible with the regions preferred by the
measured values of the lepton mixing angles. In addition, we observe that the future bounds are slightly more likely to be satisfied for light neutrino masses with NO and $m_0=0.03 \, \mbox{eV}$ 
than in the case of an IO light neutrino mass spectrum with $m_0=0.015 \, \mbox{eV}$ (compare plots in the left with those in the right in figs.~\ref{fig:Case3b1_sn_thetaL_m10seven}-\ref{fig:Case3b1_sn_thetaL_m11seven}).
 \begin{figure}[t!]
    \centering
    \includegraphics[width=\textwidth]{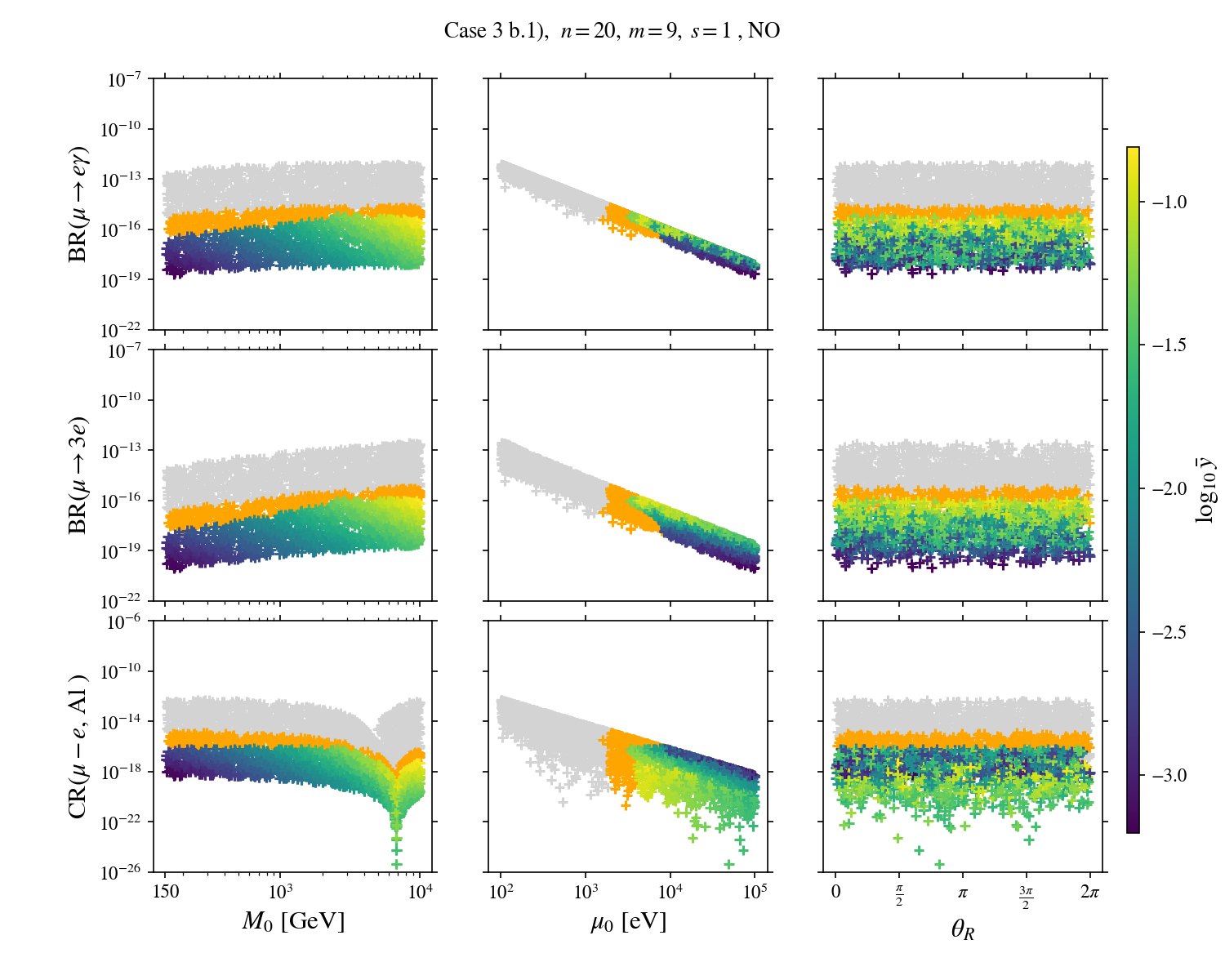}
    \caption{{\bf Case 3 b.1), \mathversion{bold}$m$ and $s$ both odd (or both even). Results of numerical scan for $\mathrm{BR} (\mu\to e \gamma)$, $\mathrm{BR} (\mu\to 3 \, e)$ and $\mathrm{CR} (\mu-e, \mathrm{Al})$ 
    varying $M_0$, $\mu_0$ and $\theta_R$
    \mathversion{normal}} in the ranges in eqs.~(\ref{eq:rangeM0}),~(\ref{eq:rangemu0}) and~(\ref{eq:rangethetaR}), respectively. As example of the group theory parameters $n=20$, $m=9$ and $s=1$ is chosen.     
    For conventions, see fig.~\ref{fig:Case1_scan}.}
\label{fig:Case3b1_scan_mseven_msodd}
\end{figure}
 When considering the dependence on $\theta_R$ for the combinations $m$ even and $s$ odd or vice versa, see especially figs.~\ref{fig:Case3b1_sn_thetaL_m9seven_thR}-\ref{fig:Case3b1_sn_thetaL_m10sodd_thR} in appendix~\ref{app:Case2Case3b1}, we note that the shape and size of the parameter space passing a certain future bound depends on the actual value of $\theta_R$ and for the
larger ones among the shown values it is usually smaller. Furthermore, the regions of parameter space that lead to an agreement of the lepton mixing angles with experimental data shift as function of $\theta_R$. It is, thus,
necessary to investigate each value of $\theta_R$ separately in order to make more quantitative statements about the constraints imposed by the future limits on the different $\mu-e$ transitions and their compatibility with the restrictions coming from
fitting the lepton mixing angles well.

 \begin{figure}[t!]
    \centering
    \includegraphics[width=\textwidth]{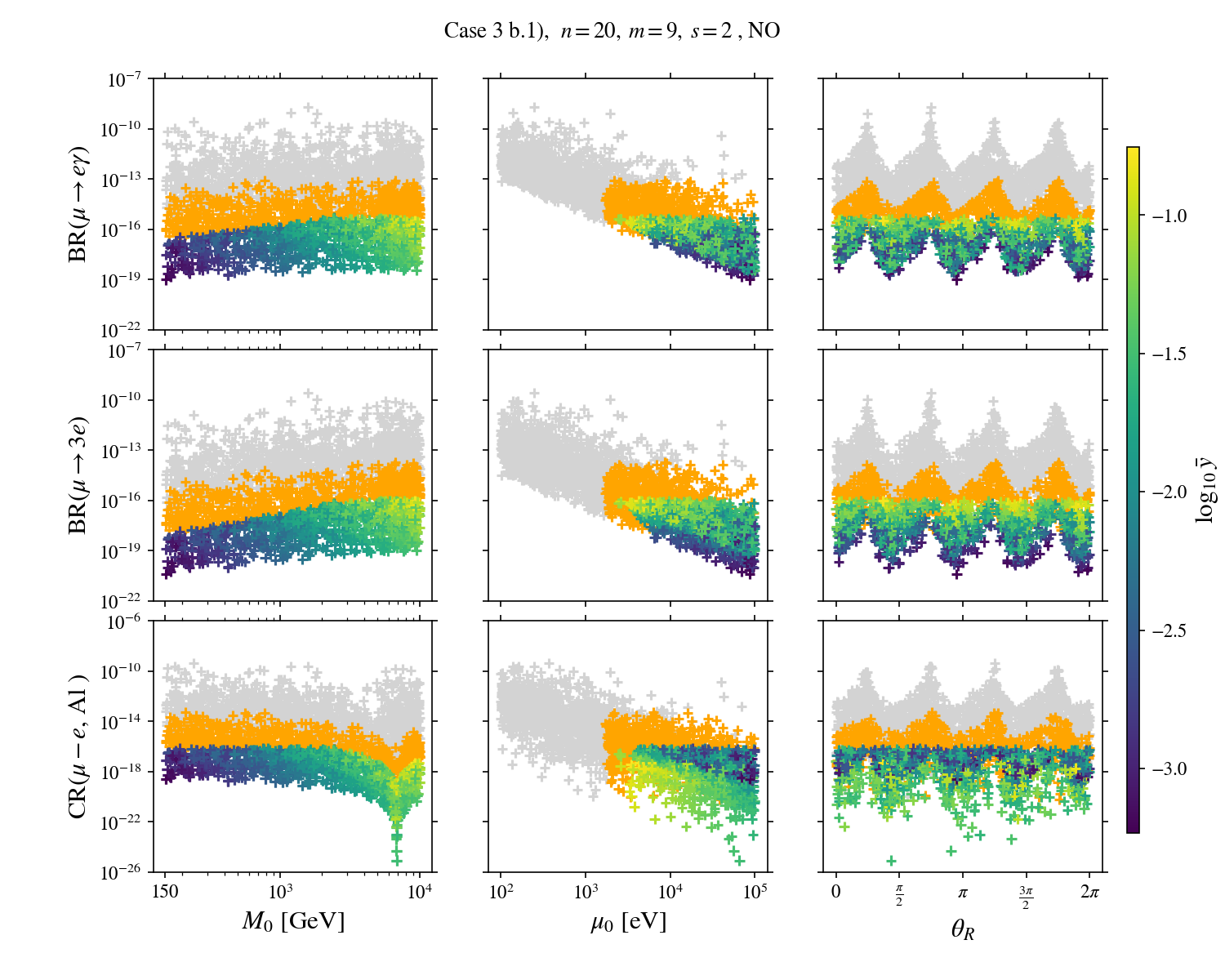}
    \caption{{\bf Case 3 b.1), \mathversion{bold}$m$ odd and $s$ even (or vice versa). Results of numerical scan for $\mathrm{BR} (\mu\to e \gamma)$, $\mathrm{BR} (\mu\to 3 \, e)$ and $\mathrm{CR} (\mu-e, \mathrm{Al})$ 
    varying $M_0$, $\mu_0$ and $\theta_R$
    \mathversion{normal}} in the ranges in eqs.~(\ref{eq:rangeM0}),~(\ref{eq:rangemu0}) and~(\ref{eq:rangethetaR}), respectively. As example of the group theory parameters $n=20$, $m=9$ and $s=2$ is chosen.     
    For conventions, see fig.~\ref{fig:Case1_scan}.
    }
\label{fig:Case3b1_scan_msmixed}
\end{figure}
To numerically estimate the BRs and CR, we first give an approximate formula for $\eta_{e \mu}$ for $m$ and $s$ both even or both odd or $m$ even and $s$ odd or vice versa, assuming $\theta_R=0$. For this, we take $m=\frac n2$, express $\theta_L$
in terms of the reactor mixing angle, compare~\cite{Hagedorn:2014wha,Drewes:2022kap}, and then arrive at 
\begin{equation}
\label{eq:etaemusquare_Case3b1}
\eta_{e \mu} = \frac{\eta_0^\prime}{3 \, \sqrt{2}} \, e^{\frac{4 \, \pi \, i}{3}} \, \left( 3 \, (e^{- 3 \, i \, \phi_s} \, \sqrt{1-3 \, \sin^2 \theta_{13}}+\sqrt{2} \, \sin\theta_{13}) \, \sin \theta_{13} \, \Delta y_{21}^2 - \sqrt{2} \, \Delta y_{31}^2 \right) \,  \; ,
\end{equation}
where $\Delta y_{21}^2$ and $\Delta y_{31}^2$ are related to the light neutrino masses as follows $\Delta y_{21}^2=\left(\frac{M_0^2}{\mu_0 \, \langle H \rangle^2}\right) \, (m_3 - m_2)$ 
and $\Delta y_{31}^2=\left(\frac{M_0^2}{\mu_0 \, \langle H \rangle^2}\right) \, (m_1 - m_2)$, since the identification of Yukawa couplings and light neutrino masses is altered compared to Case 1) through Case 3 a) due to the permutation, c.f.~eq.~(\ref{eq:mfyf_Case3b1}).
Using that $-0.6 \lesssim \cos 3 \, \phi_s \lesssim 0.71$, see~\cite{Hagedorn:2014wha}, we obtain as intervals for the BRs and CR
\begin{eqnarray}\nonumber
3.9 \times 10^{-15} \lesssim  &\mathrm{BR} (\mu\to e \gamma)& \lesssim 9.3 \times 10^{-15} \; ,
\\ \nonumber
6.3 \times 10^{-16} \lesssim  &\mathrm{BR} (\mu\to 3 \, e)& \lesssim 1.5 \times 10^{-15} \; ,
\\
\label{eq:estimatesNOCase3b1}
2.7 \times 10^{-16} \lesssim &\mathrm{CR} (\mu-e, \mathrm{Al})& \lesssim 6.6 \times 10^{-16} \;
\end{eqnarray}
for light neutrino masses with NO and $m_0=0.03 \, \mbox{eV}$, while for IO light neutrino masses wtih $m_0=0.015 \, \mbox{eV}$ we get
\begin{eqnarray}\nonumber
8.8 \times 10^{-15} \lesssim  &\mathrm{BR} (\mu\to e \gamma)& \lesssim 1.3 \times 10^{-14} \; ,
\\ \nonumber
1.4 \times 10^{-15} \lesssim  &\mathrm{BR} (\mu\to 3 \, e)& \lesssim 2.1 \times 10^{-15} \; ,
\\
\label{eq:estimatesIOCase3b1}
6.2 \times 10^{-16} \lesssim &\mathrm{CR} (\mu-e, \mathrm{Al})& \lesssim 9.1 \times 10^{-16} \; .
\end{eqnarray}
Again, we have fixed the two scales $\mu_0$ and $M_0$ to $\mu_0= 1 \, \mbox{keV}$ and $M_0=3 \, \mbox{TeV}$, respectively.
From these estimates we see that the future bound on $\mathrm{BR} (\mu\to e \gamma)$ is usually passed as well as the limit on $\mathrm{BR} (\mu\to 3 \, e)$ expected from Mu3E Phase 1,
whereas Mu3E Phase 2 can exclude this scenario and also the experiments COMET and Mu2e have this potential, since $\mathrm{CR} (\mu-e, \mathrm{Al})$ is larger than $2.7 \times 10^{-16}$
for the chosen values of the scales $\mu_0$ and $M_0$ and the assumed light neutrino mass spectra.

 \begin{figure}[t!]
    \centering
    \includegraphics[width=\textwidth]{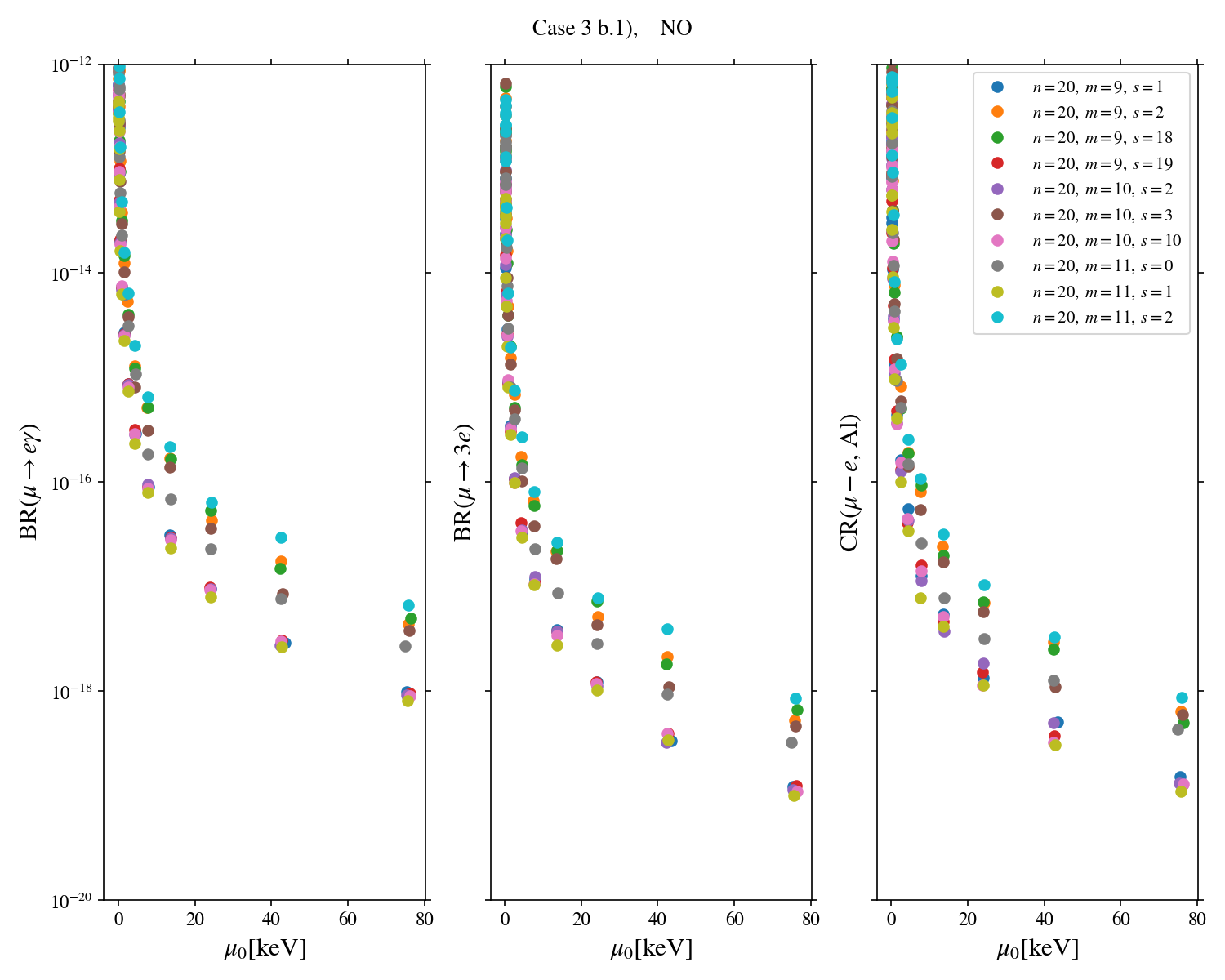}
    \caption{{\bf Case 3 b.1). Comparison of results for \mathversion{bold}$\mathrm{BR} (\mu\to e \gamma)$, $\mathrm{BR} (\mu\to 3 \, e)$ and $\mathrm{CR} (\mu-e, \mathrm{Al})$ for different combinations of $m$ and $s$}\mathversion{normal}. 
   The group index $n$ is set to $n=20$. For further conventions and details see fig.~\ref{fig:Case2_alltogether} for Case 2).}
\label{fig:Case3b1_alltogether}
\end{figure}
In a next step, we scrutinise the parameter space by performing numerical scans for different values of $m$ and $s$: for $m=10$ we have studied $s=2$, $s=3$ and $s=10$, for $m=9$ the choices of 
$s$ are $s=1$, $s=2$, $s=18$ and $s=19$, while for $m=11$ we have fixed $s$ to $s=0$, $s=1$ and $s=2$. We show examples of these scans in figs.~\ref{fig:Case3b1_scan_mseven_msodd} 
and~\ref{fig:Case3b1_scan_msmixed} for the concrete choices $m=9$ and $s=1$ as well as $m=9$ and $s=2$, respectively, assuming that light neutrino masses follow NO and $m_0=0.03 \, \mbox{eV}$. 
Since $m=9$ and $s=1$ are both odd, no dependence on $\theta_R$ is revealed, while for $m=9$ and $s=2$ a clear dependence on $\theta_R$ is observed, as expected.
 Like for Case 1), at least one Yukawa coupling is enhanced for $\cos 2 \, \theta_R \approx 0$ and, consequently, also the BRs and CR, whereas for $|\cos 2 \, \theta_R| \approx 1$ the signal strength of the 
studied cLFV processes is suppressed, compare~\cite{Drewes:2022kap} for the dependence on $\theta_R$. 
 We have checked that the other choices of $m$ and $s$ both even or both odd as well as light neutrino masses with IO and $m_0=0.015 \, \mbox{eV}$ lead to plots that are very similar to those found in fig.~\ref{fig:Case3b1_scan_mseven_msodd}.
This holds analogously for the other choices of $m$ and $s$ with one of them being even and the other one odd 
and the results displayed in fig.~\ref{fig:Case3b1_scan_msmixed}. Regarding the ranges of the BRs and CR, we usually find for the viable data points (the ones coloured
according to the colour bar in figs.~\ref{fig:Case3b1_scan_mseven_msodd} and~\ref{fig:Case3b1_scan_msmixed}) 
$10^{-19} \lesssim  \mathrm{BR} (\mu\to e \gamma) \lesssim 6 \times 10^{-16} $, $\mathrm{BR} (\mu\to 3 \, e) \gtrsim 10^{-21}$ and $\mathrm{CR} (\mu-e, \mathrm{Al}) \gtrsim 10^{-22}$ (excluding 
$4.5 \,\mbox{TeV} \lesssim M_0 \lesssim 7.5 \, \mbox{TeV}$) with $\mathrm{BR} (\mu\to 3 \, e)$ 
and $\mathrm{CR} (\mu-e, \mathrm{Al})$ saturating the expected limits from Mu3E Phase 2 and COMET, respectively.

We summarise and compare the results of all the performed numerical scans for light neutrino masses with NO and non-vanishing $m_0$ in fig.~\ref{fig:Case3b1_alltogether} using the same strategy of binning and averaging as in 
fig.~\ref{fig:Case2_alltogether}. We observe that the combinations of $m$ even and $s$ odd or vice versa lead to larger BRs and CR in average (up to a factor $25$)   
than those with $m$ and $s$ both even or both odd. Analogous to Case 2), this is due to the enhancement of the BRs and CR for certain values of the angle $\theta_R$ for $m$ even and $s$ odd or vice versa, entailing larger averages, 
which does not occur for $m$ and $s$ both even or both odd. For light neutrino masses with IO and $m_0=0.015 \, \mbox{eV}$ the plots shown in fig.~\ref{fig:Case3b1_alltogether} look very similar.

%%%%%%%%%%%%%%%%%%%%%%%%%%%%%%%%%%%%%%%%%%%%%%%%%%%%%%%
\mathversion{bold}
\section{Comments on cLFV $\tau$ decays}
\mathversion{normal}
\label{sec:taudecays}
%%%%%%%%%%%%%%%%%%%%%%%%%%%%%%%%%%%%%%%%%%%%%%%%%%%%%%%

\begin{figure}[t!]
    \centering
    \includegraphics[width=\textwidth]{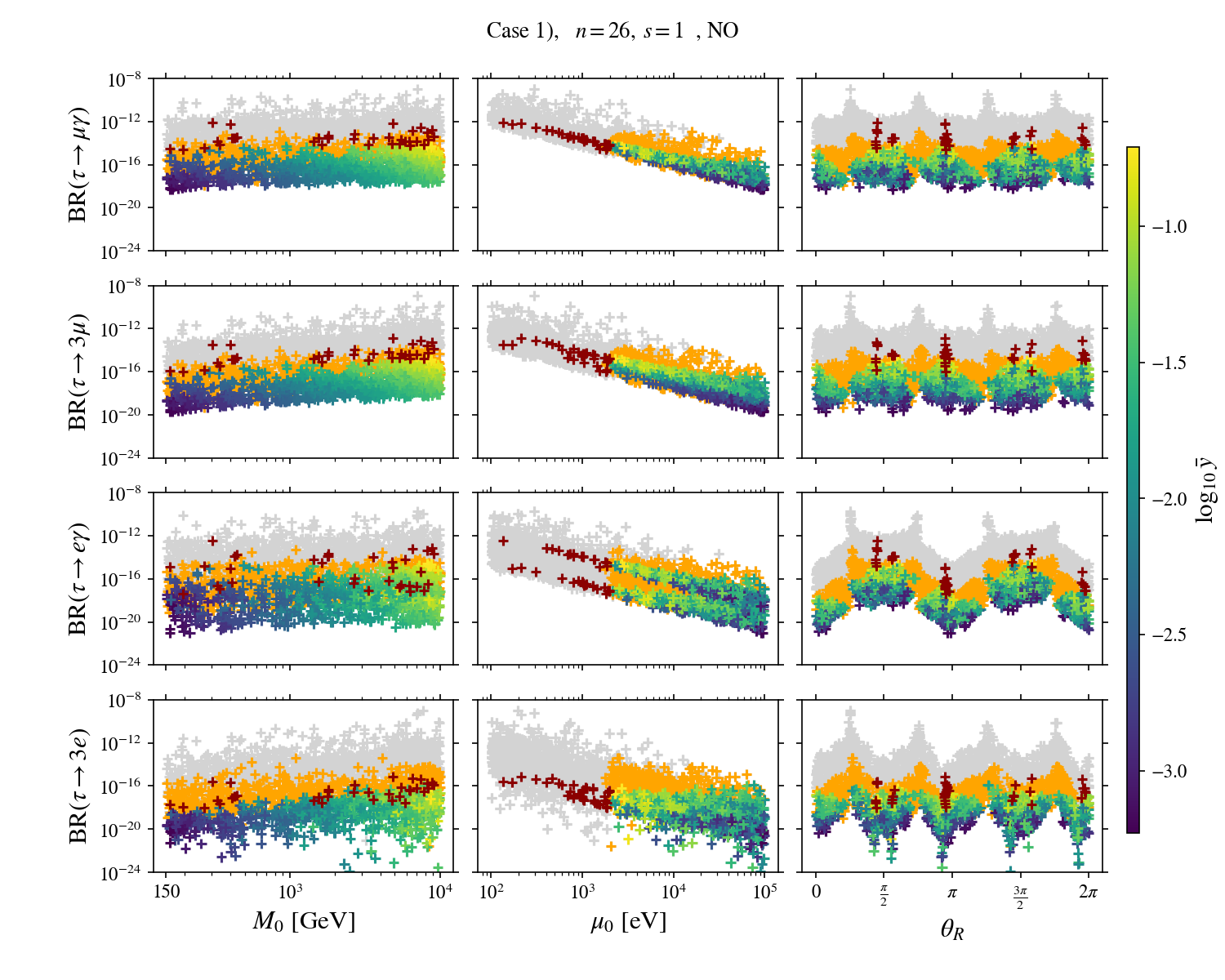}
    \caption{{\bf Case 1). Results of numerical scan for \mathversion{bold} $\mathrm{BR} (\tau\to \mu \gamma)$, $\mathrm{BR} (\tau\to 3 \, \mu)$, $\mathrm{BR} (\tau\to e \gamma)$ and $\mathrm{BR} (\tau\to 3 \, e)$ 
    varying $M_0$, $\mu_0$ and $\theta_R$
    \mathversion{normal}} in the ranges in eqs.~(\ref{eq:rangeM0}),~(\ref{eq:rangemu0}) and~(\ref{eq:rangethetaR}), respectively. The parameters $n$ and $s$ are the same as in fig.~\ref{fig:Case1_thetaL}.
    For conventions see fig.~\ref{fig:Case1_scan}.}
\label{fig:Case1_scan_taudecays}
\end{figure}

We also consider the cLFV decays of the tau lepton $\tau \to \mu \gamma$, $\tau \to e \gamma$, $\tau \to 3 \, \mu$ and $\tau \to 3 \, e$.
As current experimental bounds on the BRs of these decays we use (all at 90\% C.L.)
\begin{eqnarray}
\nonumber
&&\mathrm{BR} (\tau \to \mu \gamma) < 4.2 \times 10^{-8} \;\;\mbox{(Belle~\cite{Belle:2021ysv})}\; , \;\;
\;\;\;\; \mathrm{BR} (\tau \to e \gamma) < 3.3\times 10^{-8} \;\;\,\mbox{(Belle~\cite{ParticleDataGroup:2020ssz})}\; ,\\
\label{eq:taudecayscurrent}
&&\mathrm{BR} (\tau \to 3 \, \mu) < 1.9 \times 10^{-8} \;\;\mbox{(Belle II~\cite{Belle-II:2024sce})}\; , \;\;
\mathrm{BR} (\tau \to 3 \, e) < 2.7\times 10^{-8} \;\;\,\mbox{(Belle~\cite{ParticleDataGroup:2020ssz})}\; .
\end{eqnarray}
The prospective limits are
\begin{eqnarray}\nonumber
&&\mathrm{BR} (\tau \to \mu \gamma) < 6.9 \times 10^{-9} \;\;\;\mbox{(Belle II~\cite{Banerjee:2022xuw})}\; , \;\;
\mathrm{BR} (\tau \to e \gamma) < 9\times 10^{-9}  \,\;\;\;\;\;\mbox{(Belle II~\cite{Banerjee:2022xuw})}\, ,\\
\label{eq:taudecaysfuture}
&&\mathrm{BR} (\tau \to 3 \, \mu) < 3.6\times 10^{-10}  \;\,\mbox{(Belle II~\cite{Banerjee:2022xuw})}\; ,\;\,
\mathrm{BR} (\tau \to 3 \, e) < 4.7\times 10^{-10} \;\mbox{(Belle II~\cite{Banerjee:2022xuw})}\, .
\end{eqnarray}
Before discussing the results of the numerical scans, we give simple formulae for $\eta_{\mu \tau}$ and $\eta_{e \tau}$ for Case 1) that are relevant for the cLFV decays $\tau \to \mu \gamma$, $\tau \to 3 \, \mu$ and $\tau \to e \gamma$, $\tau \to 3 \, e$, respectively,
similar to $\eta_{e \mu}$ dominating the $\mu-e$ transitions.
They read
\begin{equation}
\label{eq:etamutau_Case1}
 \eta_{\mu\tau} = \frac{\eta_0^\prime}{6} \, \left( 2 \, \Delta y_{21}^2 - 3 \, (1-2 \, \sin^2 \theta_{13}) \, \Delta y_{31}^2 \right) 
\end{equation}
and
\begin{equation}
\label{eq:etaetau_Case1}
\eta_{e \tau} = \frac{\eta_0^\prime}{6} \, \left( 2 \, \Delta y_{21}^2 + 3 \, (\sqrt{2-3 \, \sin^2 \theta_{13}} -\sin \theta_{13}) \, \sin \theta_{13} \, \Delta y_{31}^2 \right) \; ,
\end{equation}
respectively, with $\Delta y_{ij}^2$ being defined in the vicinity of eq.~(\ref{eq:etaexpressionCase1}). We see that, in particular, $\eta_{e \tau}$ has a form similar to $\eta_{e \mu}$.
 Numerical estimates for the corresponding BRs can be derived with the help of these formulae and the approximations found in section~\ref{sec:anacon}. For $\mu_0=1 \, \mbox{keV}$ and $M_0=3 \, \mbox{TeV}$
we get
\begin{eqnarray}\nonumber
\mathrm{BR} (\tau\to \mu \gamma) \approx  2.3 \times 10^{-14} \; , && \mathrm{BR} (\tau\to 3 \, \mu) \approx 3.6 \times 10^{-15} \; ,
\\
\label{eq:tauestimate_NO_Case1}
\mathrm{BR} (\tau\to e \gamma) \approx  1.2 \times 10^{-15} \; , && \mathrm{BR} (\tau\to 3 \, e) \approx 2.0 \times 10^{-16} 
\end{eqnarray}
for light neutrino masses with NO and $m_0=0.03 \, \mbox{eV}$, while for IO light neutrino masses with $m_0=0.015 \, \mbox{eV}$ we find
\begin{eqnarray}\nonumber
\mathrm{BR} (\tau\to \mu \gamma) \approx  4.2 \times 10^{-14} \; , && \mathrm{BR} (\tau\to 3 \, \mu) \approx 6.6 \times 10^{-15} \; ,
\\
\label{eq:tauestimate_IO_Case1}
\mathrm{BR} (\tau\to e \gamma) \approx  1.4 \times 10^{-15} \; ,  && \mathrm{BR} (\tau\to 3 \, e) \approx 2.3 \times 10^{-16} \; .
\end{eqnarray}
These already indicate that the presented scenario is most likely not constrained by the current nor by the prospective limits on the studied cLFV tau lepton decays, see eqs.~(\ref{eq:taudecayscurrent}) and~(\ref{eq:taudecaysfuture}).
This observation is confirmed by the numerical scans that we have performed. As example, the results for $n=26$, $s=1$ and light neutrino masses with NO and $m_0=0.03 \, \mbox{eV}$ are shown in fig.~\ref{fig:Case1_scan_taudecays}. The colour-coding is the same as in fig.~\ref{fig:Case1_scan} for the corresponding $\mu-e$ transitions. The largest values obtained for the different BRs for the viable data points are
\begin{eqnarray}
\nonumber
&&\mathrm{BR} (\tau\to \mu \gamma) \lesssim 7 \times 10^{-15} \; , \;\; \mathrm{BR} (\tau\to3 \, \mu) \lesssim 2 \times 10^{-15} \; , \;\; \\
&&\label{eq:tauupperCase1_NO}
\mathrm{BR} (\tau\to e \gamma) \lesssim 3 \times 10^{-15} \;\, , \;\; \mathrm{BR} (\tau\to 3 \, e) \lesssim 2 \times 10^{-16}
\end{eqnarray}
and, thus, clearly inaccessible by the mentioned facilities. The observed dependence on the angle $\theta_R$ is driven by $\cos 2 \, \theta_R$, like for the $\mu-e$ transitions, and is due to the fact that for $\cos 2 \, \theta_R \approx 0$ at least one Yukawa coupling is large, compare eqs.~(\ref{eq:mfyfthRCase1})-(\ref{eq:mfyfthRCase1_strongIO}), while for $|\cos 2 \, \theta_R|$ large smaller values of the Yukawa couplings are accessible and, consequently, also smaller values of the BRs can be obtained. 
 The results for light neutrino masses with IO and $m_0=0.015 \, \mbox{eV}$ are similar and the upper limits obtained for the BRs are only slightly larger than those given in eq.~(\ref{eq:tauupperCase1_NO}). We have also performed numerical scans for $m_0=0$ and light neutrino masses with NO and IO, respectively, and obtain results that are qualitatively similar to those displayed in fig.~\ref{fig:Case1_scan_taudecays}. 

Also for the other cases, Case 2) through Case 3 b.1), the values of the BRs are of similar order as those given in eq.~(\ref{eq:tauupperCase1_NO}) 
for the viable points in the scans and, thus, cannot be tested by current and near-future experiments. We have checked that
for Case 2) the BRs of the different studied cLFV tau lepton decays do not depend on the angle $\theta_R$ for $t$ even that corresponds to $u$ even, whereas the expected dependence on $\theta_R$ is found for $t$ odd equivalent 
to $u$ odd. Furthermore, the exact value of the parameter $s$ is not directly relevant. 
For Case 3 a), we observe that the BRs of the cLFV tau lepton decays also have similar features as the studied $\mu-e$ transitions, i.e.~for 
 $m$ and $s$ being both even or both odd there is no dependence on the angle $\theta_R$, while for $m$ even and $s$ odd or vice versa such a dependence is expected and usually 
a milder one is found in the numerical scans. Lastly, we have checked that statements analogous to those made for the $\mu-e$ transitions also hold for the cLFV tau lepton decays for Case 3 b.1).

%%%%%%%%%%%%%%%%%%%%%%%%%%%%%%%%%%%%%%%%%%%%%%%%%%%%%%%
\mathversion{bold}
\section{Comment on $\beta$ and $0\nu\beta\beta$ decay}
\mathversion{normal}
\label{beta0nubbdecay}
%%%%%%%%%%%%%%%%%%%%%%%%%%%%%%%%%%%%%%%%%%%%%%%%%%%%%%%

Before closing, we briefly comment on the possible impact of the experimental limits on the quantities $m_\beta$ and $m_{\beta\beta}$ extracted from
the study of Tritium $\beta$ decay~\cite{KATRIN:2024cdt} and searches for $0\nu\beta\beta$ decay in different nuclei~\cite{GERDA:2020xhi,CUORE:2021mvw,EXO-200:2019rkq,KamLAND-Zen:2024eml}, respectively.
They can in general constrain the parameter space of the studied scenario in two ways: by delimiting the lightest neutrino mass $m_0$ and by restricting (a combination of) the CP phases. In particular, the
results for $0\nu\beta\beta$ decay have been analysed in a framework with $G_f$ and CP and their residuals $G_e$ and $G_\nu$ in the literature, see e.g.~\cite{DeltaCPothers1,DeltaCPothers3,Hagedorn:2016lva}.
As has been discussed for option 1 of the ISS scenario~\cite{Hagedorn:2021ldq}, the contribution of the heavy sterile states is negligible, since they form pseudo-Dirac pairs and all are (nearly) degenerate in mass. This is also the case for option 2, 
compare eq.~(\ref{eq:massesheavy}). Thus, the results for $m_{\beta\beta}$ are the same as those obtained in e.g.~\cite{Hagedorn:2021ldq,Hagedorn:2016lva}. 
 
Neither the constraints from $m_\beta$ and $m_{\beta\beta}$ on $m_0$ nor on the CP phases are expected to have a relevant impact on the results for the different flavour observables.
As commented in sections~\ref{sec:muetrans} and~\ref{sec:taudecays}, there is in general a minor dependence of the latter observables on $m_0$ and, especially, the order of magnitude of the BRs of the radiative and tri-lepton cLFV
decays as well as of the CR of $\mu-e$ conversion hardly depends on $m_0$. As regards a possible correlation between the size of the CP phases and the BRs and CR, we remind that for Case 1)
the only non-trivial CP phase, the Majorana phase $\alpha$, depends on the parameter $s$ which characterises the CP symmetry, compare section~\ref{sec:leptonmixing}, while we have shown at the beginning of section~\ref{sec:rescase1}
that the matrix $\eta$ and, thus, also the BRs and CR are independent of $s$. Similarly, we see for Case 2) that the Majorana phase $\alpha$ is (mainly) driven by the parameter $v$, encoding the choice of the CP symmetry, while
the matrix $\eta$ is independent of $v$, see beginning of section~\ref{sec:rescase2}. The expressions for the CP phases for Case 3 a) and Case 3 b.1) are more involved, but also for these cases no obvious
correlation between the size of the CP phases and the signal strength of the cLFV processes has been observed. In conclusion, the experiments targeting at $m_\beta$ or $m_{\beta\beta}$ and searches for cLFV 
put complementary constraints on the parameter space of the studied scenario.

%%%%%%%%%%%%%%%%%%%%%%%%%%%%%%%%%%%%%%%%%%%%%%%%%%%%%%%
\section{Summary}
\label{summ}
%%%%%%%%%%%%%%%%%%%%%%%%%%%%%%%%%%%%%%%%%%%%%%%%%%%%%%%

We have studied a scenario in which light neutrino masses are generated via the ISS mechanism and a flavour and CP symmetry determine the flavour structure. The 3+3 gauge singlet
fermions form three pairs of pseudo-Dirac states that are (nearly) degenerate in mass, because in the neutral lepton sector only the Dirac neutrino Yukawa matrix
breaks $G_f$ and CP to $G_\nu$ which is the direct product of a $Z_2$ symmetry and CP. Among charged leptons the residual symmetry is $G_e=Z_3$ and in the chosen
basis the charged lepton mass matrix is diagonal with three free parameters corresponding to the charged lepton masses.  There are four cases, Case 1) through Case 3 b.1), that lead to four different types of 
 lepton mixing patterns. In the neutral lepton sector, the parameter space is spanned by the three Yukawa couplings $y_f$, the two angles $\theta_L$ and $\theta_R$, all contained in the Dirac neutrino Yukawa matrix, 
as well as the two scales $M_0$ and $\mu_0$. The couplings $y_f$ are fixed by the light neutrino masses, the angle $\theta_L$ is adjusted such that the lepton mixing angles
are accommodated well and the two scales $M_0$ and $\mu_0$ determine the mass of the heavy sterile states and the size of lepton number breaking, respectively.
 The angle $\theta_R$ is a free parameter and varied in its entire range, $0 \leq \theta_R \leq 2 \, \pi$.
 
 We have focussed on the BRs of the cLFV decays $\mu\to e \gamma$ and $\mu\to 3 \, e$ and $\mu-e$ conversion in nuclei, both analytically and numerically.
 To study these processes numerically, we have chosen examples for each of the cases which can accommodate the lepton mixing angles well.
  While current limits on these processes do not constrain the studied parameter space, future bounds can exclude part of it.
  In particular, the expected limit on $\mu \to 3 \, e$ from Mu3E Phase 2 and the prospective bounds on $\mu-e$ conversion in aluminium from COMET and Mu2e
  set relevant restrictions. If these are passed, the size of $\mathrm{BR}(\mu\to e \gamma)$ turns out to be about two orders of magnitude smaller than the 
  future limit from MEG II. 
  
  We have also analysed the signal strength of four cLFV tau lepton decays, $\tau \to \mu \gamma$, $\tau \to e \gamma$, $\tau \to 3 \, \mu$ and $\tau \to 3 \, e$.
 If the lepton mixing angles are accommodated well and the future bounds on the three most relevant $\mu-e$ transitions are passed, the BRs of these decays
 are rather suppressed, below $10^{-14}$, meaning these are much lower than the expected limits from Belle II. 
 
 We have observed that the BRs and CR reveal different dependences on the free angle $\theta_R$, depending on the choice of the case, Case 1) through
 Case 3 b.1), and the parameters determining the residual symmetries: for Case 1) they always have such a dependence, while for Case 2) it is decisive whether 
 the parameter $t$ is even or odd, i.e.~for $t$ even there is no such dependence, whereas for $t$ odd a dependence is observed. For Case 3 a) and Case 3 b.1)
 we have found that for the parameters $m$ and $s$ both even or both odd no dependence exists, while for one of them being even and the other one odd a dependence similar
 to the one observed for Case 1) is seen. Nevertheless, it is usually milder for Case 3 a) than for the other cases.
 
 We have also briefly commented on the possible impact of the limits coming from $\beta$ and $0\nu\beta\beta$ decay experiments and have concluded that
these and searches for cLFV processes put complementary constraints on the parameter space.
 
  The phenomenology of this option, option 2, is very different from that of the already studied option 1~\cite{Hagedorn:2021ldq}, where only the Majorana mass matrix $\mu_S$ of the
  gauge singlets $S_j$, $j=1,2,3$, carries non-trivial flavour structure in the neutral lepton sector, since in the case of option 1 all cLFV signals are highly suppressed. 
   For this reason, it can be interesting to consider further options and variants of this scenario.

%%%%%%%%%%%%%%%%%%%%%%%%%%%%%%%%%%%%%%%%%%%%%%%%%%%%%%%
\section*{Acknowledgements}
%%%%%%%%%%%%%%%%%%%%%%%%%%%%%%%%%%%%%%%%%%%%%%%%%%%%%%%

We thank Miguel G.~Folgado for help with the local computer cluster and Robert H.~Bernstein for clarifications regarding the prospects of the experimental reach of searches for $\mu-e$ conversion in aluminium. This work is supported by the Spanish MINECO through the Ram\'o{}n y Cajal programme  RYC2018-024529-I, the FPI fellowship PRE2021-098730, by the national grant PID2020-113644GB-I00, by the Generalitat Valenciana through PROMETEO/2021/083 as well as by the European Union's Horizon 2020 research and innovation programme under the Marie Sk\l{}odowska-Curie grant agreement No.~860881 (HIDDe$\nu$ network) and under the Marie Skłodowska-Curie Staff Exchange grant agreement No.~101086085 (ASYMMETRY).

\appendix

%%%%%%%%%%%%%%%%%%%%%%%%%%%%%%%%%%%%%%%%%%%%%%%%%%%%%%%%%%%%%%%%%%
\mathversion{bold}
\section{Basics of group theory of $\Delta (3 \, n^2)$ and $\Delta (6 \, n^2)$}
\mathversion{normal}
\label{app:grouptheory}
%%%%%%%%%%%%%%%%%%%%%%%%%%%%%%%%%%%%%%%%%%%%%%%%%%%%%%%%%%%%%%%%%%

For completeness, the basics of the discrete groups $\Delta (3 \, n^2)$ and $\Delta (6 \, n^2)$ are presented.
The groups $\Delta (3 \, n^2)$, $n \geq 2$ integer, can be described with three generators $a$, $c$ and $d$ that satisfy the relations
\begin{equation}
\label{eq:D3n2rels}
a^3=e \; , \;\; c^n =e \; , \;\; d^n =e \; , \;\; c \, d =d \, c\; , \;\; a \, c \, a^{-1} = c^{-1} d^{-1} \; , \;\; a \, d \, a^{-1} = c \; ,
\end{equation}
where $e$ is the neutral element of the group~\cite{Luhn:2007uq}. As detailed in~\cite{Escobar:2008vc}, the groups $\Delta (6 \, n^2)$, $n \geq 2$ integer, can be generated 
by adding a further generator $b$. The relations involving this generator read
\begin{equation}
\label{eq:D6n2rels}
b^2=e \; , \;\; (a \, b)^2= e \; , \;\; b \, c \, b^{-1} = d^{-1} \; , \;\; b \, d \, b^{-1} = c^{-1} \, .
\end{equation}
For the trivial representation ${\bf 1}$ all elements of the group are represented by the character $1$.
The representation matrices of the generators, $a ({\bf 3})$, $b ({\bf 3})$, $c ({\bf 3})$ and $d ({\bf 3})$, are taken in one irreducible, faithful, complex three-dimensional representation, called ${\bf 3}$,
of $\Delta (6 \, n^2)$ as\footnote{The similarity transformation $U= \frac{1}{\sqrt{3}} \, \left(
\begin{array}{ccc}
1 & 1 & 1\\
\omega^2 & \omega & 1\\
\omega & \omega^2 & 1
\end{array}
\right)$ has to be applied to the generators given in eq.~(\ref{abcd3}) in order to obtain the form of the representation matrices
as found in~\cite{Escobar:2008vc}.}
\begin{eqnarray}
&&\label{abcd3}
a ({\bf 3}) =  \left( \begin{array}{ccc}
1 & 0 & 0\\
0 & \omega & 0\\
0 & 0 & \omega^2
\end{array}
\right)
\;\; , \;\;
b ({\bf 3}) =  \left( \begin{array}{ccc}
1 & 0 & 0\\
0 & 0 & \omega^2\\
0 & \omega & 0
\end{array}
\right)
\;\; , \\
&&\nonumber
c({\bf 3})= \frac 13 \, \left( \begin{array}{ccc}
1 + 2\cos\phi_n & 1 -\cos\phi_n - \sqrt{3} \sin \phi_n & 1-\cos\phi_n + \sqrt{3} \sin \phi_n \\
1-\cos\phi_n + \sqrt{3} \sin \phi_n &  1 + 2\cos\phi_n & 1 -\cos\phi_n - \sqrt{3} \sin \phi_n\\
1 -\cos\phi_n - \sqrt{3} \sin \phi_n &  1-\cos\phi_n + \sqrt{3} \sin \phi_n & 1 + 2\cos\phi_n
\end{array}
\right)
\end{eqnarray}
with $\omega=e^{\frac{2 \pi i}{3}}$ and $\phi_n = \frac{2 \pi}{n}$. The form of $d ({\bf 3})$ can be calculated using $d ({\bf 3})=a({\bf 3})^2 c ({\bf 3}) a ({\bf 3})$.
We note that the representation ${\bf 3}$ corresponds to ${\bf 3}_{\bf{1} \, (1)}$ in the nomenclature of~\cite{Escobar:2008vc}.

If the index $n$ of the group $\Delta (6 \, n^2)$ is even, there exists an irreducible, unfaithful, real three-dimensional representation ${\bf 3^\prime}$.
The form of the representation matrices  $a ({\bf 3^\prime})$, $b ({\bf 3^\prime})$ and $c ({\bf 3^\prime})$ is then
\begin{equation}
\label{abcd3prime}
a ({\bf 3^\prime}) = a ({\bf 3}) \;\; , \;\; b ({\bf 3^\prime}) = b ({\bf 3}) \;\; , \;\;
c ({\bf 3^\prime}) = \frac 13 \, \left(
\begin{array}{ccc}
-1 & 2 & 2\\
2 & -1 & 2\\
2 & 2 &-1
\end{array}
\right)
\end{equation}
and $d ({\bf 3^\prime}) = a({\bf 3^\prime})^2 c ({\bf 3^\prime}) a ({\bf 3^\prime})$. Note that they do not depend on the index $n$ of the group. 
 The group generated by the representation matrices $g({\bf 3^\prime})$ has 24 elements
and corresponds to the group $\Delta (24)$. The  representation ${\bf 3^\prime}$ together with the one generated by
the representation matrices $a ({\bf 3^\prime})$, $c ({\bf 3^\prime})$, $d ({\bf 3^\prime})$ and $-b ({\bf 3^\prime})$ are the only real three-dimensional representations in a generic group $\Delta (6 \,  n^2)$ with even $n$ and $3 \nmid n$. 
They are denoted as ${\bf 3}_{\bf{1} \, (n/2)}$ and ${\bf 3}_{\bf{2} \, (n/2)}$ in~\cite{Escobar:2008vc}, respectively.

%%%%%%%%%%%%%%%%%%%%%%%%%%%%%%%%%%%%%%%%%%%%%%%%%%%%%%%%%%%%%%%%%%
\mathversion{bold}
\section{Relevant matrices $\Omega ({\bf 3})$, $\Omega ({\bf 3^\prime})$ and $R_{ij} (\theta)$}
\mathversion{normal}
\label{app:matrices}
%%%%%%%%%%%%%%%%%%%%%%%%%%%%%%%%%%%%%%%%%%%%%%%%%%%%%%%%%%%%%%%%%%

As supplementary material the different forms of the matrices $\Omega ({\bf 3})$ and  $\Omega ({\bf 3^\prime})$ for the cases, Case 1) through Case 3 b.1), as well as
the form of the rotation matrices $R_{ij} (\theta_L)$ and $R_{kl} (\theta_R)$ for each combination are collected in this appendix; for further details see~\cite{Drewes:2022kap}.

\noindent For Case 1) the following form of $\Omega ({\bf 3})$ and $\Omega ({\bf 3^\prime})$ is used
\begin{eqnarray}
&&\label{eq:Case1Omega3s}
\Omega(s) ({\bf 3}) = e^{i \, \phi_s} \, U_{\mbox{\scriptsize{TB}}} \,
\left( \begin{array}{ccc}
1 & 0 & 0 \\
0 & e^{-3 \, i \, \phi_s} & 0\\
0 & 0 & -1
\end{array}
\right) \; , 
\\ 
&&\label{eq:Case1Omega3ps}
\Omega(s \, \mbox{even}) ({\bf 3^\prime}) =  U_{\mbox{\scriptsize{TB}}} 
\;\; \mbox{and} \;\;
\Omega(s \, \mbox{odd}) ({\bf 3^\prime}) =  U_{\mbox{\scriptsize{TB}}} \, \left(
\begin{array}{ccc}
i & 0 & 0\\
0 & 1 & 0\\
0 & 0 & i
\end{array}
\right) \; ,
\end{eqnarray}
where the matrix $U_{\mbox{\scriptsize{TB}}}$ encodes tri-bimaximal (TB) mixing, 
\begin{equation}
\label{eq:UTB}
U_{\mbox{\scriptsize{TB}}} =
\left( \begin{array}{ccc}
\sqrt{2/3} & \sqrt{1/3} & 0\\
-\sqrt{1/6} & \sqrt{1/3} & \sqrt{1/2} \\
-\sqrt{1/6} & \sqrt{1/3} & -\sqrt{1/2}
\end{array}
\right) \; ,
\end{equation}
and $\phi_s$ is defined as $\phi_s=\frac{\pi \, s}{n}$. The rotation matrices $R_{ij} (\theta_L)$ and $R_{kl} (\theta_R)$ read
\begin{equation}
\label{eq:R13s}
R_{13} (\theta_L) = \left(
\begin{array}{ccc}
\cos\theta_L & 0 & \sin\theta_L\\
0 & 1 & 0\\
-\sin\theta_L & 0 & \cos\theta_L
\end{array}
\right)
\;\; \mbox{and} \;\;
R_{13} (\theta_R) = \left(
\begin{array}{ccc}
\cos\theta_R & 0 & \sin\theta_R\\
0 & 1 & 0\\
-\sin\theta_R & 0 & \cos\theta_R
\end{array}
\right)\; .
\end{equation}
For Case 2) we use as matrices $\Omega ({\bf 3})$ and $\Omega ({\bf 3^\prime})$ 
\begin{eqnarray}
&&\label{eq:Omega3_Case2}
\Omega (s,t) ({\bf 3}) = \Omega (u,v) ({\bf 3}) =  e^{i \phi_v/6} \, U_{\mbox{\scriptsize{TB}}} \, R_{13} \left( -\frac{\phi_u}{2} \right) \, \left( \begin{array}{ccc}
1 & 0 & 0\\
0 & e^{- i \phi_v/2} & 0\\
0 & 0 & -i
\end{array}
\right) \; ,
\\
&&\Omega (s \, \mbox{even},t \, \mbox{even}) ({\bf 3^\prime}) = U_{\mbox{\scriptsize{TB}}} \, \left( \begin{array}{ccc}
1 & 0 & 0\\
0 & 1 & 0\\
0 & 0 & i
\end{array}
\right) \; ,
\\
&&\Omega (s \, \mbox{even},t \, \mbox{odd}) ({\bf 3^\prime}) = e^{-i \pi/4} \, U_{\mbox{\scriptsize{TB}}} \, R_{13} \left( \frac{\pi}{4} \right) \, \left(
\begin{array}{ccc}
-i & 0 & 0\\
0 & e^{- i \pi/4} & 0\\
0 & 0 & 1
\end{array}
\right) \; ,
\\
&&\Omega (s \, \mbox{odd},t \, \mbox{even}) ({\bf 3^\prime}) = U_{\mbox{\scriptsize{TB}}} \, \left( \begin{array}{ccc}
i & 0 & 0\\
0 & 1 & 0\\
0 & 0 & 1
\end{array}
\right) \; ,
\\
&&\Omega (s \, \mbox{odd},t \, \mbox{odd}) ({\bf 3^\prime}) = e^{- 3 \, i \, \pi/4} \, U_{\mbox{\scriptsize{TB}}} \, R_{13} \left( \frac{\pi}{4} \right) \, \left(
\begin{array}{ccc}
-i & 0 & 0\\
0 & e^{i \, \pi/4} & 0\\
0 & 0 & 1
\end{array}
\right) \; .
\end{eqnarray}
Here, $\phi_u$ and $\phi_v$ are $\phi_u=\frac{\pi \, u}{n}$ and $\phi_v=\frac{\pi \, v}{n}$, respectively. The rotation matrices $R_{ij} (\theta_L)$ and $R_{kl} (\theta_R)$
act both in the (13)-plane, see eq.~(\ref{eq:R13s}).

\noindent The matrices $\Omega ({\bf 3})$ and $\Omega ({\bf 3^\prime})$ for Case 3 a) and Case 3 b.1) are taken to be of the form 
\begin{eqnarray}
&&\Omega (s, m) ({\bf 3}) =e^{i \, \phi_s} \,  \left( \begin{array}{ccc}
1 & 0 & 0\\
0 & \omega & 0 \\
0 & 0 & \omega^2
\end{array}
\right) \, U_{\mbox{\scriptsize{TB}}} \,
\left( \begin{array}{ccc}
1 & 0 & 0\\
0 & e^{-3 \, i \, \phi_s} & 0 \\
0 & 0 & -1
\end{array}
\right) \, R_{13} \left( \phi_m \right) \; ,
\\
&&\Omega (s \, \mbox{even}) ({\bf 3^\prime}) = \left(
\begin{array}{ccc}
1 & 0 & 0\\
0 & \omega & 0\\
0 & 0 & \omega^2
\end{array}
\right) \, U_{\mbox{\scriptsize{TB}}} \, \left(
\begin{array}{ccc}
1 & 0 & 0\\
0 & 1 & 0\\
0 & 0 & -1
\end{array}
\right) \; ,
\\
&&\Omega (s \, \mbox{odd}) ({\bf 3^\prime}) = \left(
\begin{array}{ccc}
1 & 0 & 0\\
0 & \omega & 0\\
0 & 0 & \omega^2
\end{array}
\right) \, U_{\mbox{\scriptsize{TB}}} \, \left(
\begin{array}{ccc}
i & 0 & 0\\
0  & -1 & 0\\
0 & 0 & -i
\end{array}
\right) \; ,
\end{eqnarray}
where $\phi_s$ and $\phi_m$ are given by $\phi_s=\frac{\pi \, s}{n}$ and $\phi_m=\frac{\pi \, m}{n}$, respectively, and $\omega=e^{\frac{2 \pi i}{3}}$.
The rotation matrix $R_{ij} (\theta_L)$ is 
\begin{equation}
R_{12} (\theta_L) =  \left(
\begin{array}{ccc}
\cos\theta_L & \sin\theta_L & 0\\
-\sin\theta_L & \cos\theta_L & 0\\
0 & 0 & 1
\end{array}
\right) \; ,
\end{equation}
while the form of $R_{kl} (\theta_R)$ depends on whether $m$ is even or odd. For $m$ even it is given by
\begin{equation}
R_{kl} (\theta_R)=R_{12} (\theta_R) \;\; \mbox{and} \;\; R_{kl} (\theta_R)=R_{23} (\theta_R)=\left(
\begin{array}{ccc}
1 & 0 & 0\\
0 &  \cos\theta_R & \sin\theta_R\\
0 & -\sin\theta_R & \cos\theta_R
\end{array}
\right)
\end{equation}
for $m$ odd. In the latter situation the permutation matrix $P^{ij}_{kl}$ is also needed and it reads
\begin{equation}
P^{12}_{23}=P_{13}= \left(
\begin{array}{ccc}
0 & 0 & 1\\
0 & 1 & 0\\
1 & 0 & 0
\end{array}
\right) \, .
\end{equation}

%%%%%%%%%%%%%%%%%%%%%%%%%%%%%%%%%%%%%%%%%%%%%%%%%%%%%%%%%%%%%%%%%%
\section{Generators of residual symmetries, CP transformations and form of mixing matrices}
\label{app:residualsCPUPMNS}
%%%%%%%%%%%%%%%%%%%%%%%%%%%%%%%%%%%%%%%%%%%%%%%%%%%%%%%%%%%%%%%%%%

We repeat, for convenience, the explicit form of the generators of the residual symmetries, the CP transformations as well as the form of the mixing patterns
for each of the four different cases, Case 1) through Case 3 b.1). This information is taken from e.g.~\cite{Hagedorn:2014wha,Drewes:2022kap}.
For all cases, the residual symmetry among charged leptons, $G_e$, is a $Z_3$ symmetry, as stated in section~\ref{sec2}. We, thus, focus in this appendix
on $G_\nu$ and the CP transformations.

\paragraph{Case 1)} The generator of the residual $Z_2$ symmetry, contained in $G_\nu$, is chosen as $Z= c^{n/2}$, indicating that the index $n$ of the 
flavour group has to be even. Furthermore, the CP transformation $X$ is given by $X = a \, b \, c^s \, d^{2 \, s} \, X_0$ with $0 \leq s \leq n-1$, such that
we have in general $n$ different CP transformations at our disposal. We remind that the explicit form of $X_0$ also depends on the representation
of the flavour group according to which the particles transform and, e.g.~the form of $X_0 ({\bf 3})$ and $X_0 ({\bf 3^\prime})$ reads
\begin{equation}
\label{eq:formX0}
X_0 ({\bf 3}) = X_0 ({\bf 3^\prime}) =  \left(
\begin{array}{ccc}
1 & 0 & 0\\
0 & 0 & 1\\
0 & 1 & 0
\end{array}
\right) \; .
\end{equation}
The form of the mixing pattern is
\begin{equation}
U_{\mathrm{mix}} = e^{i \, \phi_s} \, U_{\mbox{\scriptsize{TB}}} \,
\left( \begin{array}{ccc}
1 & 0 & 0 \\
0 & e^{-3 \, i \, \phi_s} & 0\\
0 & 0 & -1
\end{array}
\right) \, \left(
\begin{array}{ccc}
\cos\theta & 0 & \sin\theta\\
0 & 1 & 0\\
-\sin\theta & 0 & \cos\theta
\end{array}
\right) \, K_\nu \; ,
\end{equation}
with $K_\nu$ ensuring that light neutrino masses are positive semi-definite  (c.f.~section~\ref{sec2}); for the matrices see also information given in appendix~\ref{app:matrices}.

\paragraph{Case 2)} The generator of the residual $Z_2$ symmetry is the same as for Case 1), $Z=c^{n/2}$. The CP transformation $X$ depends on the two different
parameters $s$ and $t$, i.e.~$X= c^s \, d^t \, X_0$ with $0 \leq s, t \leq n-1$ and $X_0$ given as in eq.~(\ref{eq:formX0}) for the three-dimensional representations ${\bf 3}$ and ${\bf 3^\prime}$.
As mentioned in section~\ref{sec:leptonmixing}, a more convenient choice of parameters than $s$ and $t$ are $u$ and $v$, see eq.~(\ref{eq:defuv}). The mixing matrix takes the form
\begin{equation}
U_{\mathrm{mix}} =
e^{i \phi_v/6} \, U_{\mbox{\scriptsize{TB}}} \, R_{13} \left( -\frac{\phi_u}{2} \right) \, \left( \begin{array}{ccc}
1 & 0 & 0\\
0 & e^{- i \phi_v/2} & 0\\
0 & 0 & -i
\end{array}
\right) \, \left(
\begin{array}{ccc}
\cos\theta & 0 & \sin\theta\\
0 & 1 & 0\\
-\sin\theta & 0 & \cos\theta
\end{array}
\right)
 \, K_\nu \; ,
\end{equation}
with $K_\nu$ determined as described above.

\paragraph{Case 3 a)} For this case the generator of the residual $Z_2$ symmetry depends on an integer $m$ with $0 \leq m \leq n-1$, $Z=b \, c^m \, d^m$.
It is, thus, not necessary that the index $n$ of the flavour group is even. However, now the group has to belong to the series $\Delta (6 \, n^2)$, since only these
groups contain $b$ as generator, compare appendix~\ref{app:grouptheory}. The CP transformation reads $X= b \, c^s \, d^{n-s} \, X_0$ with $0 \leq s \leq n-1$
and hence also depends on one integer parameter. For the form of the mixing matrix we have
\begin{equation}
U_{\mathrm{mix}} = e^{i \, \phi_s} \,  \left( \begin{array}{ccc}
1 & 0 & 0\\
0 & \omega & 0 \\
0 & 0 & \omega^2
\end{array}
\right) \, U_{\mbox{\scriptsize{TB}}} \,
\left( \begin{array}{ccc}
1 & 0 & 0\\
0 & e^{-3 \, i \, \phi_s} & 0 \\
0 & 0 & -1
\end{array}
\right) \, R_{13} \left( \phi_m \right)  \,  \left(
\begin{array}{ccc}
\cos\theta & \sin\theta & 0\\
-\sin\theta & \cos\theta & 0\\
0 & 0 & 1
\end{array}
\right) \, K_\nu \; ,
\end{equation} 
see also appendix~\ref{app:matrices}. As argued in section~\ref{sec:leptonmixing}, the smallness of the reactor mixing angle strongly restricts the viable choice of the parameter $m$,
i.e.~the choice of the residual $Z_2$ symmetry. 

\paragraph{Case 3 b.1)} The residual $Z_2$ symmetry and the CP transformation are the same for Case 3 a) and Case 3 b.1). The mixing matrix differs, since
the columns are permuted through the matrix $P_{\mathrm{cyc}}$,
\begin{equation}
P_{\mathrm{cyc}}= \left(
\begin{array}{ccc}
0 & 1 & 0 \\
0 & 0 & 1 \\
1 & 0 & 0
\end{array}
\right) \; ,
\end{equation}
such that the mixing matrix takes the form
\begin{equation}
U_{\mathrm{mix}} = e^{i \, \phi_s} \,  \left( \begin{array}{ccc}
1 & 0 & 0\\
0 & \omega & 0 \\
0 & 0 & \omega^2
\end{array}
\right) \, U_{\mbox{\scriptsize{TB}}} \,
\left( \begin{array}{ccc}
1 & 0 & 0\\
0 & e^{-3 \, i \, \phi_s} & 0 \\
0 & 0 & -1
\end{array}
\right) \, R_{13} \left( \phi_m \right)  \,  \left(
\begin{array}{ccc}
\cos\theta & \sin\theta & 0\\
-\sin\theta & \cos\theta & 0\\
0 & 0 & 1
\end{array}
\right) \, P_{\mathrm{cyc}} \, K_\nu \; .
\end{equation}
For this case, the parameter $m$ is mainly constrained by the requirement to accommodate well the solar mixing angle, compare section~\ref{sec:leptonmixing}.

%%%%%%%%%%%%%%%%%%%%%%%%%%%%%%%%%%%%%%%%%%%%%%%%%%%%%%%%%%%%%%%%%%
\section{Description of numerical scan}
\label{app:numerics}
%%%%%%%%%%%%%%%%%%%%%%%%%%%%%%%%%%%%%%%%%%%%%%%%%%%%%%%%%%%%%%%%%%

In this appendix, we provide details on the numerical analysis. We use as free parameters the two mass scales $\mu_0$ and $M_0$ and the angle $\theta_R$.
These are varied according to the description found in section~\ref{sec:scan}. Having generated values for these three, we impose a certain light neutrino
mass ordering, NO or IO, and fix the lightest neutrino mass $m_0$ to one of the benchmark values, $m_0^{\mathrm{bm}}$, mentioned in section~\ref{sec:expdata}. With this information,
we determine the three Yukawa couplings $y_f$ using the analytic form of the eigenvalues of the light neutrino mass matrix $m_\nu$ at leading order, see eq.~(\ref{eq:mnuform}) and
 for examples of the eigenvalues eqs.~(\ref{eq:mfyf}) and~(\ref{eq:mfyfthRCase1}) (we remind that their exact form depends in general on the considered case, the group theory parameters, the choice of the 
 CP symmetry, the light neutrino mass ordering, the value of the lightest neutrino mass $m_0$ and potentially on the value of $\theta_R$). This is accomplished by minimising the $\chi^2$-function
\begin{equation}
\label{eq:chi2m}
\chi^2_m (y_f) = \left( \frac{\Delta m_{\mathrm{sol}}^2 (y_f) - (\Delta m_{\mathrm{sol}}^2)^{\mathrm{bf}}}{\sigma_{\Delta m_{\mathrm{sol}}^2}} \right)^2
+ \left( \frac{\Delta m_{\mathrm{atm}}^2 (y_f) - (\Delta m_{\mathrm{atm}}^2)^{\mathrm{bf}}}{\sigma_{\Delta m_{\mathrm{atm}}^2}} \right)^2
+w \, \left( m_0 (y_f) - m_0^{\mathrm{bm}} \right)^2
\end{equation}
with $(\Delta m_{\mathrm{sol}}^2)^{\mathrm{bf}}$ and $(\Delta m_{\mathrm{atm}}^2)^{\mathrm{bf}}$ being the best-fit values of the solar and the atmospheric mass squared difference as well as
$\sigma_{\Delta m_{\mathrm{sol}}^2}$ and $\sigma_{\Delta m_{\mathrm{atm}}^2}$ being the corresponding $1 \, \sigma$ errors~\cite{Esteban:2020cvm}.\footnote{In case the $1 \, \sigma$ errors for e.g.~$\Delta m_{\mathrm{sol}}^2 (y_f) > (\Delta m_{\mathrm{sol}}^2)^{\mathrm{bf}}$
and for $\Delta m_{\mathrm{sol}}^2 (y_f) < (\Delta m_{\mathrm{sol}}^2)^{\mathrm{bf}}$ are different, we use the smaller one of the two.}  
Furthermore, the quantity $w$ is a weighting factor introduced in order to achieve one of the benchmark values of $m_0$ and, for concreteness, we take $w=1000$.
This minimisation is performed with the function \verb1differential_evolution1 of the Python package \verb3NumPy3.
In the next step, we consider the nine-by-nine mass matrix ${\cal M}_{\mathrm{Maj}}$, see eq.~(\ref{eq:MMaj}), with the computed values of the Yukawa couplings and the chosen values of $\mu_0$, $M_0$ and $\theta_R$. 
We fix $\theta_L$ such that the $\chi^2$-function 
\begin{equation}
\label{eq:chi2theta}
\chi^2_\theta (\theta_L) =  \left( \frac{\sin^2 \theta_{12} (\theta_L) - (\sin^2 \theta_{12})^{\mathrm{bf}}}{\sigma_{\sin^2 \theta_{12}}} \right)^2
+ \, \left( \frac{\sin^2 \theta_{13} (\theta_L) - (\sin^2 \theta_{13})^{\mathrm{bf}}}{\sigma_{\sin^2 \theta_{13}}} \right)^2
+  \,\left( \frac{\sin^2 \theta_{23} (\theta_L) - (\sin^2 \theta_{23})^{\mathrm{bf}}}{\sigma_{\sin^2 \theta_{23}}} \right)^2
\end{equation}
is minimised. The quantities $(\sin^2 \theta_{ij})^{\mathrm{bf}}$ denote the experimental best-fit values and $\sigma_{\sin^2 \theta_{ij}} $ the corresponding $1 \, \sigma$ errors~\cite{Esteban:2020cvm}. 
In the case of the ISS mechanism, i.e.~for $|\mu_S| \ll |m_D| \ll |M_{NS}|$ corresponding to $\mu_0 \ll y_f \, \langle H \rangle \ll M_0$, the eigenvalues of the matrix ${\cal M}_{\mathrm{Maj}}$ are strongly hierarchical (three are of order sub-eV
and six of order sub-TeV to TeV). This requires a high precision in the diagonalisation of this matrix and the extraction of the unitary matrix ${\cal U}$ and, consequently, $\tilde{U}_\nu$, see eqs.~(\ref{eq:UMdiag}) and~(\ref{eq:formU}). 
 It can be achieved by setting the precision to 60 digits using the library \verb1mpmath1.\footnote{Using an even higher precision does not affect the results.} The diagonalisation itself is performed with the function 
\verb3eighe3 of this library. We have checked that the resulting light neutrino masses are consistent with the chosen benchmark value of $m_0$ and the best-fit values of the mass squared differences such that 
$\chi^2_m (y_f) \lesssim 0.1$. The Yukawa couplings $y_f$ and the angle $\theta_L$ could be fixed in one step by diagonalising the matrix ${\cal M}_{\mathrm{Maj}}$, extracting the eigenvalues and the matrix ${\cal U}$,
and minimising $\chi^2_\mathrm{tot}  (y_f, \theta_L) = \chi^2_m (y_f, \theta_L) + \chi^2_\theta (y_f, \theta_L)$ with $\chi^2_m$ and $\chi^2_\theta$ defined analogously as in eqs.~(\ref{eq:chi2m}) and~(\ref{eq:chi2theta}), respectively.
However, this requires an increased time of computation and, at the same time, leads to results very similar to those obtained by the described procedure.

After having adjusted $y_f$ and $\theta_L$ all parameters of the neutral lepton sector are determined and the BRs and CR are computed with the formulae found in~\cite{Ilakovac:1994kj,Alonso:2012ji}.

%%%%%%%%%%%%%%%%%%%%%%%%%%%%%%%%%%%%%%%%%%%%%%%%%%%%%%%
\section{Supplementary plots for Case 2) and Case 3 b.1)}
\label{app:Case2Case3b1}
%%%%%%%%%%%%%%%%%%%%%%%%%%%%%%%%%%%%%%%%%%%%%%%%%%%%%%%

\begin{figure}
    \centering
    \includegraphics[width=\textwidth]{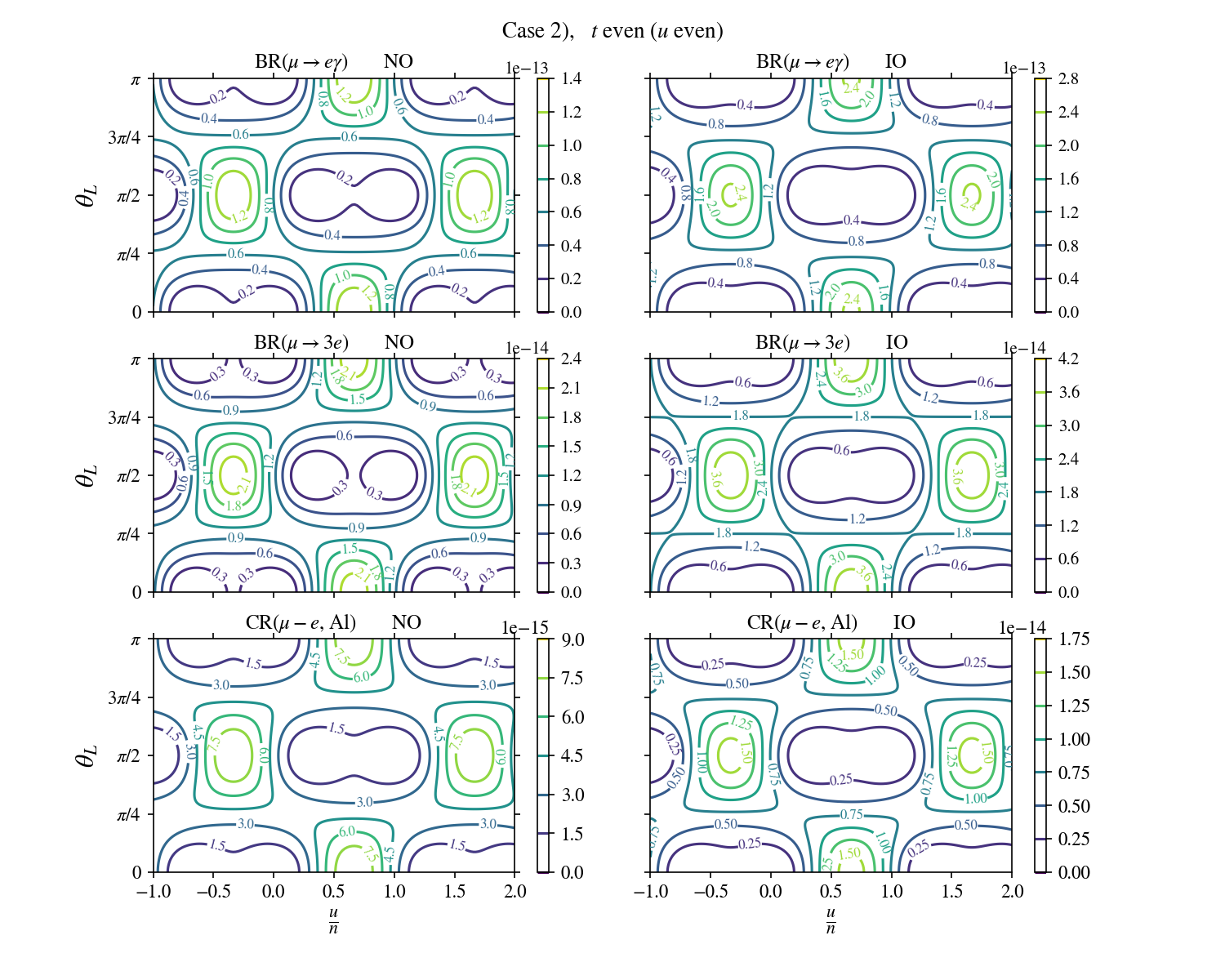}
    \caption{{\bf Case 2), \mathversion{bold}$t$ even ($u$ even). Predictions for $\mathrm{BR} (\mu\to e \gamma)$, $\mathrm{BR} (\mu\to 3 \, e)$ and $\mathrm{CR} (\mu-e, \mathrm{Al})$ in the $\frac un-\theta_L$-plane} in the upper, 
    middle and lower row.\mathversion{normal}  Left (right) plots are for light neutrino masses with NO (IO)
  and $m_0= 0.03 \, (0.015) \, \mathrm{eV}$. The scales $\mu_0$ and $M_0$ are set to $\mu_0=1 \, \mbox{keV}$ and $M_0=3 \, \mbox{TeV}$, respectively. In this simplified version of fig.~\ref{fig:Case2_un_thetaL_teven} we only display
  the contour lines for the different flavour observables.}
\label{fig:Case2_un_thetaL_teven_simplified_contours}
\end{figure}

\begin{figure}
    \centering
    \includegraphics[width=\textwidth]{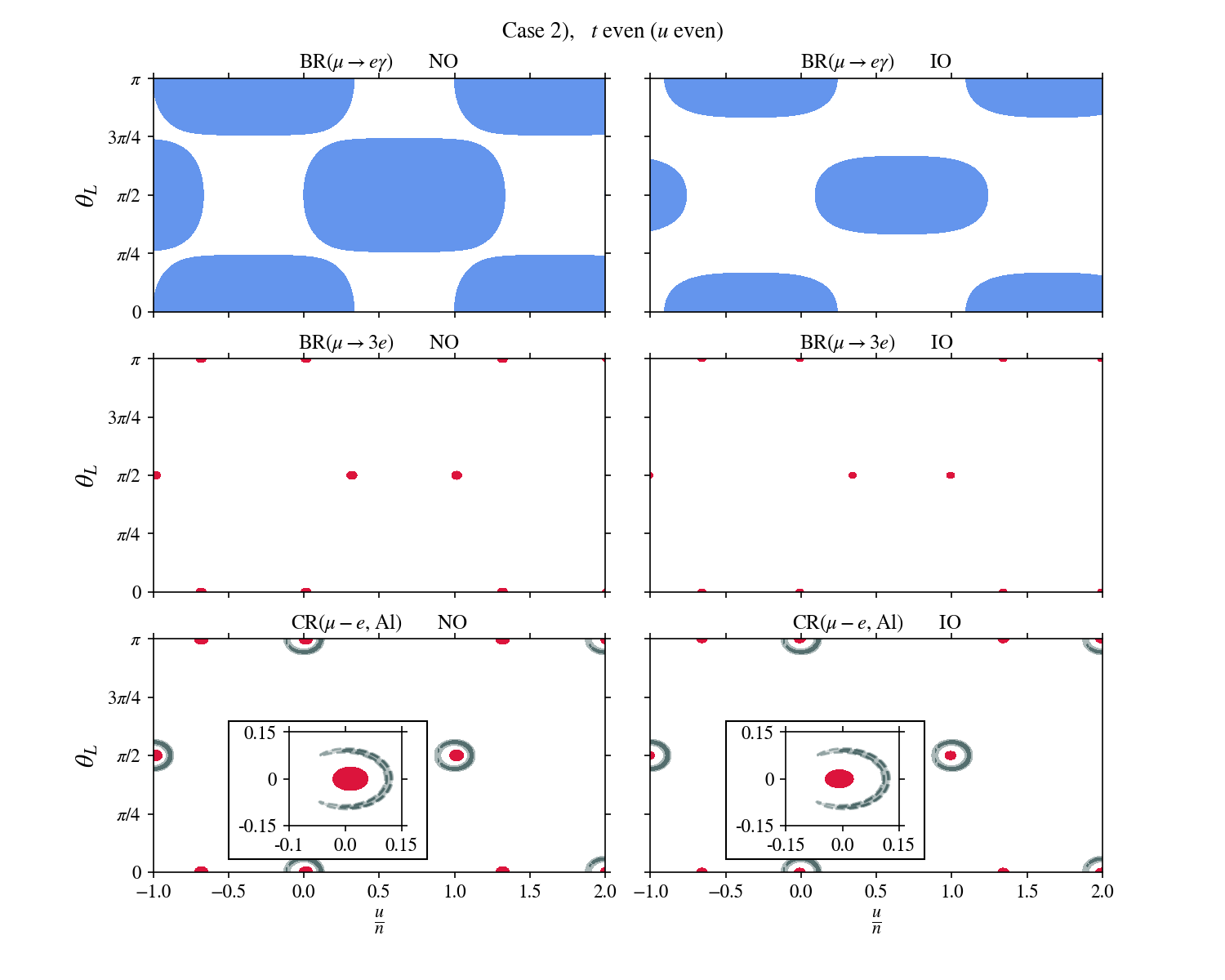}
    \caption{{\bf Case 2), \mathversion{bold}$t$ even ($u$ even). Predictions for $\mathrm{BR} (\mu\to e \gamma)$, $\mathrm{BR} (\mu\to 3 \, e)$ and $\mathrm{CR} (\mu-e, \mathrm{Al})$ in the $\frac un-\theta_L$-plane} in the upper, 
    middle and lower row.\mathversion{normal}  Left (right) plots are for light neutrino masses with NO (IO)
  and $m_0= 0.03 \, (0.015) \, \mathrm{eV}$. The scales $\mu_0$ and $M_0$ are set to $\mu_0=1 \, \mbox{keV}$ and $M_0=3 \, \mbox{TeV}$, respectively. In this simplified version of fig.~\ref{fig:Case2_un_thetaL_teven} we only show
  the regions compatible with the strongest experimental prospective bounds, together with the constraints from lepton mixing in the plots for $\mu-e$ conversion in aluminium. The colour-coding is the same as in fig.~\ref{fig:Case2_un_thetaL_teven}.}
\label{fig:Case2_un_thetaL_teven_simplified_limits}
\end{figure}

\begin{figure}
    \centering
    \includegraphics[width=\textwidth]{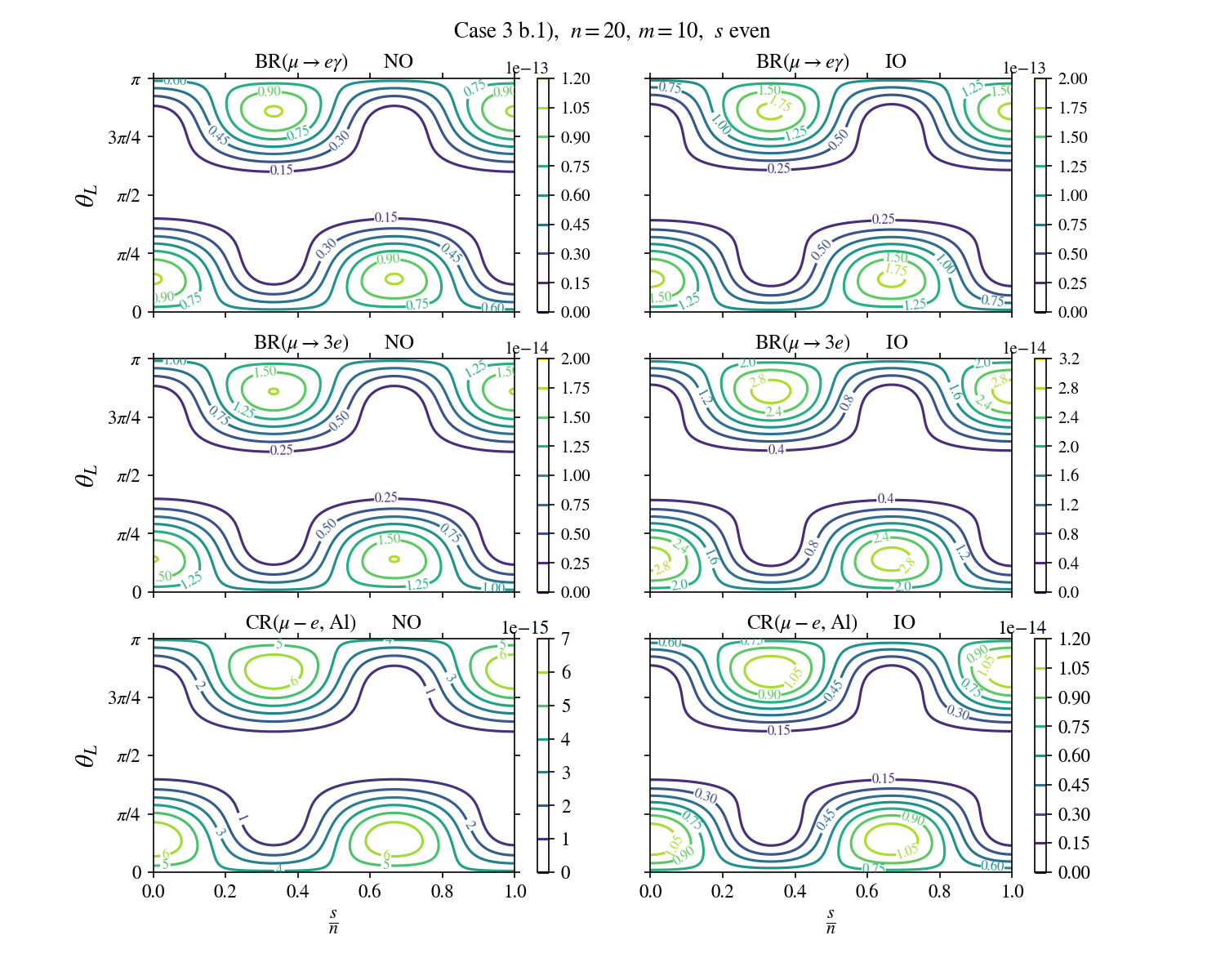}
    \caption{{\bf Case 3 b.1), \mathversion{bold}$m$ and $s$ both even. Predictions for $\mathrm{BR} (\mu\to e \gamma)$, $\mathrm{BR} (\mu\to 3 \, e)$ and $\mathrm{CR} (\mu-e, \mathrm{Al})$ in the $\frac sn-\theta_L$-plane} in the upper, 
    middle and lower row.\mathversion{normal}  The group theory parameters are chosen as $n=20$ and $m=10$. In this simplified version of fig.~\ref{fig:Case3b1_sn_thetaL_m10seven} we only display
  the contour lines for the different flavour observables.}
\label{fig:Case3b1_sn_thetaL_m10seven_simplified_contours}
\end{figure}

\begin{figure}
    \centering
    \includegraphics[width=\textwidth]{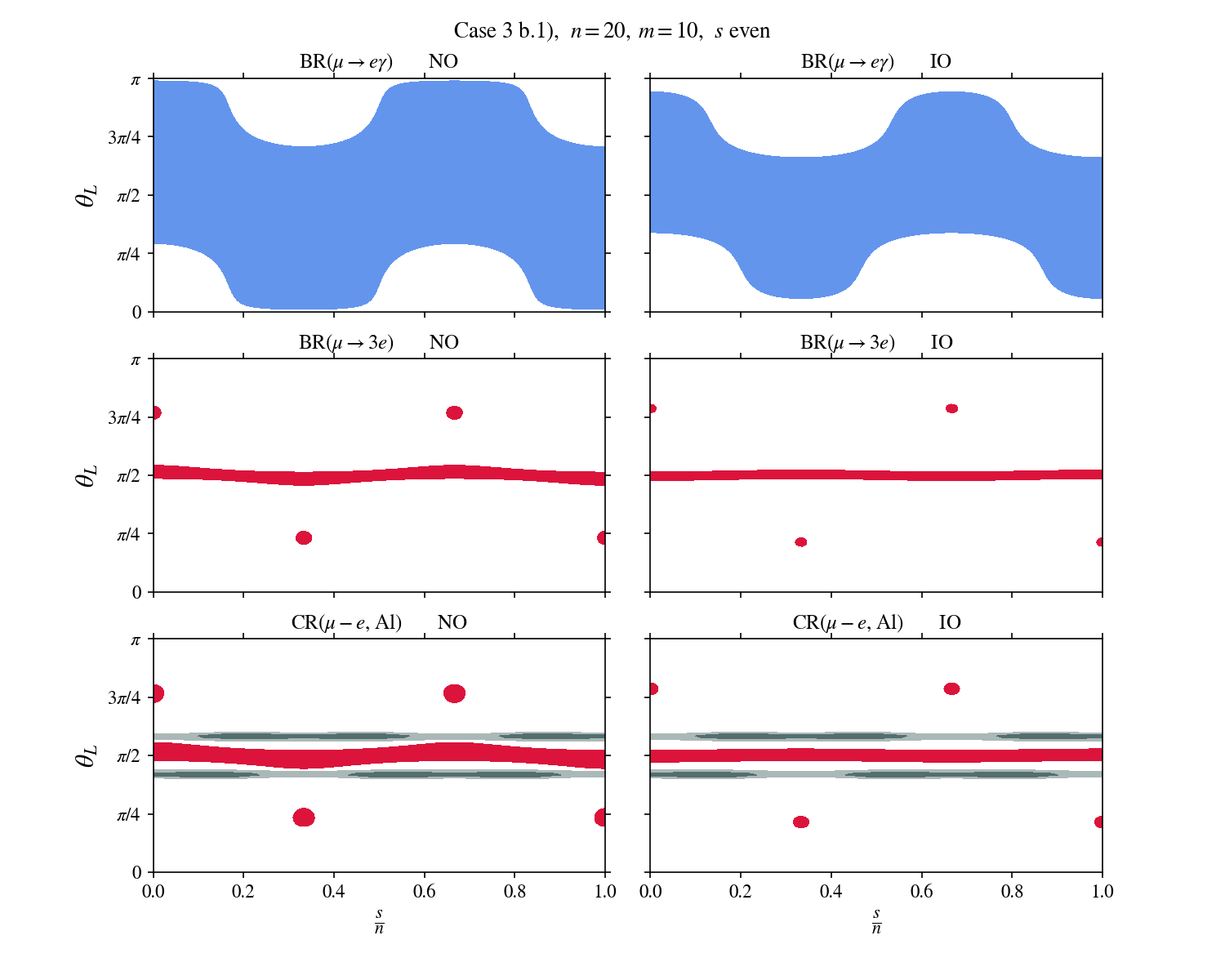}
    \caption{{\bf Case 3 b.1), \mathversion{bold}$m$ and $s$ both even. Predictions for $\mathrm{BR} (\mu\to e \gamma)$, $\mathrm{BR} (\mu\to 3 \, e)$ and $\mathrm{CR} (\mu-e, \mathrm{Al})$ in the $\frac sn-\theta_L$-plane} in the upper, 
    middle and lower row.\mathversion{normal}  The group theory parameters are chosen as $n=20$ and $m=10$. In this simplified version of fig.~\ref{fig:Case3b1_sn_thetaL_m10seven} we only show
  the regions compatible with the strongest experimental prospective bounds, together with the constraints from lepton mixing in the plots for $\mu-e$ conversion in aluminium. The colour-coding is the same as in fig.~\ref{fig:Case3b1_sn_thetaL_m10seven}.}
\label{fig:Case3b1_sn_thetaL_m10seven_simplified_limits}
\end{figure}

\begin{figure}
    \centering
    \includegraphics[width=\textwidth]{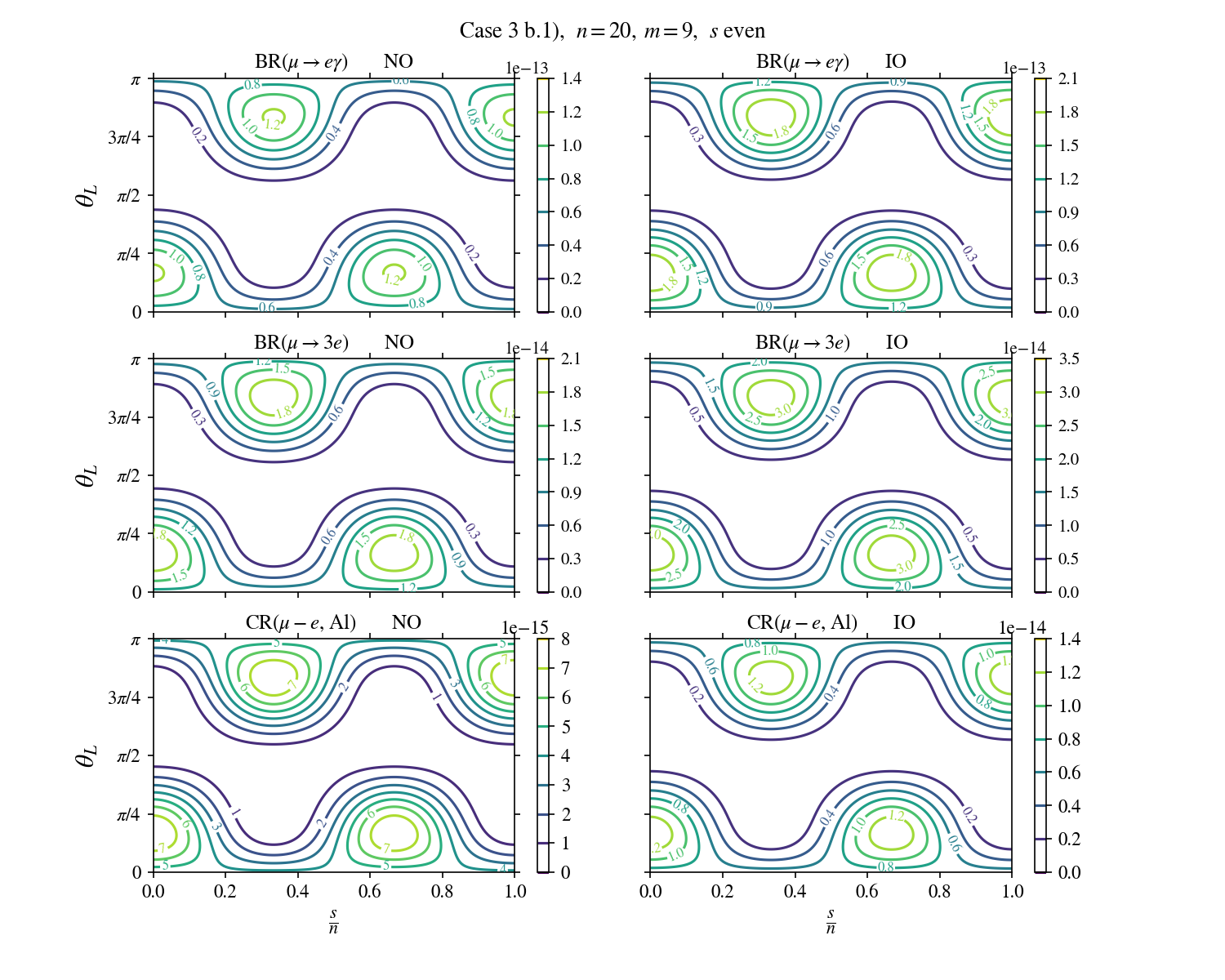}
    \caption{{\bf Case 3 b.1), \mathversion{bold}$m$ odd, $s$ even and $\theta_R=0$. Predictions for $\mathrm{BR} (\mu\to e \gamma)$, $\mathrm{BR} (\mu\to 3 \, e)$ and $\mathrm{CR} (\mu-e, \mathrm{Al})$ in the $\frac sn-\theta_L$-plane} in the upper, 
    middle and lower row.\mathversion{normal}  The group theory parameters are chosen as $n=20$ and $m=9$. In this simplified version of fig.~\ref{fig:Case3b1_sn_thetaL_m9seven} we only display
  the contour lines for the different flavour observables.}
\label{fig:Case3b1_sn_thetaL_m9seven_simplified_contours}
\end{figure}

\begin{figure}
    \centering
    \includegraphics[width=\textwidth]{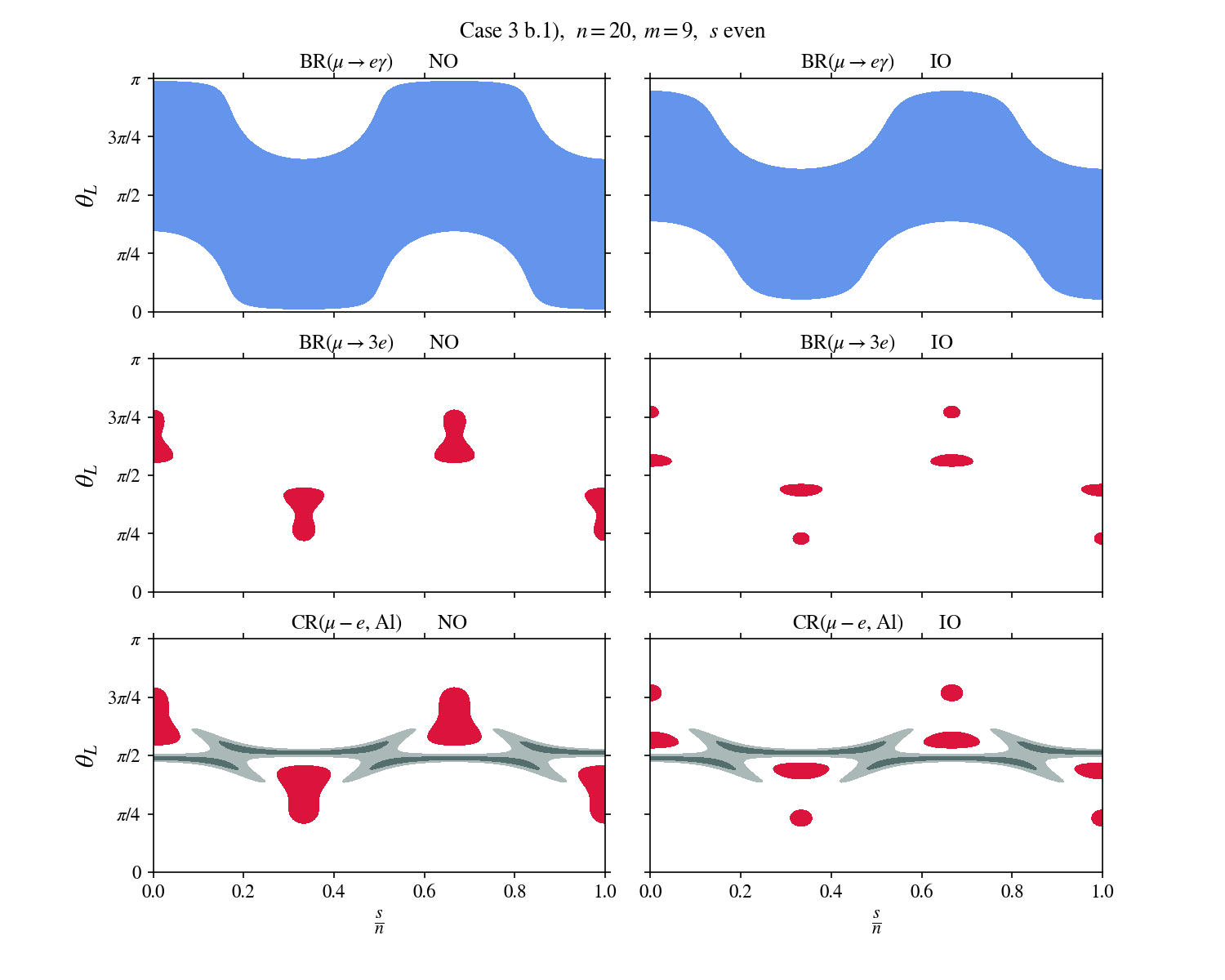}
    \caption{{\bf Case 3 b.1), \mathversion{bold}$m$ odd, $s$ even and $\theta_R=0$. Predictions for $\mathrm{BR} (\mu\to e \gamma)$, $\mathrm{BR} (\mu\to 3 \, e)$ and $\mathrm{CR} (\mu-e, \mathrm{Al})$ in the $\frac sn-\theta_L$-plane} in the upper, 
    middle and lower row.\mathversion{normal}  The group theory parameters are chosen as $n=20$ and $m=9$. In this simplified version of fig.~\ref{fig:Case3b1_sn_thetaL_m9seven} we only show
  the regions compatible with the strongest experimental prospective bounds, together with the constraints from lepton mixing in the plots for $\mu-e$ conversion in aluminium. The colour-coding is the same as in fig.~\ref{fig:Case3b1_sn_thetaL_m9seven}.}
\label{fig:Case3b1_sn_thetaL_m9seven_simplified_limits}
\end{figure}

\begin{figure}
    \centering
    \includegraphics[width=\textwidth]{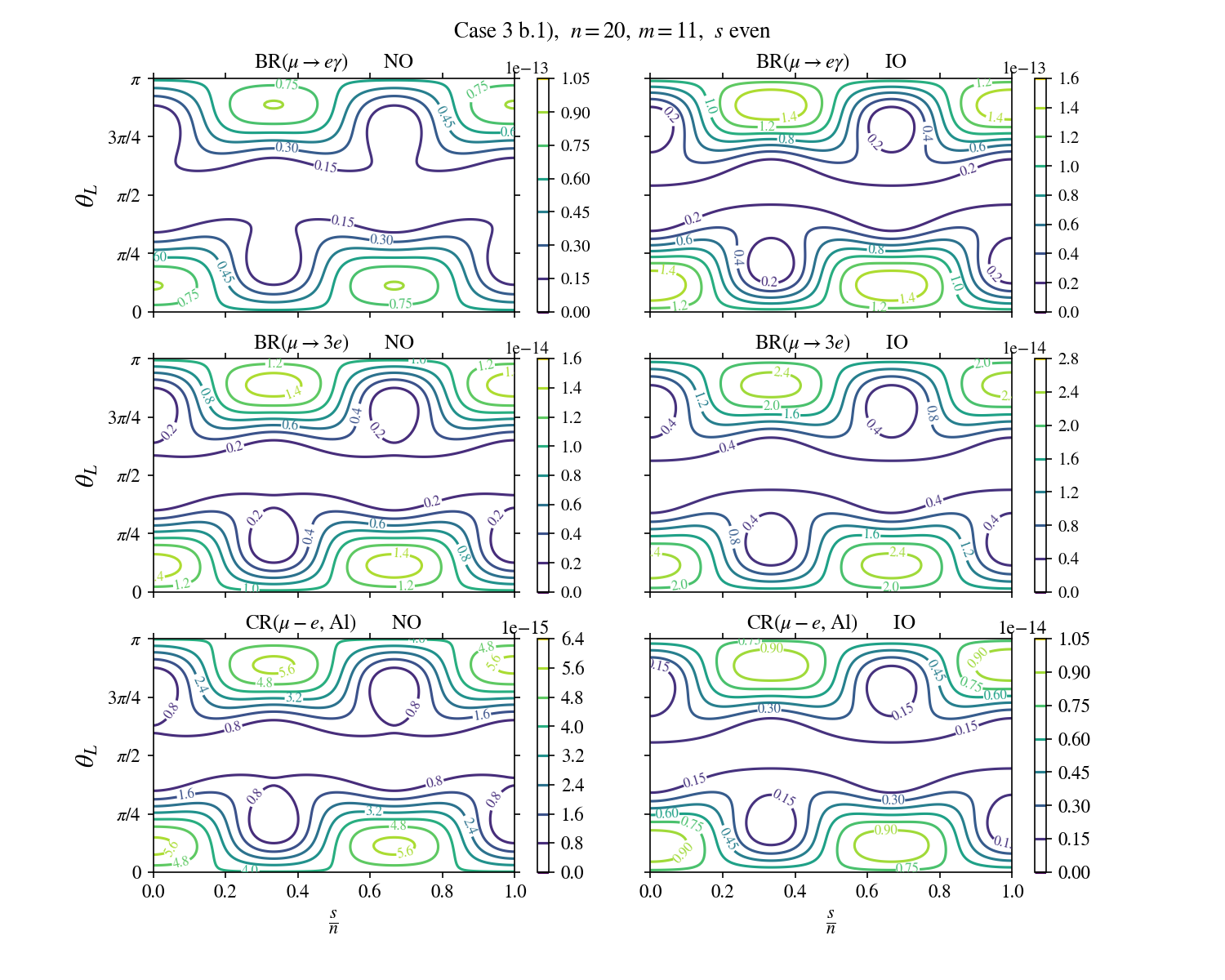}
    \caption{{\bf Case 3 b.1), \mathversion{bold}$m$ odd, $s$ even and $\theta_R=0$. Predictions for $\mathrm{BR} (\mu\to e \gamma)$, $\mathrm{BR} (\mu\to 3 \, e)$ and $\mathrm{CR} (\mu-e, \mathrm{Al})$ in the $\frac sn-\theta_L$-plane} in the upper, 
    middle and lower row.\mathversion{normal}  The group theory parameters are chosen as $n=20$ and $m=11$. In this simplified version of fig.~\ref{fig:Case3b1_sn_thetaL_m11seven} we only display
  the contour lines for the different flavour observables.}
\label{fig:Case3b1_sn_thetaL_m11seven_simplified_contours}
\end{figure}

\begin{figure}
    \centering
    \includegraphics[width=\textwidth]{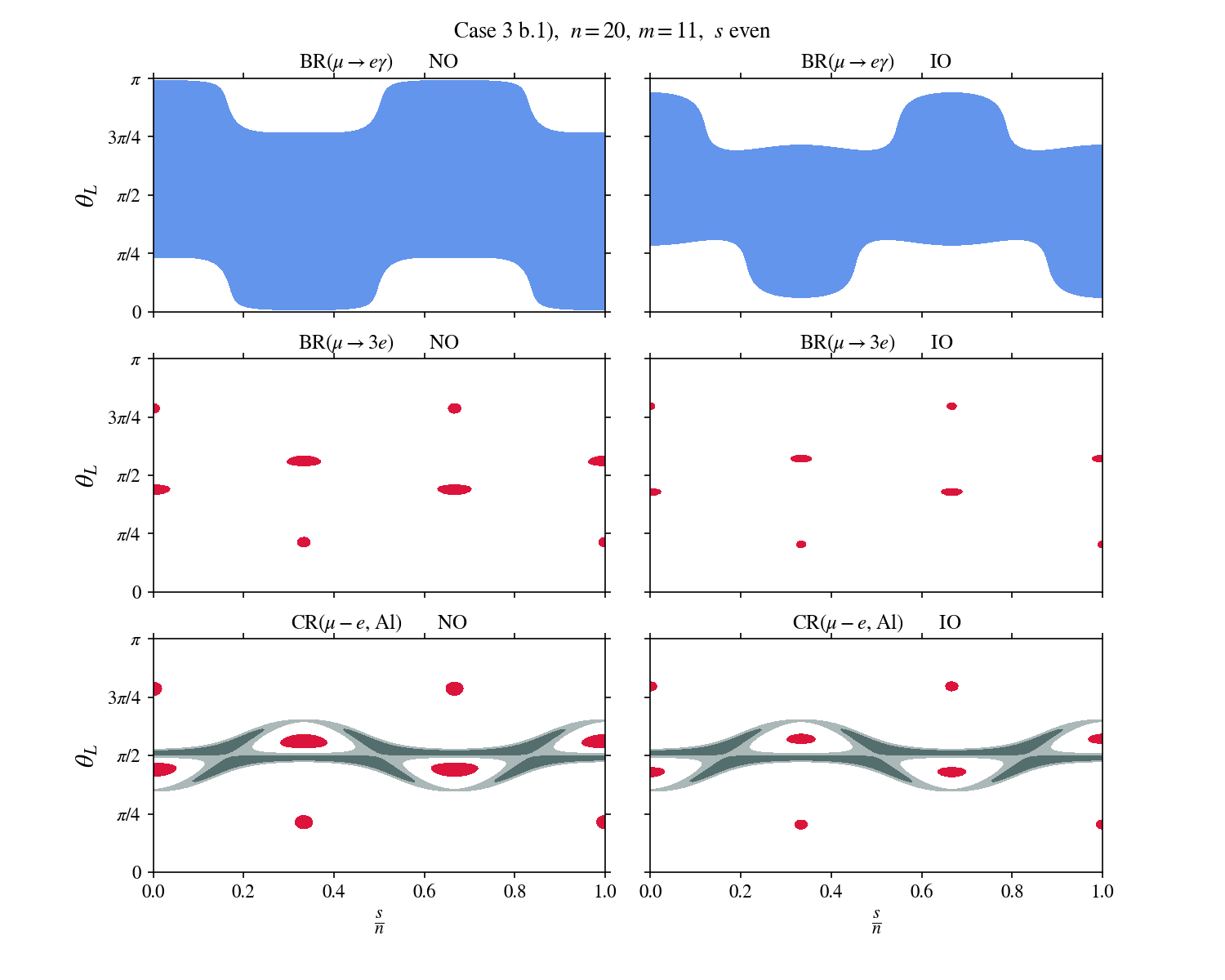}
    \caption{{\bf Case 3 b.1), \mathversion{bold}$m$ odd, $s$ even and $\theta_R=0$. Predictions for $\mathrm{BR} (\mu\to e \gamma)$, $\mathrm{BR} (\mu\to 3 \, e)$ and $\mathrm{CR} (\mu-e, \mathrm{Al})$ in the $\frac sn-\theta_L$-plane} in the upper, 
    middle and lower row.\mathversion{normal}  The group theory parameters are chosen as $n=20$ and $m=11$. In this simplified version of fig.~\ref{fig:Case3b1_sn_thetaL_m11seven} we only show
  the regions compatible with the strongest experimental prospective bounds, together with the constraints from lepton mixing in the plots for $\mu-e$ conversion. The colour-coding is the same as in fig.~\ref{fig:Case3b1_sn_thetaL_m11seven}.}
\label{fig:Case3b1_sn_thetaL_m11seven_simplified_limits}
\end{figure}

\begin{figure}
    \centering
    \includegraphics[width=\textwidth]{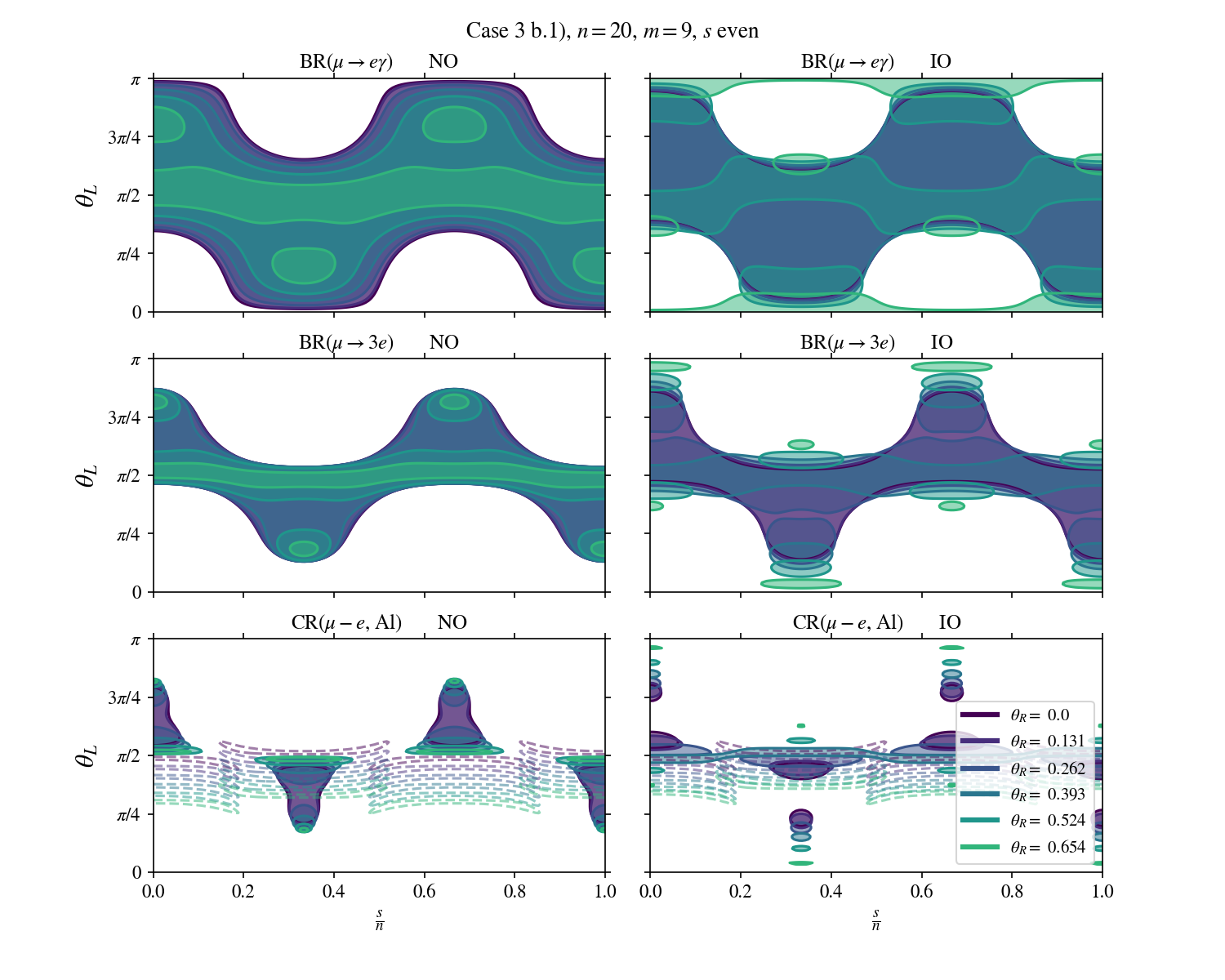}
    \caption{{\bf Case 3 b.1), \mathversion{bold}$m$ odd and $s$ even. Predictions for $\mathrm{BR} (\mu\to e \gamma)$, $\mathrm{BR} (\mu\to 3 \, e)$ and $\mathrm{CR} (\mu-e, \mathrm{Al})$ in the $\frac sn-\theta_L$-plane} in the upper, 
    middle and lower row for six different values of $\theta_R$.\mathversion{normal} The group theory parameters are chosen as $n=20$ and $m=9$. Conventions are analogous as in fig.~\ref{fig:Case2_un_thetaL_todd} for Case 2) and $t$ odd ($u$ odd).
    In fig.~\ref{fig:Case3b1_sn_thetaL_m9seven} in the main text the plots for $\theta_R=0$ are displayed.}
\label{fig:Case3b1_sn_thetaL_m9seven_thR}
\end{figure}

\begin{figure}
    \centering
    \includegraphics[width=\textwidth]{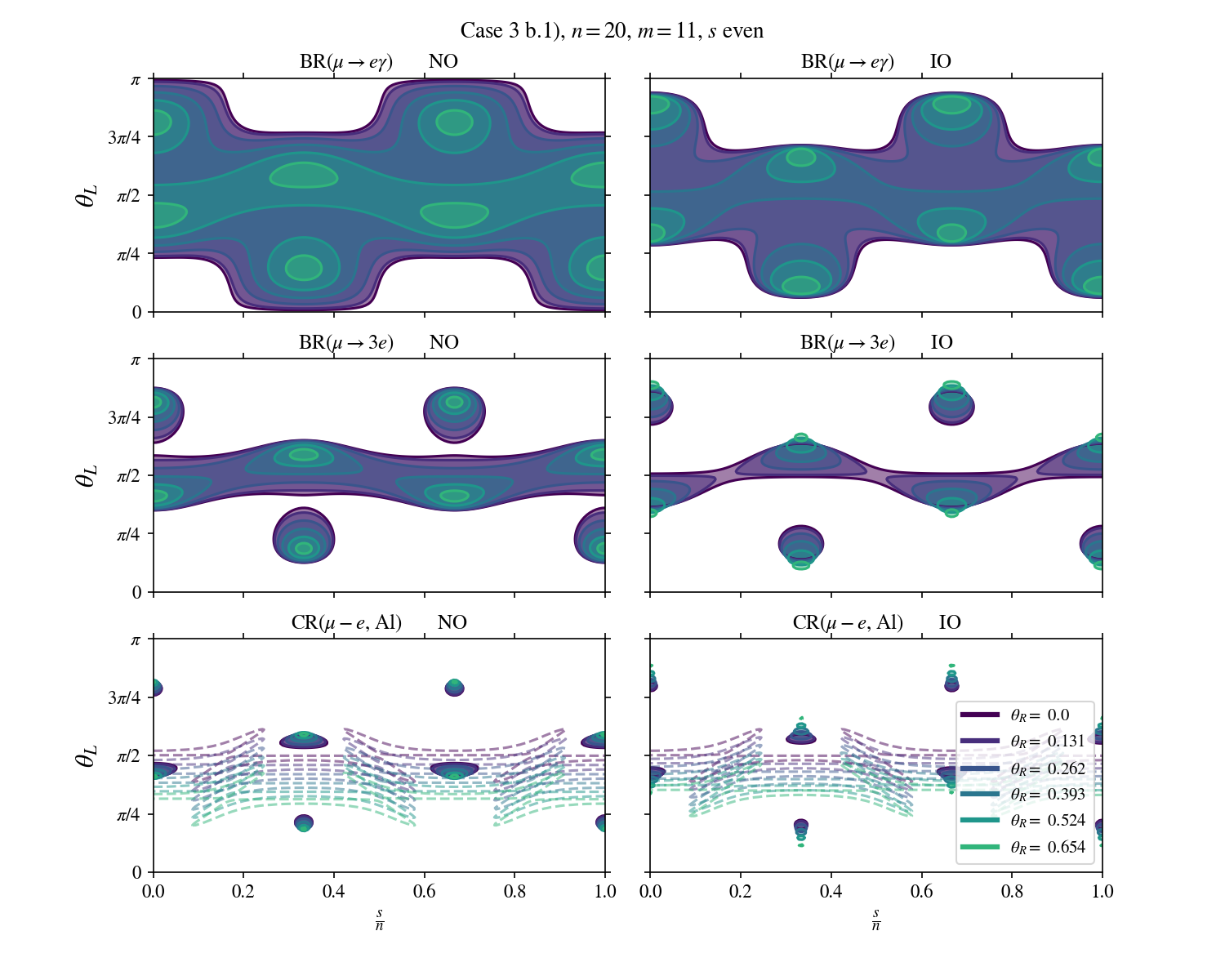}
    \caption{{\bf Case 3 b.1), \mathversion{bold}$m$ odd and $s$ even. Predictions for $\mathrm{BR} (\mu\to e \gamma)$, $\mathrm{BR} (\mu\to 3 \, e)$ and $\mathrm{CR} (\mu-e, \mathrm{Al})$ in the $\frac sn-\theta_L$-plane} in the upper, 
    middle and lower row for six different values of $\theta_R$.\mathversion{normal} The group theory parameters are chosen as $n=20$ and $m=11$. For conventions see fig.~\ref{fig:Case3b1_sn_thetaL_m9seven_thR}.
    In fig.~\ref{fig:Case3b1_sn_thetaL_m11seven} in the main text the plots for $\theta_R=0$ are displayed.}
\label{fig:Case3b1_sn_thetaL_m11seven_thR}
\end{figure}

\begin{figure}
    \centering
    \includegraphics[width=\textwidth]{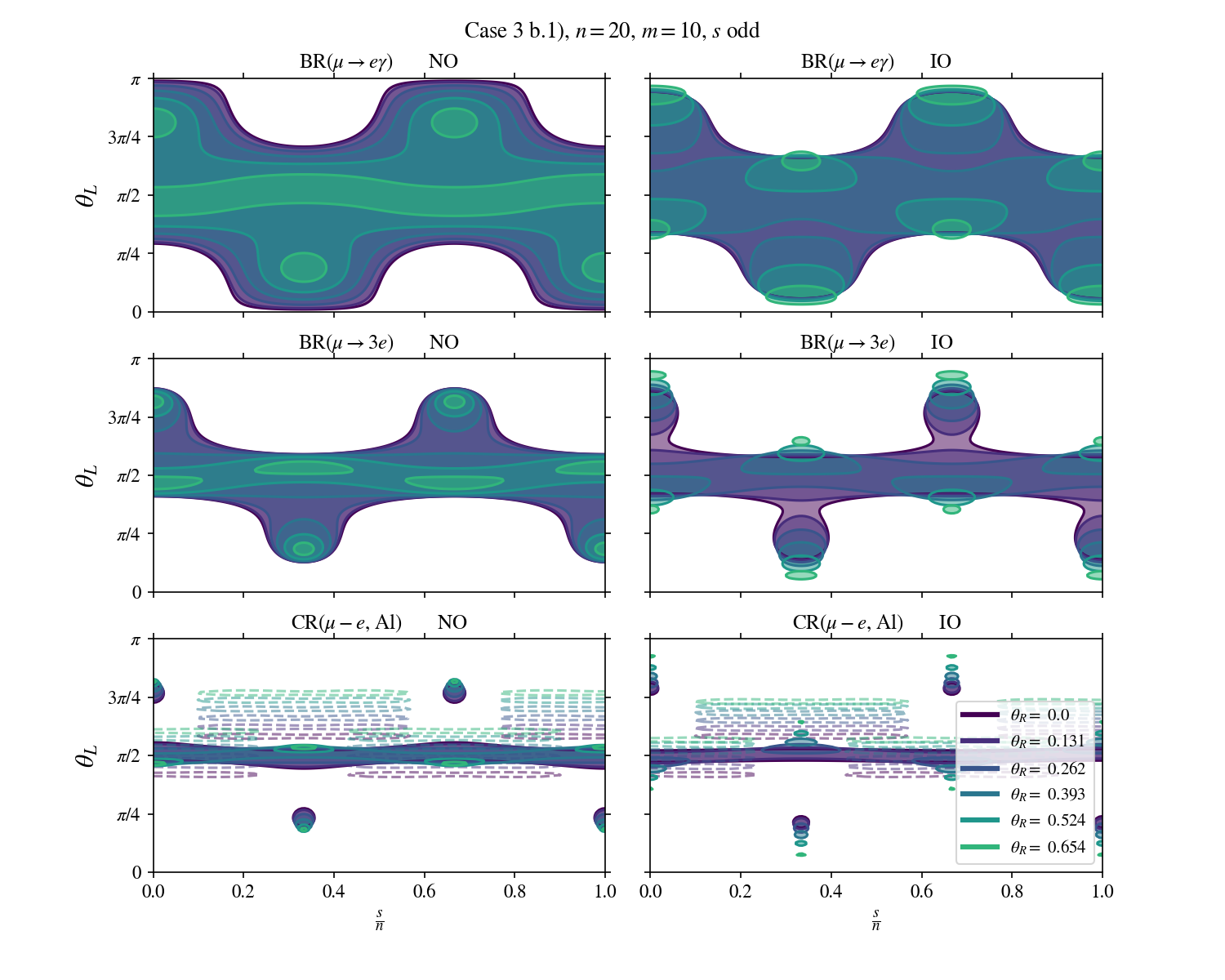}
    \caption{{\bf Case 3 b.1), \mathversion{bold}$m$ even and $s$ odd. Predictions for $\mathrm{BR} (\mu\to e \gamma)$, $\mathrm{BR} (\mu\to 3 \, e)$ and $\mathrm{CR} (\mu-e, \mathrm{Al})$ in the $\frac sn-\theta_L$-plane} in the upper, 
    middle and lower row for six different values of $\theta_R$.\mathversion{normal} The group theory parameters are chosen as $n=20$ and $m=10$. For conventions see fig.~\ref{fig:Case3b1_sn_thetaL_m9seven_thR}.
    The plots shown in fig.~\ref{fig:Case3b1_sn_thetaL_m10seven} in the main text are very similar to the ones obtained for the specific choice $\theta_R=0$ for $m=10$ and $s$ odd.}
\label{fig:Case3b1_sn_thetaL_m10sodd_thR} 
\end{figure}

Here, we display further plots. On the one hand, simplified versions of figs.~\ref{fig:Case2_un_thetaL_teven},~\ref{fig:Case3b1_sn_thetaL_m10seven},~\ref{fig:Case3b1_sn_thetaL_m9seven} and~\ref{fig:Case3b1_sn_thetaL_m11seven} can be found in which either only the contour lines are displayed, see figs.~\ref{fig:Case2_un_thetaL_teven_simplified_contours},~\ref{fig:Case3b1_sn_thetaL_m10seven_simplified_contours},~\ref{fig:Case3b1_sn_thetaL_m9seven_simplified_contours} and~\ref{fig:Case3b1_sn_thetaL_m11seven_simplified_contours}, 
or the regions are highlighted in which the prospective limit on $\mathrm{BR} (\mu\to e \gamma)$ of MEG II is passed at the $1 \, \sigma$ level in the upper row, 
the future bound expected on $\mathrm{BR} (\mu\to 3 \, e)$ from Mu3E Phase 2 is fulfilled at the $1 \, \sigma$ level in the middle row as well as the expected limit on $\mathrm{CR} (\mu-e, \mathrm{Al})$ from COMET is respected at the $1 \, \sigma$ level in
 the lower row, together with the constraints from lepton mixing, see figs.~\ref{fig:Case2_un_thetaL_teven_simplified_limits},~\ref{fig:Case3b1_sn_thetaL_m10seven_simplified_limits},~\ref{fig:Case3b1_sn_thetaL_m9seven_simplified_limits} and~\ref{fig:Case3b1_sn_thetaL_m11seven_simplified_limits}. Note that we use the same colour-coding as in the figures in the main text.
On the other hand, three further figures for Case 3 b.1) are shown that evidence the dependence of the BRs and CR on the free angle $\theta_R$ in the $\frac sn-\theta_L$-plane for examples of the combinations $m$ odd and $s$ even, see figs.~\ref{fig:Case3b1_sn_thetaL_m9seven_thR} and~\ref{fig:Case3b1_sn_thetaL_m11seven_thR}, or vice versa, see fig.~\ref{fig:Case3b1_sn_thetaL_m10sodd_thR}.


\end{document}